\UseRawInputEncoding
\documentclass{aastex63}

\newcommand{\revision}[1]{\textcolor{black}{#1}}
\submitjournal{ApJ}

\shorttitle{Distributions of crust models}
\shortauthors{Balliet et al.}
\graphicspath{{./}{figures/}}
\usepackage{comment}
\usepackage{amsmath}
\begin{document}

\title{Prior probability distributions of neutron star crust models}

\correspondingauthor{William Newton}
\email{william.newton@tamuc.edu}

\author{Lauren E. Balliet}
\affiliation{Department of Physics and Astronomy, Texas A\&M University-Commerce, Commerce, TX 75429-3011, USA}

\author{William G. Newton}
\affiliation{Department of Physics and Astronomy, Texas A\&M University-Commerce, Commerce, TX 75429-3011, USA}

\author{Sarah Cantu}
\affiliation{Department of Physics and Astronomy, Texas A\&M University-Commerce, Commerce, TX 75429-3011, USA}

\author{Srdan Budimir}
\affiliation{Department of Physics and Astronomy, Texas A\&M University-Commerce, Commerce, TX 75429-3011, USA}




\begin{abstract}

To make best use of multi-faceted astronomical and nuclear data-sets,  probability distributions of neutron star models that can be used to propagate errors consistently from one domain to another are required. We take steps toward a consistent model for this purpose, highlight where model inconsistencies occur and assess the resulting model uncertainty. Using two distributions of nuclear symmetry energy parameters - one uniform, the other based on pure neutron matter theory, we prepare two ensembles of neutron star inner crust models. We use an extended Skyrme energy-density functional within a compressible liquid drop model (CLDM). We fit the surface parameters of the CLDM to quantum 3D Hartree-Fock calculations of crustal nuclei. All models predict more than 50\% of the crust by mass and 15\% of the crust by thickness comprises pasta with medians of around 62\% and 30\% respectively. We also present 68\% and 95\% ranges for the crust  composition as a function of density. We examine the relationships between crust-core boundary and pasta transition properties, the thickness of the pasta layers, the symmetry energy at saturation and sub-saturation densities and the neutron skins of $^{208}$Pb  and $^{48}$Ca. We quantify the correlations using the maximal information coefficient, which can effectively characterize non-linear relationships. Future measurements of neutron skins, information from nuclear masses and giant resonances, and theoretical constraints on PNM will be able to place constraints on the location of the pasta and crust-core boundaries and the amount of pasta in the crust.

\end{abstract}



\section{Introduction} \label{sec:intro}

There is very strong observational evidence that neutron stars have a solid crust with a thickness of order 10\% the radius and a hydrostatic structure and heat capacity determined predominantly by a pure superfluid neutron gas permeating a nuclear lattice \citep{Brown:2009aa,Chamel:2008aa}. These crust properties are consistent with theoretical predictions of condensed matter models using nuclear interactions \citep{Baym:1971aa,Buchler:1971aa,Negele:1973aa,Lorenz:1993aa,Douchin:2001aa}. The models also robustly predict the existence of a layer of exotic nuclear shapes at the base of the crust called nuclear pasta \citep{Ravenhall:1983aa,Hashimoto:1984lq,Oyamatsu:1993aa,Lorenz:1993aa,Maruyama:1998om,Magierski:2002fk} which might occupy 50\% of the mass of the crust and span 20\% of its thickness \citep{Lorenz:1993aa,Newton:2013sp,Grill:2014vp}. While robust observational evidence for nuclear pasta is not yet forthcoming, over the past decade some tantalizing observational hints of its existence have begun to emerge \citep{Pons:2013ly,Horowitz:2015rt}. The existence of the crust as a whole, and the possible nuclear pasta phases within, is consistent with a number of observed neutron star behaviors \citep{Chamel:2008aa,Newton:2014pb} and may have an important role to play in their explanations. Astrophysical phenomena sensitive to the amount of crust and the amount of nuclear pasta within include pulsar glitches \citep{Andersson:2012aa,Piekarewicz:2014aa,Newton:2015fe}, crust cooling \citep{Brown:2009aa,Horowitz:2015aa}, neutrino cooling in the nuclear pasta phases \citep{Newton:2013dz}, magnetic field decay and its effect on the rotational evolution of puslars \citep{Pons:2013ly}, crust shattering just prior to binary neutron star merger \citep{Tsang:2012aa,Neill:2021tg}, crust oscillations \citep{Steiner:2009wo,Gearheart:2011tg,Sotani:2012oz} and damping of core oscillations \citep{Wen:2012aa,Vidana:2012aa}. 

As the era of multi-messenger astronomy matures, the possibilities afforded by ever more multi-faceted data-sets for determining the structure and dynamics of neutron stars and constraining the behavior of fundamental forces at extreme densities is forcing the nature of modeling to evolve. There has been a shift from precise calculations of neutron star properties using specific fundamental interaction models \citep{Baym:1971aa,Buchler:1971aa,Negele:1973aa,Lorenz:1993aa,Douchin:2001aa,Pearson:2018aa}, through systematic surveys of models and the correlations between fundamental physics parameters and properties of neutron star crusts \citep{Horowitz:2001aa,Kubis:2007aa,Oyamatsu:2007aa,Fattoyev:2010aa,Newton:2013sp,Bao:2015aa,Zhang:2018aa}, to assigning probability distributions to ensembles of models and making explicit the weights we are putting on the model parameters and the physics they represent \citep{Steiner:2015aa,Carreau:2019aa,Beloin:2019aa}. This shift is necessary for the consistent propagation of theoretical, experimental and observational errors across domains \citep{Schunck:2015aa,Editors:2011aa,Horowitz:2014aa} demanded as our data sets improve in quantity and fidelity. This process is made explicit in Bayesian inference in which making explicit your input probability distributions of models - priors - plays an essential role. For crust models, model probability distributions have been presented only for the crust-core transition properties \citep{Carreau:2019aa}. Here we prepare distributions of models for the whole inner crust, and present the distributions of properties at the transitions to the pasta layers and the core, and of the composition of the crust.

Given the central role the interstitial pure neutron gas over the density region $10^{-4}$-$10^{-1}$fm$^{-3}$ plays in determining both the global structure, dynamics and composition of the crust, the pure neutron matter (PNM) equation of state (EOS) and it theoretical uncertainties are very important to capture in ensembles of neutron star crust models. Over the past decade, sophisticated computations of the PNM EOS coupled with theoretical advances such as chiral effective field theory (EFT)  \citep{Gandolfi:2009qf,Gezerlis:2010fu,Hebeler:2010kb,Tews:2013yg,Gezerlis:2013lr,Kruger:2013fv,Gandolfi:2014qv,Gandolfi:2015nr,Sammarruca:2015lr,Tews:2016ty,Lynn:2016sf, Holt:2017lr} have led us to the point where rigorous theoretical errors can be placed on the PNM EOS \citep{Drischler:2020aa,Drischler:2020ab}. At the same time, our best knowledge of the PNM EOS has been incorporated into crust models \citep{Hebeler:2013zl, Lim:2017aa, Tews:2017aa}.

Given the relatively tight constraints on the symmetric nuclear matter EOS at saturation density, a common way to parameterize the PNM EOS and the isospin-asymmetric EOS in general is via the symmetry energy. Here we define it as the second term in the Taylor expansion of the nuclear matter EOS with respect to isospin asymmetry, which, if the higher order terms in the expansion are small is roughly the difference between the energy of symmetric nuclear matter and pure neutron matter (although those terms could be very important, so it is important to be very aware of what definition of the EOS we are using \citep{Chen:2009aa}).

\begin{equation}
    E(n,\delta) = E_{\rm SNM} + S(n) \delta^2 + \dots \;\;\;\;\;\;\;\;\;\;\;\;\; J(n) = \frac{\partial^2 E(n,\delta)}{\partial \delta^2}\bigg|_{\delta=0},
\end{equation}

\noindent where $\delta = (n_{\rm n}-n_{\rm p})/n$ is the isospin asymmetry. Expanding the symmetry energy about saturation density gives the first three symmetry energy parameters $J$, $L$, $K_{\rm sym}$ - characterizing the magnitude, slope and curvature.

\begin{equation}
S(\rho) = J + \chi L + {\frac{1}{2}} \chi^2 K_{\rm sym} + \dots
\end{equation}

\noindent Here $\chi = {(\rho - \rho_0) / 3\rho_0}$. The symmetry energy $J$ largely determines the proton fraction in neutron stars at a given density, while its slope ($L$) largely determines the pressure around nuclear saturation density in neutron stars. Our aim in this paper is to construct two ensembles of models parameterized explicitly by two choices of probability distributions of symmetry energy parameters, one of which is informed by PNM constraints. We will model nuclear matter using an extended Skyrme energy-density functional (EDF) \citep{Lim:2017aa}. This allows us to take steps towards making contact with nuclear properties and crust properties consistently. In particular, we calculate the neutron skins of $^{208}$Pb and $^{48}$Ca and analyze their correlations with crust properties, so we might incorporate the results of the parity-violating electron scattering experiments to measure these quantities at Jefferson lab (the PREX-II and CREX experiments)  \citep{Roca-Maza:2011aa,Horowitz:2012aa} and at Mainz energy-recovering superconducting accelerator (the upcoming MREX experiment) \citep{Becker:2018aa,Thiel:2019aa}.

The symmetry energy parameters and their correlations with crust properties and nuclear observables such as neutron skins have been the subjected to systematic surveys with a wide range of EDFs \citep{Lattimer:2001aa, Horowitz:2001aa, Steiner:2005aa, Lattimer:2007aa, Kubis:2007aa, Oyamatsu:2007aa, Fattoyev:2010aa, Ducoin:2011aa, Piekarewicz:2014aa, Newton:2013sp, Bao:2015aa, Steiner:2015aa, Fortin:2016aa, Pais:2016aa}. An analysis of many studies reveals the importance of making clear any implicit probability distributions of parameters that may influence the correlations that may be present \citep{Fattoyev:2010aa,Ducoin:2011aa,Carreau:2019aa}. We use the maximal information coefficient (MIC) to characterize the strength of the correlations; this has the advantage of capturing non-linear relationships. As we shall see, some of the strong correlations we find are strongly non-linear.

Calculating a large ensemble of crust models requires a balance between the simplicity of the model used, so that it is computationally feasible to calculate a large number of models, and the sophistication required to capture the crust physics of interest. In this paper, we use a compressible liquid drop model (CLDM). This introduces additional model parameters with large uncertainties - namely those that characterize the all important surface and curvature energies of crustal nuclei and their dependence on isospin asymmetry. Previous works have fit the parameters to properties of finite nuclei \citep[e.g.][]{Lim:2017aa} (although the isospin dependence of the parameters at large isospin asymmetry is not well constrained in that case), or varied the surface energy parameters over an underlying probability distribution \citet{Carreau:2019aa}. In this paper, we try a new approach by fitting the surface parameters of the model to the results of microscopic 3D Hartree-Fock calculations of deep inner crust nuclei over a range of isospin asymmetries and densities. \revision{Although we will fit the surface energy of the CLDM to microscopic calculations, it is not consistent with the microscopic surface energy derived from gradient terms in the EDF. In a simplified way, these gradient terms are included in the dynamic spinodal method, while gradient terms are absent from the thermodynamic spinodal. We will probe the this systematic model dependence of the surface energy by comparing the crust-core transition properties obtained using the CLDM, thermodynamic and dynamic spinodal methods.}

We prepare of order $10^3$ models of the entire inner crust at a reasonable computational cost (of order $10^2$ CPU hours for each ensemble of 1000 models). \citet{Carreau:2019aa}, using the CLDM to calculate only the crust-core transition, calculate of order $10^6$ models. We make up for the smaller number of models by rigorously estimating the sampling error from this smaller ensemble using the bootstrap method \citep{Pastore:2019aa}. This gives us reliable estimates for the standard errors on the mean and the percentiles that give the credible limits for a non-Gaussian distribution. Some models that give predictions for the crust-core boundary do not give stable inner crusts due to pathologies at lower densities such as negative PNM pressure. We filter these models out of our distributions. 

To summarize the purpose of this study, we aim to prepare two ensembles of crust models suitable for use in statistical inference of nuclear and/or neutron star properties.  We extend the work by \citet{Carreau:2019aa} by (i) take steps to reduce the number of model parameters by fitting the surface parameters of the CLDM to results of 3D Hartree-Fock calculations of nuclei in neutron rich matter, and assessing the systematic uncertainty between the surface energy fit in the CLDM, dynamic and thermodynamic spindoal models (ii) calculating the resulting distributions of composition of the crust and the properties of nuclear pasta (iii) taking steps towards a consistent propagation of nuclear experimental uncertainty into the astrophysical domain by using calculating nuclear observables and neutron star crusts with the same EDF and (iv) using the mutual information coefficient to characterize the (possibly non-linear) correlations between variables. We calculate two distinct ensembles of crust models: one characterized by a uniform distribution of $J$, $L$, and $K_{\rm sym}$ parameters, and the other characterized by a probability distribution of $J$, $L$ and $K_{\rm sym}$ informed by our best knowledge of the pure neutron matter EOS which includes information from chiral EFT calculations. We will refer to these distributions as our priors, indicating their possible use in Bayesian studies, and note that the uniform priors are often referred to as uninformative priors in Bayesian parlance.

In section~2 we outline the method: (i) the input distributions of symmetry energy parameters (ii) the extended Skyrme energy-density functional, (iii) the calculations of neutron skins, (iv) The CLDM and the determination of the surface parameters, (v) comparison with the spinodal methods and (vi) the bootstrap method to estimate sampling uncertainties and the maximal information coefficient to analyze correlations.

In section~3 we present the resulting distributions of crust-core and pasta transition parameters, including the mass and radius fractions of pasta. In section~4 we give detailed visualizations of the relationships between the model parameters, crust properties and neutron skins of $^{208}$Pb and $^{48}$Ca, and also the strength of the correlations between them. In section~5 we show the composition of the inner crust and present a summary and discussion of our results in section~6.


\section{Methods}

In this section we construct two ensembles of crust models step by step, motivated by the need to make explicit our modeling choices and probability distributions for model parameters. The following are considerations that go into the construction of our crust models, and which all contribute uncertainties.

\begin{enumerate}
    \item \emph{The nuclear matter EoS (section 2.1)}. The most uncertain part of this can be characterized by uncertainties in the symmetry energy parameters. We characterize each crust model by a value of $J$, $L$, $K_{\rm sym}$, and we create ensembles using two different probability distributions over these parameters. Including up to 2nd order ($K_{\rm sym}$) represents an improvement on previous studies of the crust that use EDFs such as the Skyrme model. \revision{It has been shown, however, that the next order symmetry energy parameter $Q_{\rm sym}$ may also be important in determining crust properties \citep{Antic:2019tk}, and that will be correlated with $J$, $L$, $K_{\rm sym}$ may affect the correlations between nuclear matter and neutron star crust parameters we find.}

    \item \emph{The energy density functional (section 2.2)}. In order to self-consistently calculate the nuclear matter EOS, finite nuclear properties and properties of nuclei in the crust, we need to choose an energy-density functional (EDF) that has sufficient flexibility in its density dependence to allow independent variation of all three symmetry energy parameters we want to use to characterize out models, and whose density dependence is consistent with microscopic calculations of the PNM EOS. Although we choose a model with relatively flexible density dependence, this is one place where additional unaccounted-for model dependence may enter our results.

    \item \emph{The calculation of neutron skins (section 2.3)}. We systematically create Skyrme parameterizations over the chosen distributions $J$, $L$, $K_{\rm sym}$ \revision{but do not refit the remaining model parameters - which include gradient terms relevant to neutron skin and crust-core calculations - to finite nuclear masses and radii.} We argue that the model uncertainty this inconsistency introduces is small, but this needs to be explored more thoroughly.

    \item \emph{The crust model (section 2.4 and 2.5)}. We will use the compressible liquid drop model (CLDM) whose surface energy parameters are determined usually by fits to properties of finite nuclei or calculations of semi-infinite nuclear matter, or simply left as free parameters. Our approach is to fit the CLDM surface parameters to microscopic 3D Hartree-Fock calculations of the system we are modeling, inner crustal nuclei plus an external neutron fluid. \revision{Although the aim is to fit surface parameters to simulations of the crust itself in order to increase physical consistency, modeling uncertainty still exists. The 3DHF calculations are computationally intensive so we fit the CLDM to calculations of only 3 EDFs, two of which are not fit to finite nuclei. We will assess the model uncertainty this introduces by comparing the crust-core transition properties calculated with the CLDM to two more approximate estimations: the dynamic spinodal (which directly includes the gradient terms from the EDF) and the thermodynamic spinodal (which includes no gradient terms).}
    
\end{enumerate}

Secondly we will equip ourselves with two statistical tools to help analyze our results (section 2.6):

\begin{enumerate}
    \item We want to quantify the strengths of correlations between parameters. We do this using the Maximal Information Coefficient (MIC).

    \item We sample only a limited number ($\sim 1000$) model, so we employ the bootstrap method to estimate the standard error on the mean and variance of the predicted crust properties.
\end{enumerate}

\subsection{Symmetry energy parameter space: two probability distributions}

Our crust models will be characterized by unique values of the first three parameters in the density expansion of the symmetry energy at saturation density, $J$, $L$ and $K_{\rm sym}$. We will create two ensembles of models, characterized by two different probability distributions for these parameters. Note that it has been shown that the next parameter in the expansion, $Q_{\rm sym}$ is also important to vary systematically \citep{Carreau:2019aa,Antic:2019tk}. We do not do that here: in our models $Q_{\rm sym}$ is implicitly a function of $J$, $L$ and $K_{\rm sym}$,  

Our first ensemble will be drawn from a uniform distribution in all three parameters, in the ranges $J$=24 to 36 MeV, $L$=-10 to 100 MeV and $K_{\rm sym}$ = -440 to 120 MeV. Thinking of our sample models as Bayesian priors, these uniform probability distributions would be uninformative priors. We draw 12x13x9 models from the uniform grid of points, giving us a potential total of 1404 models. These are shown as the red points in Fig.~1, with the corresponding uniform probability distribution shown as the red histogram.

\begin{figure}[t!]
    \centering
     \includegraphics[width=0.33\linewidth]{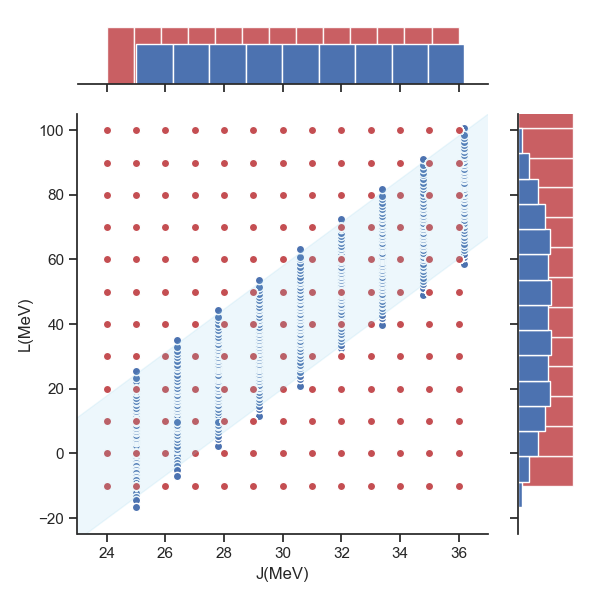}\includegraphics[width=0.33\linewidth]{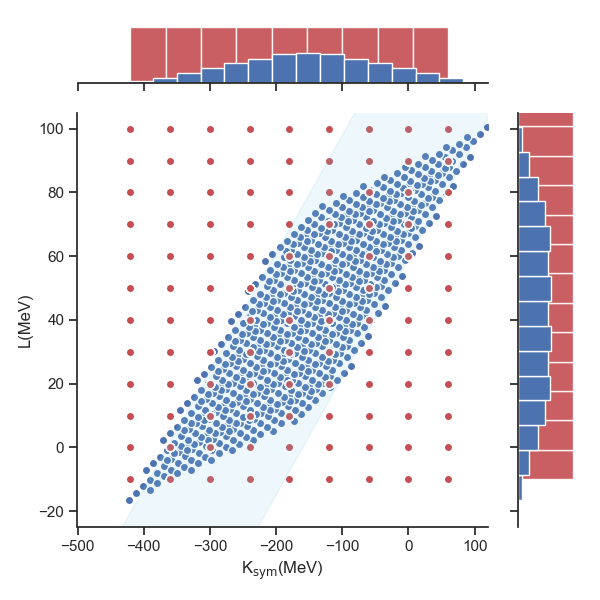}\includegraphics[width=0.33\linewidth]{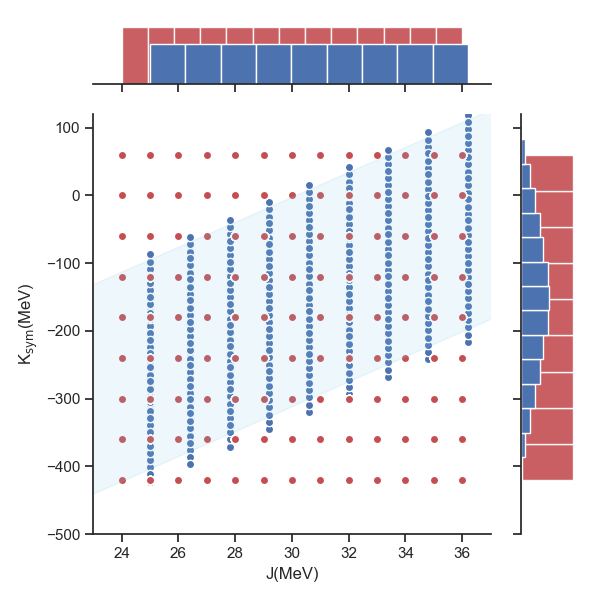}
    \caption{Our two input distributions for $J$, $L$ and $K_{\rm sym}$. We show the distributions in the $J$ vs $L$ (left), $L$ vs $K_{sym}$ (middle) and $J$ vs $K_{sym}$ (right) planes (every point corresponds to one model), and corresponding marginalized 1-d distributions as histrograms. The bands obtained from \citet{Holt:2018aa} are shown in light blue.}
    \label{fig:1}
\vspace{11pt}
\vspace{11pt}
\vspace{11pt}
\end{figure}

Our second probability distribution is over a distribution of symmetry energy parameters consistent with chiral EFT calculations of PNNM. \citet{Holt:2018aa} showed that from general arguments about the neutron-neutron interaction from Fermi Liquid theory, there exist correlations between $J$, $L$ and $K_{\rm sym}$. The correlations can be written in the following form, in terms of $J$: 
\begin{equation}
    L = 6.7J + C_{L},
\end{equation}

\begin{equation}
    K_{sym} = 18.4J + C_{K_{\rm sym}},
\end{equation} 

\noindent where $C_{\rm L}$ and $C_{\rm K_{sym}}$ are given by

\begin{equation}
    C_{L} = -19.47a_{0}+1.56b_{12}-59.22,
\end{equation}

\begin{equation}
    C_{K_{\rm sym}} = 5(C_{L}+50.22)+7.79b_{12}-258.3.
\end{equation}

\noindent $a_0$ and $b_{12}$ are related to the Fermi-liquid parameters that characterize the two-neutron interaction at a particular reference density \citep{Holt:2018aa,Holt:2018ab}; see also the description in \citet{Newton:2020aa}. Varying that reference density over a range subject to constraints from chiral-EFT, \citep{Holt:2018aa} obtain $a_0$ = 5.53-6.41 MeV fm$^3$ and $b_{\rm 12}$= 0 to 16 MeV fm$^2$ \citep{Holt:2018aa}. This represents a conservative range that encompasses all current PNM EOSs predicted by chiral-EFT and their uncertainties.

\begin{figure}[t!]
    \centering
     \includegraphics[width=0.7\linewidth]{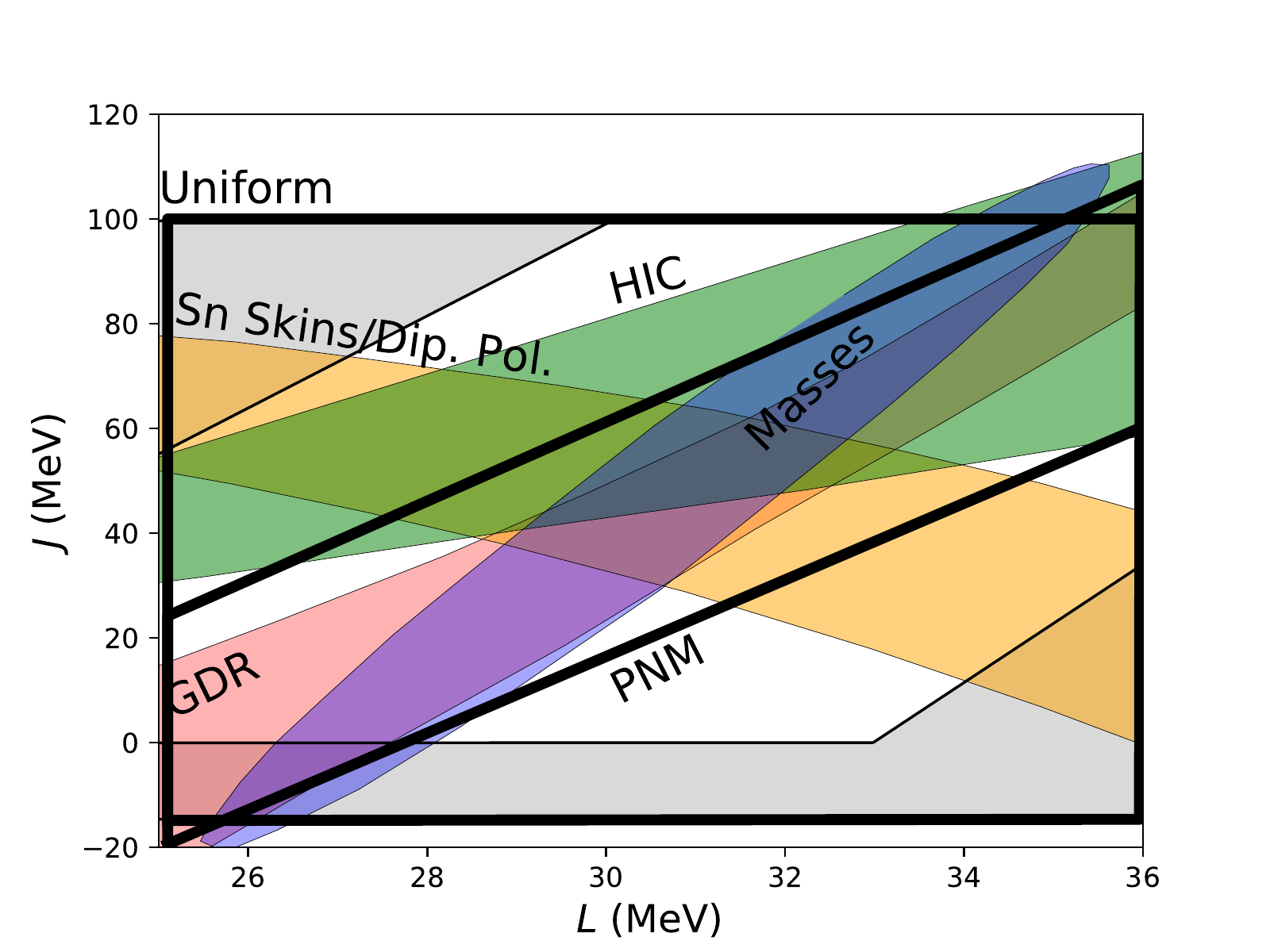}
    \caption{Comparison of the range of $J$ vs $L$ parameter space covered by our uniform and PNM ensembles of crust models (thick black lines) with experimental constraints \citep{Tsang:2012aa,Lattimer:2013aa} from giant dipole resonances (GDR), nuclear mass fits (Masses), heavy ion collisions (HIC), and neutron skins of tin isotopes and dipole polarizability measurements (Sn Skins/Dip. Pol.). The range of the uniform distribution of symmetry energy parameters encompasses a region containing all experimental constraints. The grayed-out regions in the top left corner, bottom right and below $L=0$ indicate the regions filtered out of our distributions with the additional requirement that our models give a stable crust.}
    \label{fig:2}
\vspace{11pt}
\vspace{11pt}
\vspace{11pt}
\end{figure}

For our PNM ensemble of models, we take a uniform probability distribution over $J$, $a_0$ and $b_{\rm 12}$ and translate that to a probability distribution for $J$, $L$ and $K_{\rm sym}$. We draw 9x9x9 models uniformly from the ranges of $J$, $a_0$ and $b_{\rm 12}$, giving us a potential total of 729 models. These are shown as the blue points in Fig~1, and for comparison the blue shaded region shows the region of parameter space consistent with chiral-EFT calculations \citep{Holt:2018aa}.

In Fig.~2 we show how the region of $J$-$L$ parameter space spanned by our initial parameter sets compare with constraints from a variety of nuclear experimental constraints \citep{Tsang:2012aa, Lattimer:2013aa}. The uniform ensemble covers a sufficiently wide range to encompass all current experimental inferences of $L$ and $J$.


\noindent
\subsection{Modeling nuclear matter: the Skyrme Energy Density Functional (EDF)}

In order to model nuclear matter in a way that allows consistent connection to nuclear properties, we use the Skyrme Energy Density Functional \citep{Vautherin:1972aa,Bender:2003aa,Lim:2017aa}. For uniform nuclear matter, the energy per particle is given as

\begin{align}
\frac{E}{N} = & \frac{1}{4} t_{0} n[(2+x_{0}) - (2x_{0}+1)(y_{p}^{2}+y_{n}^{2})] 
\\ \notag 
& + \sum_{i=3}^4 \frac{1}{24} t_{i}n^{(\alpha_{i}+1)}[(2+x_{i}) - (2x_{i}+1)(y_{p}^{2}+y_{n}^{2})] 
\\ \notag & + \frac{1}{8}[t_{1}(2+x_{1})+t_{2}(2+x_{2})]\tau
\\ \notag & + \frac{1}{8}[t_{1}(2x_{1}+1)+t_{2}(2x_{2}+1)](\tau_{p}y_{p}+\tau_{n}y_{n}).
\end{align}

\noindent where $\rho_i$ and $\tau_i$ ($i=p,n$) are the density and kinetic energy density respectively. $y_{\rm p}$ and $y_{\rm n}$ are the proton and neutron fractions respectively.

The most widely used version of the Skyrme EDF functional contains nine parameters $x_{0-3}, t_{0-3} $ and $\alpha_3$ that determine the nuclear matter EOS. We add additional degrees of freedom to the density dependence by including the $t_4$, $x_4$ and $\alpha_4$ parameters \citep{Lim:2017aa}. Although there are a number of other ways of extending the Skyrme EDF \citep{Agrawal:2006aa,Erler:2010aa}, this is a simple modification that is (i) consistent with the range of possible sub-saturation density dependences of the PNM EOS predicted by chiral-EFT calculations, and (ii) also allows us to vary $J$, $L$ and $K_{\rm sym}$ independently. The purely isovector parameters $x_0$, $x_3$ and $x_4$ can be related directly to $J$, $L$ and $K_{\rm sym}$; the details of these relations are given in\citet{Newton:2020aa}. The variation of these three isovector parameters do not change the symmetric nuclear matter EOS, and only change the binding energies and charge radii of doubly magic nuclei by an average of 1-2\% \citep{Newton:2020aa}. We take the remaining parameters from the Sk$\chi$450 parameter set from  Table~1 of \citet{Lim:2017aa}, which has been fit to the properties of doubly magic nuclei. 

Starting from this baseline model, for each choice of $J$, $L$ and $K_{\rm sym}$, we obtain a unique Skyrme model using the relations with $x_0$, $x_3$ and $x_4$. We then use these to calculate properties of infinite nuclear matter, finite nuclei, and nuclei in the crust.

\subsection{The neutron skins}

\revision{Neutron skins of $^{208}$Pb and $^{48}$Ca are calculated using the Sky3D code \citep{Maruhn:2014aa}. More details of the neutron skin calculations are given in \citep{Newton:2020aa}. Our aim is to consistently connect neutron skins to neutron star properties within the same EDF. There is still a notable inconsistency in our method which is a source of model uncertainty. We vary $J$, $L$ and $K_{\rm sym}$ through the parameters $x_0$,$x_3$ and $x_4$. However, we do not refit the remaining model parameters to finite nuclear properties. This should eventually be done since we are calculating the neutron skins of finite nuclei. In particular, the parameters $x_1,$,$t_1$,$x_2$, and $t_2$ determine the coefficients in the isoscalar and isovector density gradient terms $G_{\rm s} [\nabla (\rho_{\rm n} +\rho_{\rm p})]^2$ and $G_{\rm v} [\nabla (\rho_{\rm n}-\rho_{\rm p})]^2$, so neglecting the refit may impact the resulting properties of the nuclear surface. These refits will be carried out and reported in a later publication. However, even without the refits there are a number of reasons why it is useful to analyze the relationships between the neutron skins and neutron star crust properties obtained with the present set of models.}

\revision{Firstly, sensitivity analyses provide good evidence that the neutron skin of $^{208}$Pb is much less sensitive to the gradient terms than to the symmetry energy parameters \citep{Chen:2010aa,Zhang:2013uv}. The dependence of the neutron skin of $^{208}$Pb on the crust-core transition pressure from a set of RMF models in which $L$ was varied systematically without refitting the remaining parameters \citep{Fattoyev:2010aa} is similar to that obtained from a similar set of RMF models in which the parameters were refit \citep{Pais:2016aa}. All this supports the claim that even without refitting the parameters, the relationship between neutron skin and neutron star properties will be qualitatively the same as with a refit. Of course the claim should and will be tested in the near future.}

\revision{Secondly, a number of other studies analyzing neutron skins a range of symmetry energy parameter space share the same inconsistency, e.g. \citep{Chen:2010aa,Fattoyev:2010aa,Piekarewicz:2014aa}. Widely used constraints on the symmetry energy have been obtained with this method. It is useful to compare our results to the existing results in the literature on the same footing.}

\revision{We are taking a step forward towards a consistent connection between nuclear and neutron star properties, which is worth exploring on its own merits. If nothing else, if it turns out the gradient terms do have an effect, our results play a role in disentangling the various contributions from the EDF to the relationships between neutron skins and neutron star properties.}


\noindent
\subsection{Modeling the crust: the Compressible Liquid Drop Model (CLDM)}
 
\begin{figure}[t!]
    \centering
     \includegraphics[width=0.5\linewidth]{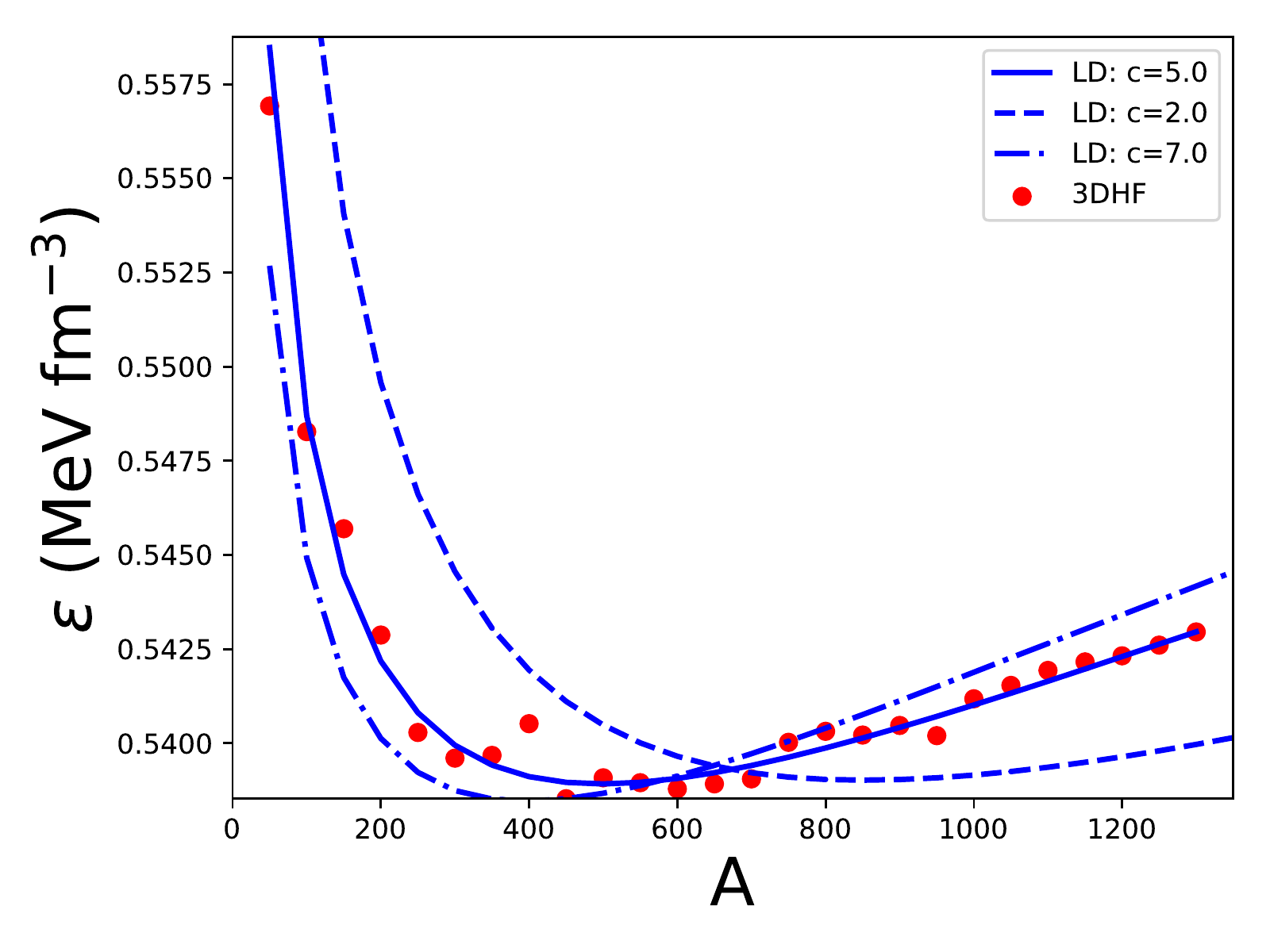}\includegraphics[width=0.5\linewidth]{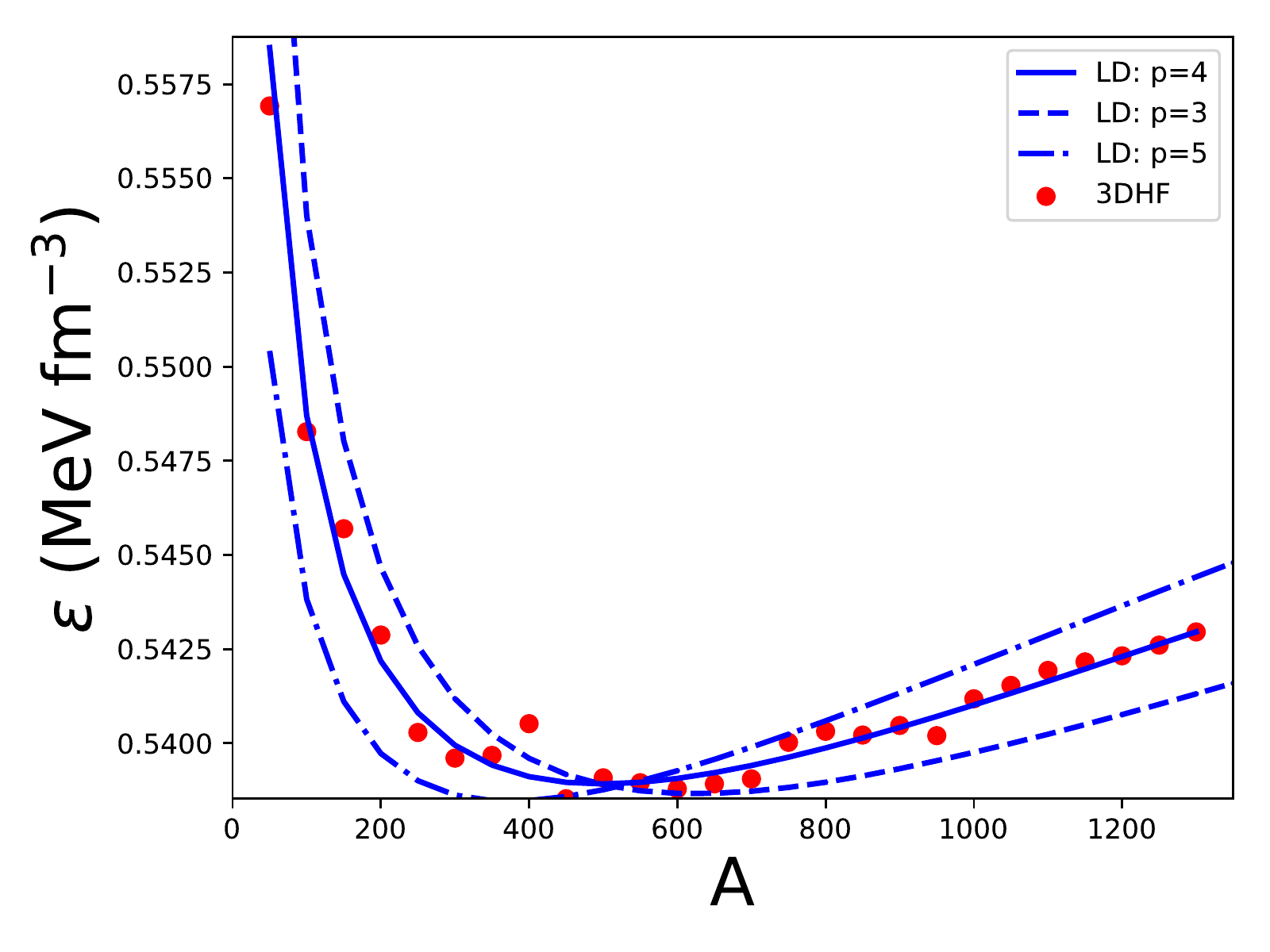}
    \caption{Example of the effect of varying the parameter $c$ (left) which controls the correlation between the surface and bulk symmetry energies, and $p$ (right) which controls the strength of the surface tension at very low proton fractions. We compare the results of the CLDM and 3DHF calculations for a density of 0.04fm$^{-3}$, a proton fraction of 0.2, and the NRAPR Skyrme EDF \citep{Steiner:2005aa} with $J$=32.8 MeV and $L$=60 MeV.}
    \label{fig:3}
\vspace{11pt}
\vspace{11pt}
\vspace{11pt}
\end{figure}

\begin{figure}[t!]
    \centering
     \includegraphics[width=1\linewidth]{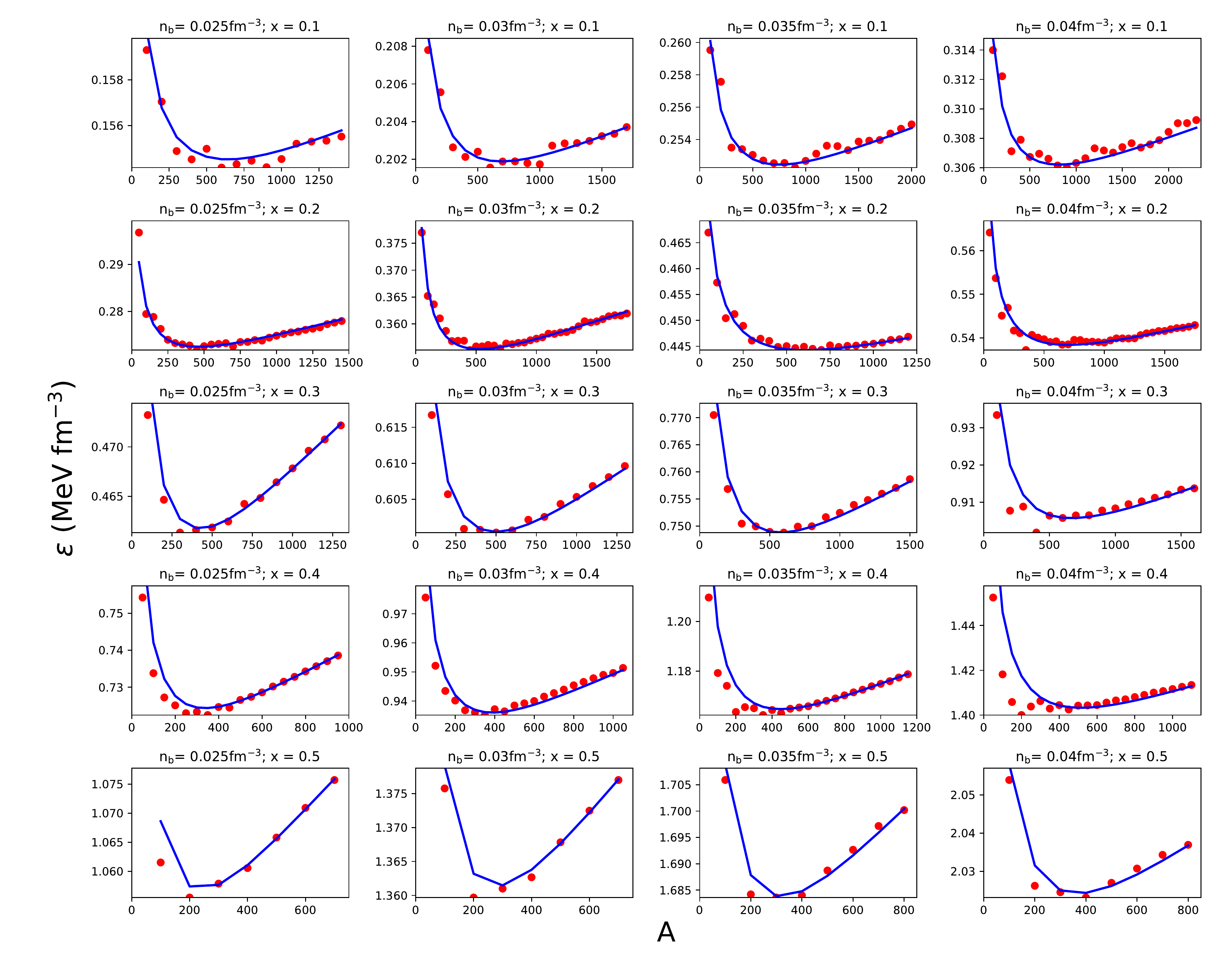}
    \caption{Fit of CLDM to results of 3D DFT calculations for the SkIUFSU30 Skyrme EDF \citep{Fattoyev:2012hc} with $J$=28.3 MeV and $L$=30 MeV.}
    \label{fig:4}
\vspace{11pt}
\vspace{11pt}
\vspace{11pt}
\end{figure}

\begin{figure}[t!]
    \centering
     \includegraphics[width=1\linewidth]{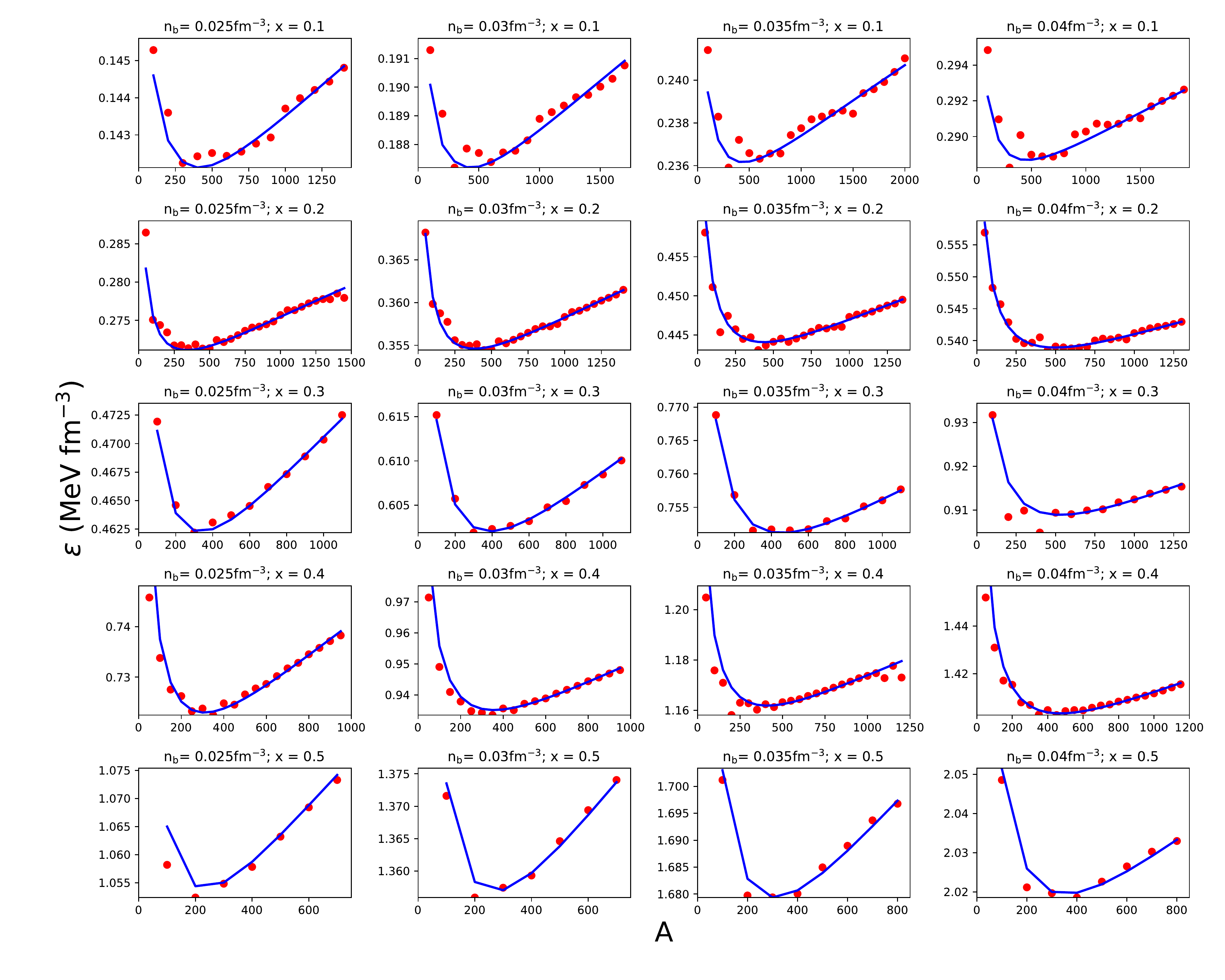}
    \caption{Fit of CLDM to results of 3D DFT calculations for the NRAPR Skyrme EDF \citep{Steiner:2005aa} with $J$=32.8 MeV and $L$=60 MeV.}
    \label{fig:5}
\vspace{11pt}
\vspace{11pt}
\vspace{11pt}
\end{figure}

\begin{figure}[!ht]
    \centering
     \includegraphics[width=1\linewidth]{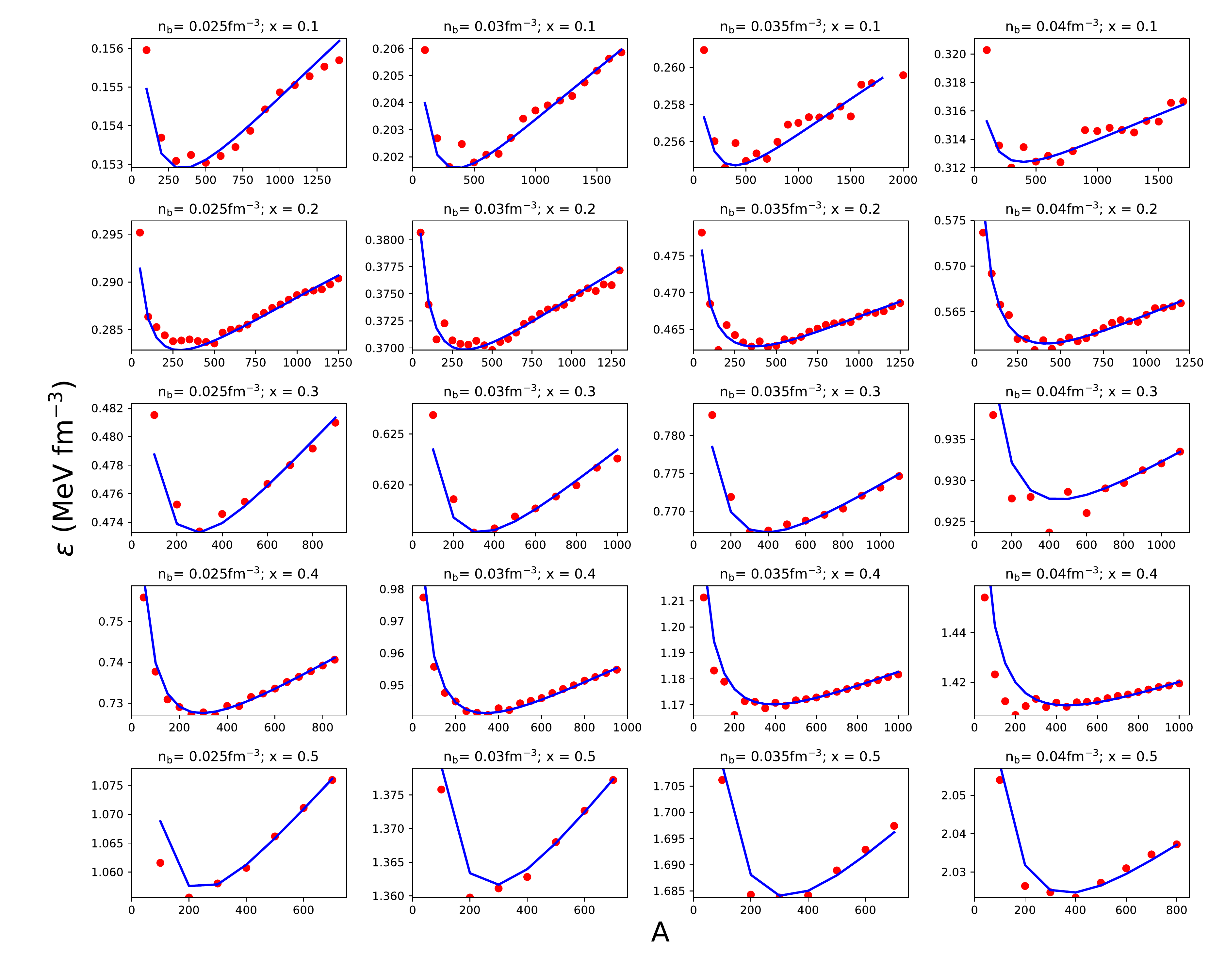}
    \caption{Fit of CLDM to results of 3D DFT calculations for SkIUFSU90 Skyrme EDF \citep{Fattoyev:2012hc} with $J$=38.3 MeV and $L$=90 MeV.}
    \label{fig:6}
\vspace{11pt}
\vspace{11pt}
\vspace{11pt}
\end{figure}

In order to generate ensembles of crust models, we need a method that is time efficient yet is able to incorporate much of the important physics that determines the crust composition and structure. We use the compressible liquid drop model (CLDM), an extension of the nuclear droplet model for neutron star crusts \citep{Baym:1971aa} which aims to describe the bulk and surface properties of nuclei immersed in a neutron gas while ignoring quantum effects such as shell effects and pairing. In the CLDM, we approximate the composition of the crust at a given depth by a single species of nucleus or segment of nuclear pasta in a Wigner-Seitz cell. The energy density is a function of cell size $r_{\rm c}$, proton fraction $y_{\rm p}$, neutron gas density $n_{\rm n}$ and nuclear density $n$ \citep{Newton:2013sp}:

\begin{align}
    \epsilon_{cell}(r_{c},y_{\rm p},n,n_{n}) = &v(n_{\rm n}, n)[nE_{\rm Sky}(n,y_{\rm p}) + \epsilon_{exch}] + [1-v(n_{\rm n}, n)]n_{n}E_{\rm Sky}(n_{n},0) +\notag \\ & u(n_{\rm n}, n)(\epsilon_{surf} 
    + \epsilon_{curv}) +\notag \\ & u(n_{\rm n}, n)\epsilon_{\rm C+L}  + \epsilon_{e}(n_{e}).
\end{align}\\

\noindent The Skyrme EOS gives the bulk energy of nuclei and the neutron gas. The term $\epsilon_{\rm exch}$ gives the exchange energy of protons. The second line accounts for the energy of the interface between the nuclei and neutron gas and is decomposed into a surface and curvature energy $\epsilon_{\rm surf}$, and $\epsilon_{\rm curv}$. We describe how we determine the surface and curvature tensions below. The third line gives the Coulomb energy $\epsilon_{\rm Coul}$ of the nucleus and the lattice, and the electron energy $\epsilon_{e}(n_{e}$. $u(n_{\rm n},n)$ is the volume fraction of nuclear matter, and $v$ is defined to be equal to $u$ for ordinary nuclei immersed in a neutron gas and equal to $1-u$ for the case at the highest densities when the structure of the unit cell inverts to bubbles of neutron matter surrounded by nuclear matter.

\revision{We account for nuclear pasta using the dimensionality parameter $d$ which enters into the Coulomb energy density:}

\begin{equation}\label{eq:ecl}
    \epsilon_{\rm (C+L)} = 2\pi (exnr_{N})^{2}f_{d}(u); \;\;\;  f_{d}(u) = \frac{1}{d+2} \left[\frac{2}{d-2}\left(\frac{ 1- du^{1-2/d} }{2} \right) + u\right].
\end{equation}

\revision{Planar, Cylindrical, and spherical geometries are specified by $d=1,2$ and 3 respectively. At each density we calculate the energy density for each phase of pasta as well as for uniform matter, and select the lowest energy phase. The density at which the cylindrical phase becomes energetically preferred over the spherical phase marks the onset of pasta and the density at which the uniform phase becomes energetically preferred over the spherical hole phase marks the crust-core transition.}

In the CLDM, the main uncertainty aside from the uniform nuclear matter EOS is the energy associated with the interface between nucleus and dripped neutrons. Given the radius of the nucleus or characteristic width of the pasta structure $r_{\rm N}$, 

\begin{equation}
    \epsilon_{\rm surf} = \frac{d \sigma_{\rm s}}{r_{N}}; \;\;\;  \epsilon_{\rm curv} = \frac{d (d-1) \sigma_{\rm c}}{r_{N}^2}.
\end{equation}

We take the surface and curvature tensions to be parameterized as follows (essentially the interfacial equivalent to choosing a particular form for the EDF):

\begin{equation} \label{eqn:surf}
    \sigma_{\rm s}(x)= \sigma_0 \frac{ 2^{p+1} + b}{ \frac{1}{x^p} + b + \frac{1}{(1-x)^p} }; \;\;\; \sigma_{\rm c}(x) = \sigma_{\rm s}(x) \frac{\sigma_{\rm 0,c}}{\sigma_{0}}. 
\end{equation}

\noindent Analogously to uniform nuclear matter, the surface and curvature tension may also be expanded about their isospin symmetric values. For example,

\begin{equation} \sigma_{\rm s}(x) = \sigma_0+ \sigma_{\delta} \delta^2, \end{equation}

\noindent where $\sigma_{\delta} = \frac{1}{2} \partial^2 \sigma_{\rm s} / \partial \delta^2 |_{x=0.5}$ is the surface symmetry tension. Our surface parameterization \ref{eqn:surf} then gives an expression for the surface symmetry tension

\begin{equation} \label{eqn:sdelta}
    \sigma_{\delta} = \sigma_0 \frac{2^p p(p+1)}{2^{p+1} + b}.
\end{equation}

The parameters of the surface energy are therefore $\sigma_0, \sigma_{\rm 0,c}, b, p$. These parameters should be calculated consistently with the bulk nuclear matter EOS - i.e. be determined uniquely by the EDF we have chosen. This has generally been achieved by either conducting a fit to finite nuclear masses \citep[e.g.][]{Lim:2017aa} or by fitting to the properties of semi-infinite nuclear matter \citep{Schneider:2017aa}. By doing this for a variety of different bulk EDFs, one can probe the dependence of the surface parameters on the bulk EDF, and hence, for example, the symmetry energy. In \citet{Newton:2013sp} we use previous examples of this to fit a relation between the surface symmetry tension and bulk symmetry energy $J$:

\begin{equation} \label{surf_corr} \sigma_{\delta} = \frac{J n_{\rm s}^{2/3}}{ (36 \pi)^{1/3}} [ (0.046 \; {\rm MeV}^{-1} c + 0.01\; {\rm MeV}^{-1} ) J - c ]. \end{equation}

The slope of the correlation between $\sigma_{\delta}$ and $J$ is given by $c$ and we can take this to be one of the parameters to fit in place of the parameter $b$ in equation~\ref{eqn:sdelta}. Specifying $c$ (along with the value of $J$ from our bulk EDF) gives $\sigma_{\delta}$. Specifying $p$ and $\sigma_0$ allows us to determine $b$ from \ref{eqn:sdelta}. Thus the interface energy is now determined from the four parameters $\sigma_0, \sigma_{\rm 0,c}, c, p$.

We adopt a slightly different method of determining these parameters. We fit the surface parameters to three dimensional Hartree-Fock calculations of inner crust neutron star matter at densities in the range $0.025-0.04$fm$^{-3}$ and over a range of proton fractions 0.1-0.5. The densities are high enough that we are sensitive to the gas of neutrons external to the nucleus, but low enough that complications of disentangling the effects of different nuclear pasta shapes do not overwhelm the problem. Of course, the validity of the best fit parameters we find when applied to the pasta phases should be explored in future works. We use the 3DHF code developed in \citep{Newton:2009eu}. 

The surface parameters are sensitive to the proton fraction and symmetry energy, so in order to obtain reliable fits we fit the CLDM to 3DHF calculations conducted at proton fractions from 0.1 to 0.5 and for 3 different Skyrme EDFs. At lower proton fractions than these, spurious shell effects in the 3DHF calculations arising from the discretization of the neutron gasÕ spectrum become sufficiently large that the fits become unreliable.

We omitted the surface thickness energy introduced by \citep{Baym:1971aa}, as done by recent works, since including it decreased the quality of fits. In addition, the 3DHF calculations include a simple zero-range pairing interaction, necessary to obtain good convergence at zero temperature, not included in the CLDM. We found that a small constant correction was necessary to obtain an agreement in the \emph{average} energy across $A$ at each density and proton fraction, but this correction approaches zero as $x\to0$ and $x\to0.5$. 

The surface parameters affect the shape of the curves of energy density versus nucleon number A. We illustrate this by two examples shown in Fig.~3. The red points are the results of the 3DHF calculations; the small fluctuations are a reflection of the oscillating shell energies \citep{Magierski:2002fk}. In Fig~3a we show the effect of varying $c$, the correlation between the surface symmetry energy and the bulk symmetry energy, and in Fig.~3b we show the effect of varying $p$. Note that we make the model choice of placing all the bulk EOS dependence of the surface energy in the parameter $c$, and we assume one can take $p$ to be EOS independent. There is support for this from our fits: for the three EDFs, we obtain similar best fit values for $p$. However, three \emph{is} a small number, so this should be noted as another source of model uncertainty.

Decreasing $c$ and $p$ move the minimum value of $A$ (and thus the equilibrium value of the cell size) to higher values and flattens the curvature. However, $c$ affects the degree to which that happens from EOS to EOS, while $p$ has an EOS independent effect, and so the combination of fitting to data on different EOSs and proton fractions is sufficient to well determine these parameters.

We use the Skyrmes SkIUFSU30 ($J=28.3$MeV, $L=30$MeV), NRAPR ($J=32.8$MeV, $L=60$MeV) and SkIUFSU90 ($J=38.3$MeV, $L=90$MeV) \citep{Fattoyev:2012hc,Steiner:2005aa}. 

\begin{table}[!t]
\caption{\label{tab:table1} Ranges varied and best fit values for the surface parameters of the CLDM.}
\begin{ruledtabular}
\begin{tabular}{ccc}
 parameter&range&best fit value\\
 \hline
 $\sigma_0$ (MeV fm$^{-2}$) &0.8 - 1.3 & 1.1 \\
 $\sigma_{\rm 0,c}$(MeV fm$^{-1}$)& 0 - 1.0 & 0.6 \\
 $c$&2.0 - 7.0 & 5.0 \\
 $p$&2.0 - 5.0& 3.8 \\
\hline
\end{tabular}
\end{ruledtabular}
\end{table}

We varied the parameters as shown in Table~1. We started out at a proton fraction of 0.5, so that we were probing only the isospin symmetric surface parameters. We conducted a $\chi$-squared fit to obtain $\sigma_0$ and $\sigma_{0,c}$. We found the best fit parameters of $\sigma_0$=1.1 and $\sigma_{0,c}$ = 0.6.

Next, fixing $\sigma_0$ and $\sigma_{0,c}$ at these values, we fit $p$ and the surface-bulk symmetry energy correlation coefficient $c$ to results of 3DHF calculations at proton fractions $y_{\rm p}$ of 0.4, 0.3, 0.2 and 0.1 ($\delta=1-2y_{\rm p}$ = 0.2, 0.4, 0.6 and 0.8). We find best fit values $p=3.8\pm0.2$ and $c=5.0\pm0.5$ with 1-$\sigma$ errors quoted. We found some evidence that allowing an EOS dependence to be incorporated into $p$ will lead to a better fit, but the effect was small. The results of the energy density versus the total nucleon number in the unit cell for the CLDM for the best fit parameters are shown in Figs. 5-7.  $c=5.0$ is in the the middle of the range explored in \citet{Newton:2013sp} and thus consistent with previous fits of the surface symmetry energy. $p$ is higher than has generally been obtained before. For example, recently, \citep{Lim:2017aa} obtained $p=3.4$ and \citep{Schneider:2017aa} obtained values between 3 and 3.5. However, these were obtained by fits to finite nuclei, which do not probe the extremely isospin asymmetric regime that $p$ is sensitive too, or semi-infinite nuclear matter which assumes a particular shape for the nuclear surface. This is the first time these parameters have been obtained by a full microscopic calculation of nuclei in a neutron gas. \revision{Indeed, a recent study of nuclei immersed in neutrons using the extended Thomas-Fermi model found a best fit value of $p=3.8$ in agreement with ours \citep{Jose-Furtado:2020vb}.}

\revision{Although we are attempting to move towards a consistent description of the surface and bulk terms in the liquid drop model, there is still some way to go. There is model uncertainty in the fit, in the limited number of EDFs used in the fit, and in the functional form of the surface energy used in the model. In particular, we assume only the parameter $c$ depends on the bulk nuclear matter parameters through $J$. There is evidence, however, that the functional shape of the surface tension itself depends on bulk nuclear matter parameters, and that a fully consistent fit could alter the relationship between nuclear properties and the bulk parameters \citep{Chatterjee:2017wu}. This is unlikely to be capture in our model.}

In this work, we will use the best fit values of $p$ and $c$ obtained here.


\noindent
\subsection{Locating the crust-core transition: comparing the CLDM with the thermodynamic and dynamic Spinodal Methods}

Although we will determine the crust-core transition using the CLDM, a simple, frequently used method to compute the location of the crust-core boundary is to locate the nuclear matter spinodal, where uniform nuclear matter becomes unstable to short wavelength density fluctuations. Including the stabilizing effects of the Coulomb force obtains the dynamic spinodal, while including only bulk nuclear matter contributions obtains the thermodynamic spinodal, which gives an upper limit on the crust-core transition density. \revision{We will use this methods to make an assessment of the model error associated with the surface energy.}

Following the original formalism of \citep{Baym:1971aa}, which has been widely used \citep{Oyamatsu:2007aa,Hebeler:2013aa,Lim:2017aa}, the locus of points in the $(n,y_{\rm p})$ plane that satisfying the following conditions defines the spinodal:

\begin{align}
    &v_{0} + 2(4\pi e^{2}\beta)^{1/2} - \beta k_{FT}^{2} > 0 \hspace{2cm}  \text{dynamic spinodal} \notag \\
    &v_0 > 0 \hspace{5cm} \text{thermodynamic spinodal}
\end{align}

The various terms are defined as follows. With $\mu_{n}$ being the neutron chemical potential and $\mu_{p}$ being the proton chemical potential, we define

\begin{equation}
    \beta = 2(Q_{pp} + 2Q_{np}\zeta + Q_{nn}\zeta^{2}) ,\;\;\;\;\;\;\; \zeta=\frac{\partial\mu_{p}/\partial\rho_{n}}{\partial\mu_{n}/\partial\rho_{n}},
\end{equation}

\noindent which accounts for the gradient terms in the energy density functional, parameterized by $Q_{pp}$, $Q_{np}$, $Q_{nn}$ which, for Skyrme models, are given by

\begin{equation}
Q_{\rm nn} = Q_{\rm pp} = \frac{1}{3}[t_1(1-x_1) - t_2(1+x_2)],
\end{equation}

\begin{equation}
Q_{\rm np} = Q_{\rm pn} = \frac{1}{16}[3t_1(2-x_1) - t_2(2+x_2)].
\end{equation}

The Thomas-Fermi wave number, $k_{TF}$ is:

\begin{equation}
    k_{TF}^2 = \frac{4e^{2}}{\pi}k_{e}^{2}, \;\;\;\;\;\; k_{e}^{2} = (3\pi^{2}\rho_{p})^{1/3},
\end{equation}

\noindent and finally, the bulk contribution to the effective interaction is

\begin{equation}
    v_{0} = \frac{\partial\mu_{p}}{\partial n_{p}} - \frac{(\partial\mu_{p}/\partial n_{n})^{2}}{(\partial\mu_{n}/\partial n_{n})}.
\end{equation}


\subsection{Analyzing the two ensembles of crust models}

\begin{figure}[t!]
    \centering
    \includegraphics[width=0.32\linewidth]{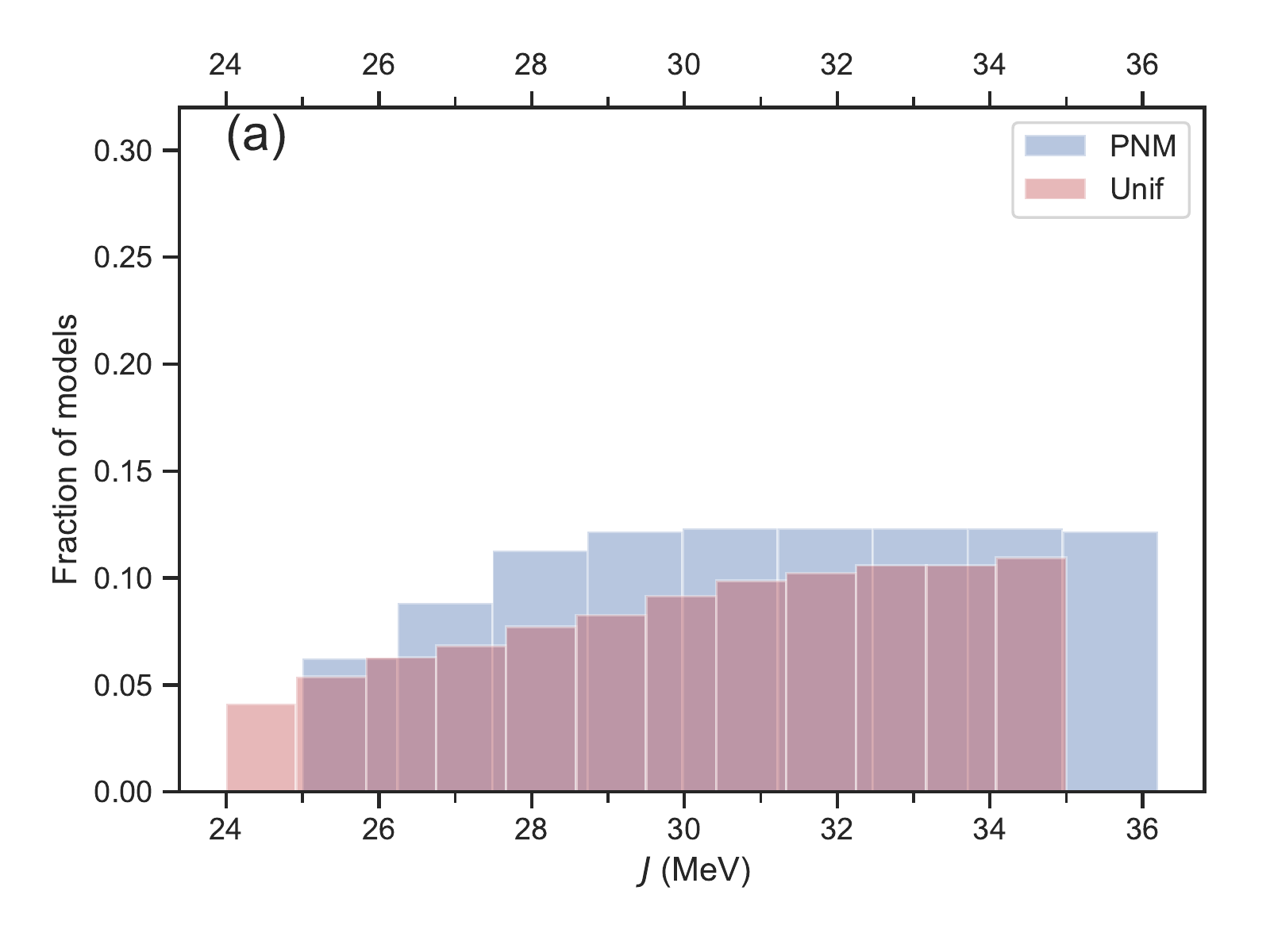}\includegraphics[width=0.32\linewidth]{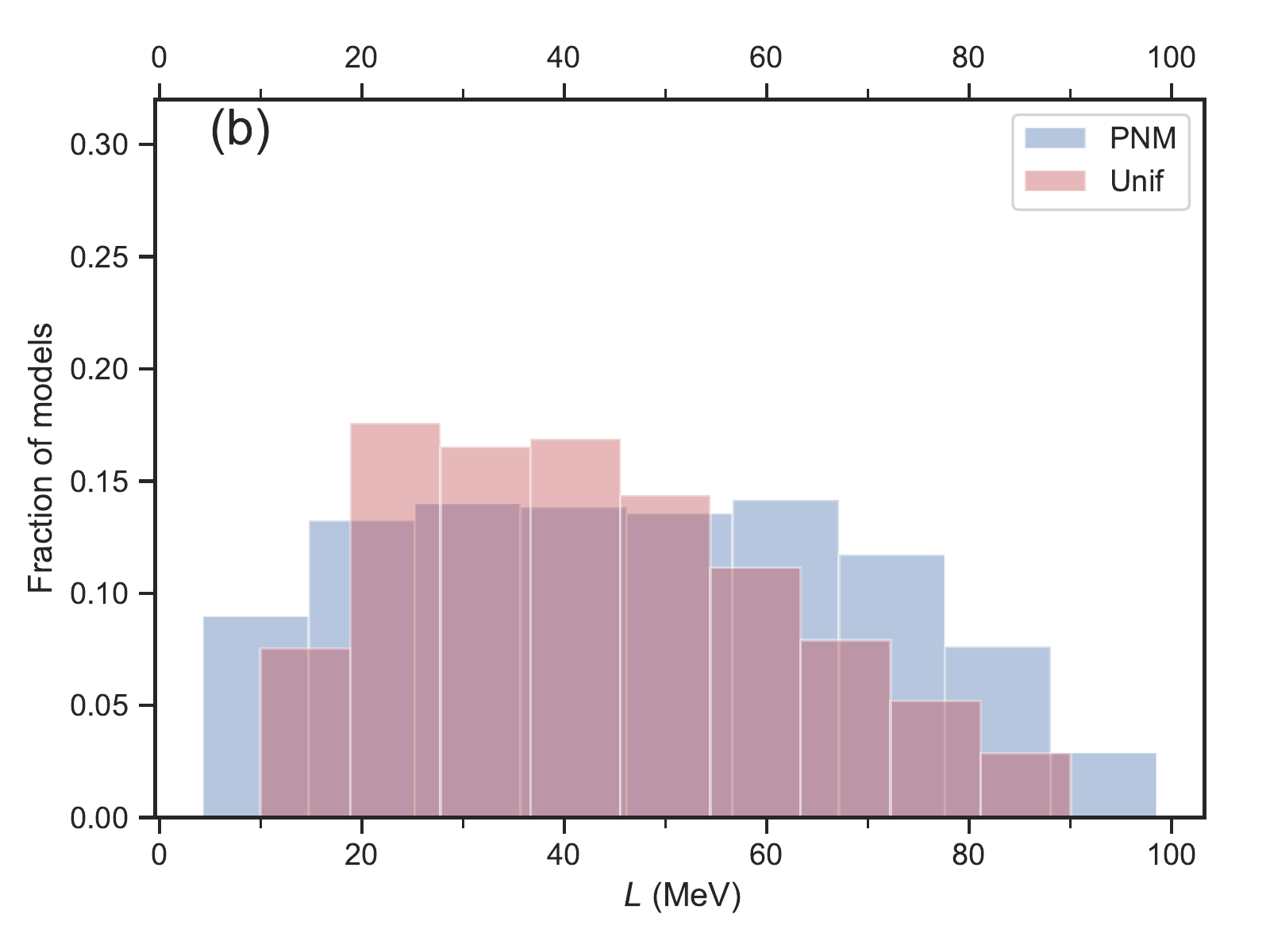}\includegraphics[width=0.32\linewidth]{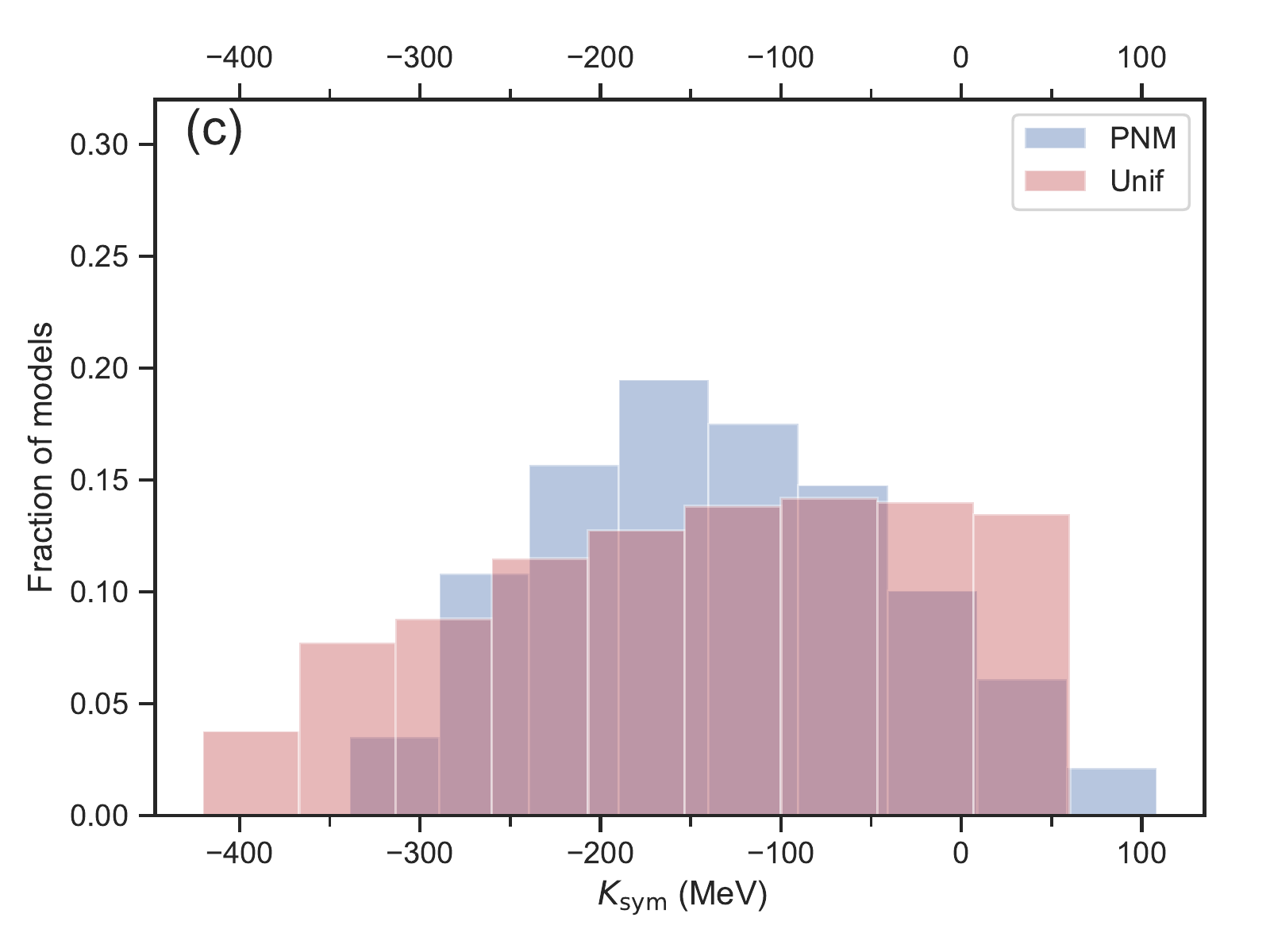}
    \caption{Histograms of the first three symmetry energy parameters $J$ (a), $L$ (b) and $K_{\rm sym}$ (c) resulting in its density expansion at saturation density for our two initial sets of prior distributions: uniform distributions (Unif) and those drawn from pure neutron matter EOSs (PNM). The distributions have now been further filtered by excluding those parameters that do not give rise to stable crust models, most of which exhibit negative neutron pressure at sub-saturation density.}
    \label{fig:7}
\end{figure}

\begin{figure}[!ht]
    \centering
     \includegraphics[width=0.45\linewidth]{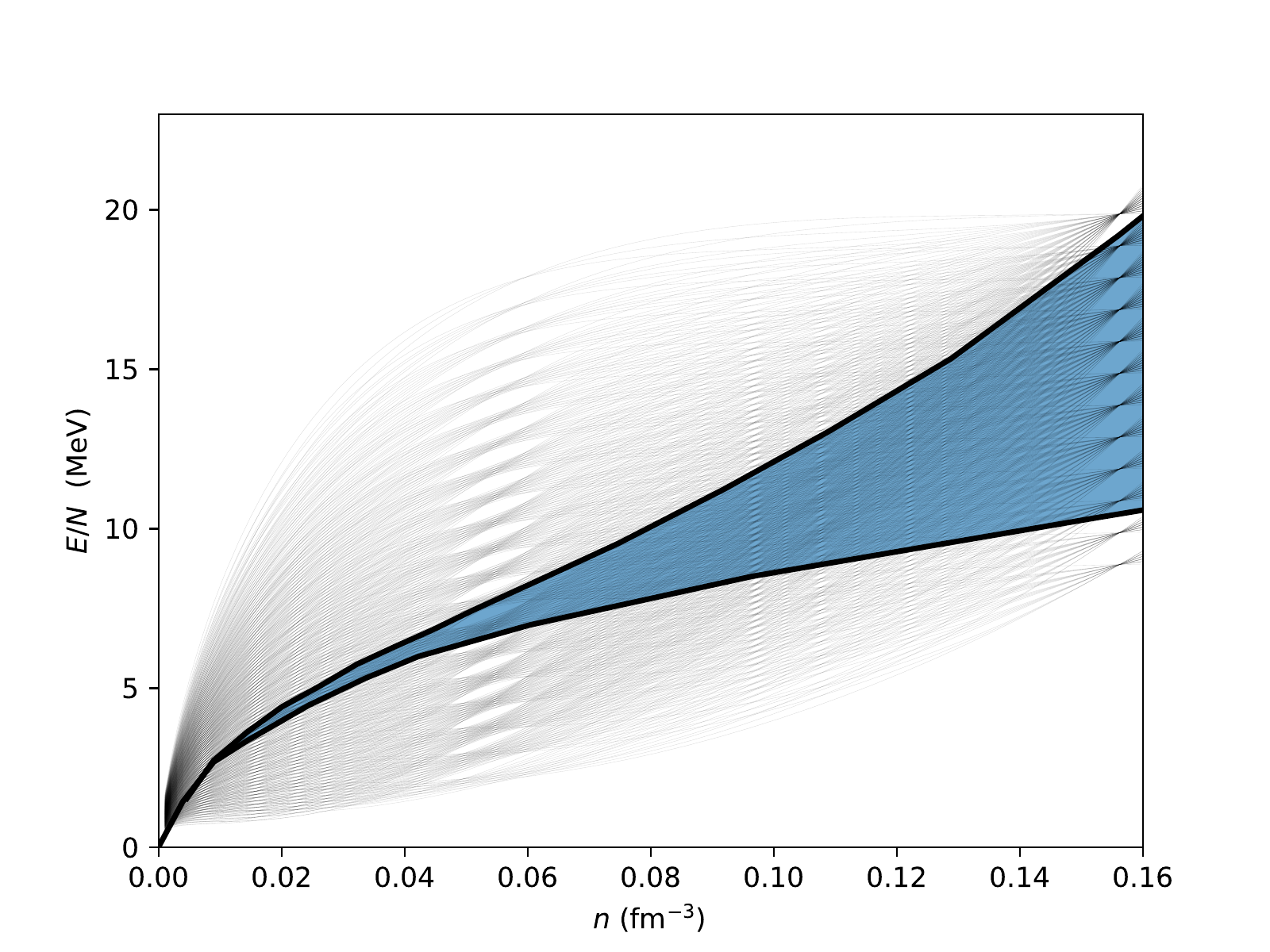}\includegraphics[width=0.45\linewidth]{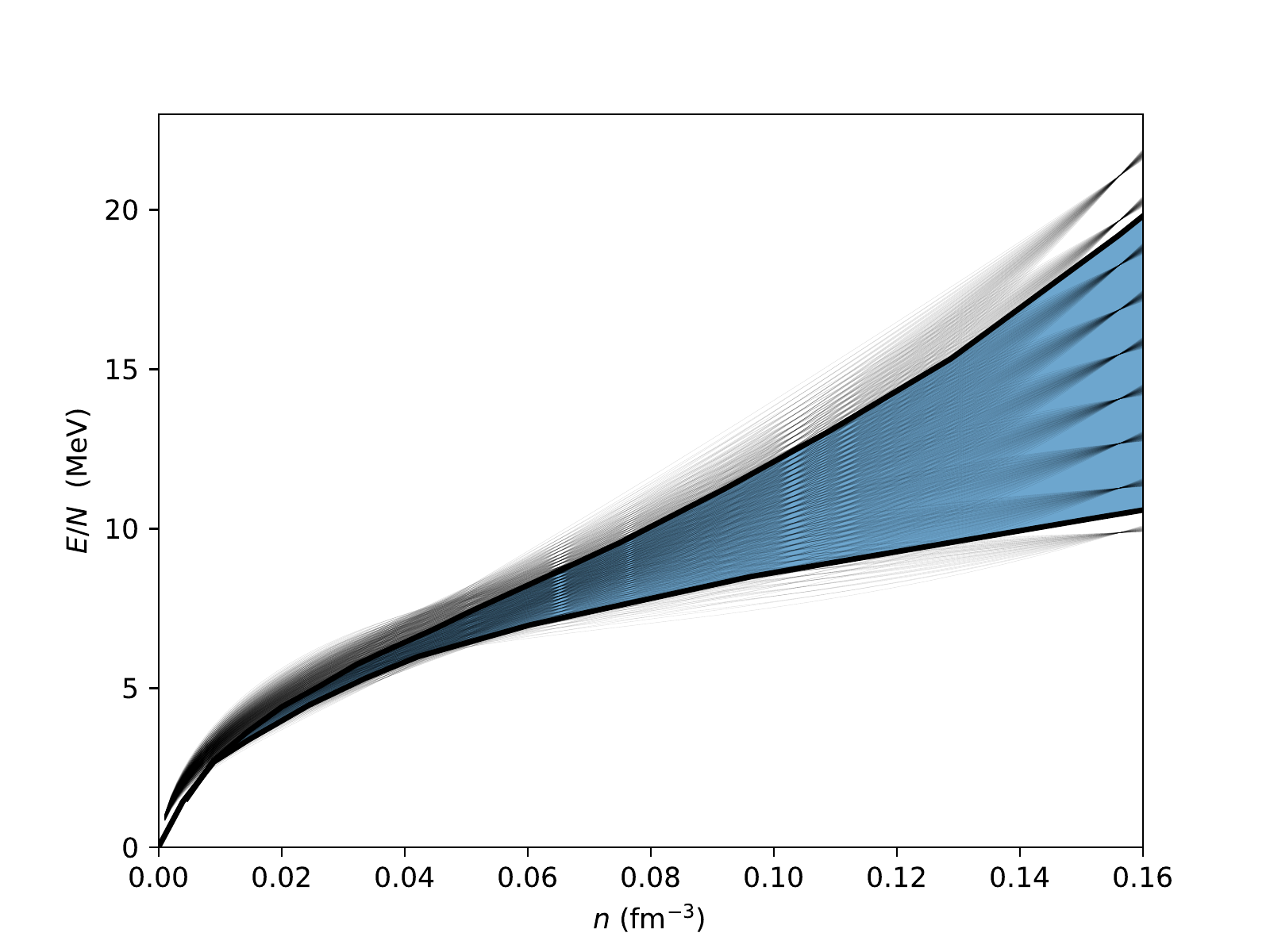}\\
     \includegraphics[width=0.45\linewidth]{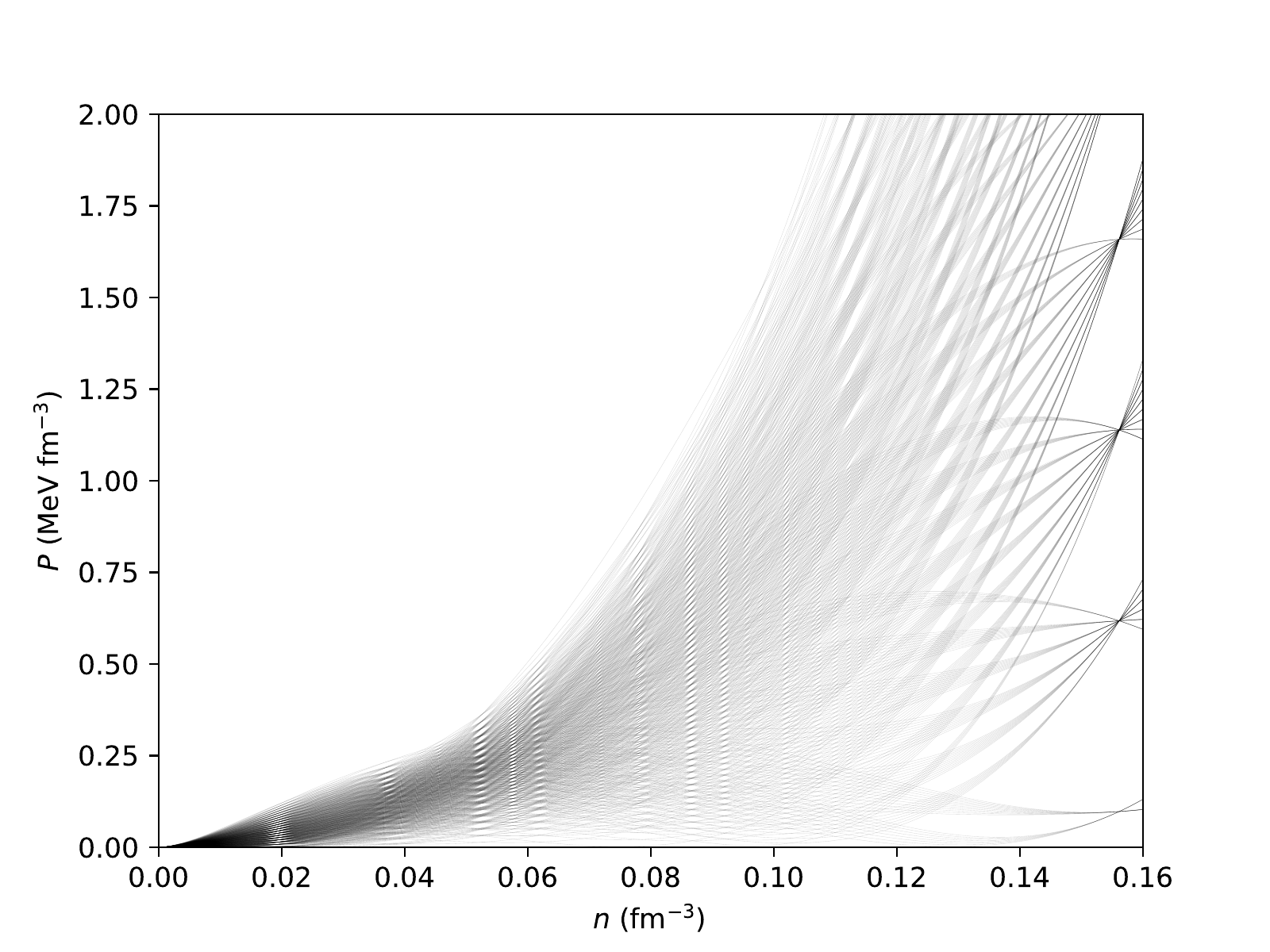}\includegraphics[width=0.45\linewidth]{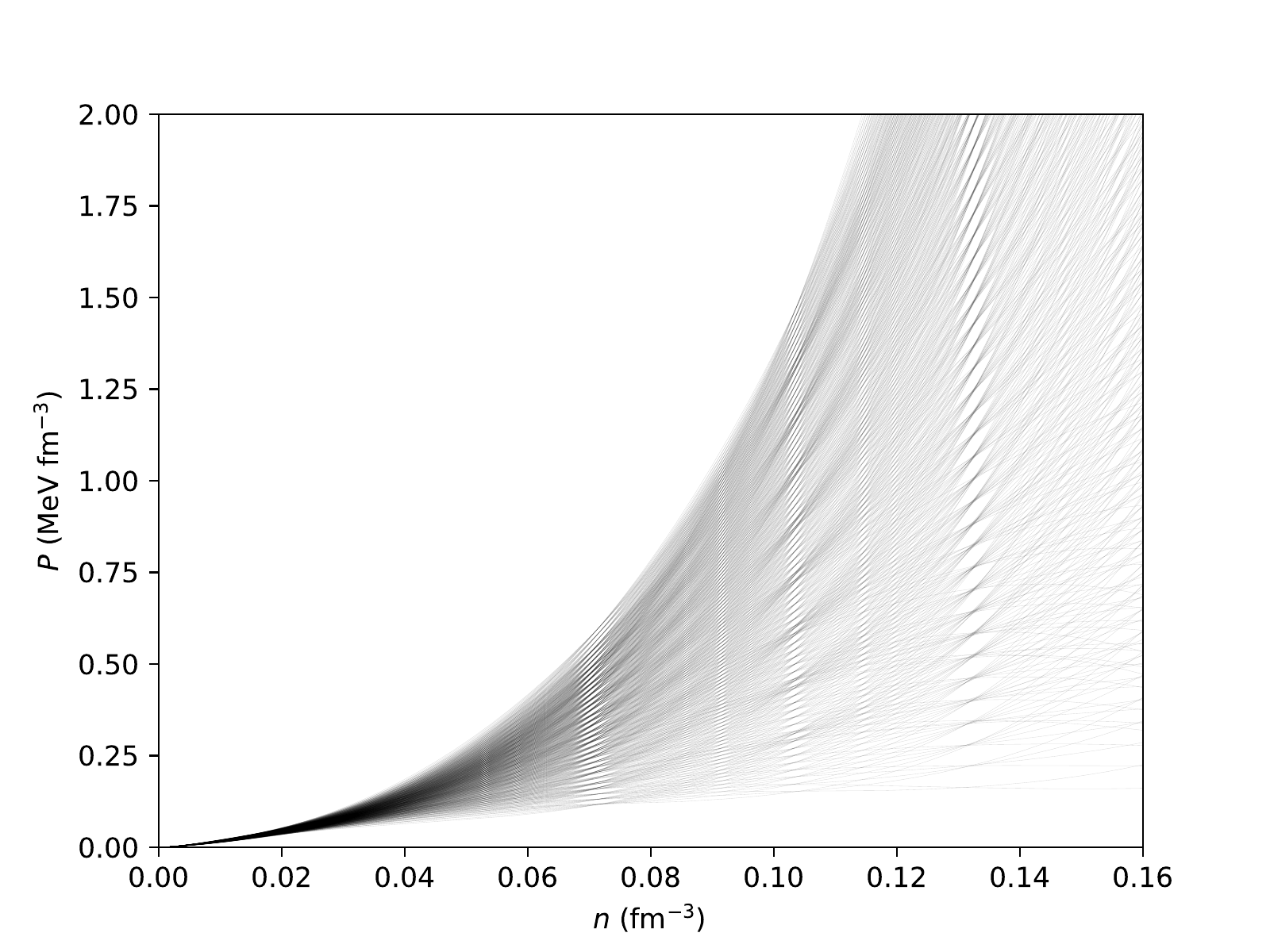}
    \caption{Depiction of the range of uncertainty of PNM (blue/grey band) and where the original Skyrme EOS fall on that band. It shows energy density per number density in relation to density up to nuclear saturation density. It is clearly shown that the models do not cover much of where the expected EOS could be. EOS of the 480 extended Skyrme models as compared to the (blue/grey) band of expected PNM. It shows energy density per number density in relation to density up to nuclear saturation density. The extended Skyrme cover a vast mast majority of the band of PNM up to nuclear saturation density.}
    \label{fig:8}
\vspace{11pt}
\vspace{11pt}
\vspace{11pt}
\end{figure}

We summarize here the process of creating our two ensembles of crust models. They are based around two samplings of the symmetry energy parameters. The first is a uniform sampling in $J$, $L$ and $K_{\rm sym}$ (in Bayesian language, uninformative priors). The ensemble of models based on this will be referred to as our ``uniform priors''. Secondly, we sample uniformly from the parameters $J$ and $a_0$ and $b_{12}$ defined above across a range that conservatively spans the predictions of chiral-EFT. These are referred to as our ``PNM priors''.

For the purposes examining the properties of the resulting crust models, we created 729 nuclear matter models from our PNM priors and 1404 from our uniform prior. We created the corresponding crust models using the CLDM with the best-fit surface parameters given in table~1. We then further filter our EOSs to rule out those that do not give a stable crust EOS - for the most part, those EOSs whose PNM pressure becomes negative in some region of the crust. After filtering out those EOSs, we end up with 658 models in our PNM prior ensemble and 558 in our uniform ensemble. The filter rules out many fewer EOSs from our PNM prior ensemble because incorporating our best knowledge of the PNM EOS generally guarantees, for the most part, a well-behaved PNM EOS at sub-saturation densities. The regions in the $J$-$L$ that are filtered out by this process are shown by the grey regions in figure~2. After the filter, figure~7 shows the resulting two distributions of $J$, $L$ and $K_{\rm sym}$. In Fig~8, we plot the PNM EOSs predicted by our two ensembles of models after unstable crust models have been filtered out: the uniform priors are shown on the left and the PNM priors on the right. The top plots are the energy per neutron and the bottom plots are the pressure, as a function of density. The blue bands shown in the energy per neutron plots are the uncertainty band from chiral-EFT \citep{Tews:2016ty}. The PNM priors systematically cover the uncertainty band, while the uniform priors cover a much larger range of PNM EOS space.

\subsubsection{The bootstrap method}

Because we are sampling only of order ~1000 models from each set of prior distributions of nuclear matter parameters, we need a way of estimating the corresponding sampling error on the resulting crust properties. We use the bootstrap resampling method \citep{Efron:1979aa,Pastore:2019aa}. This is a simple example of resampling with replacement: for example, for our PNM ensemble (658 models) we randomly select 658 models at random, returning each randomly selected model to the ensemble before drawing another. We repeat this of order 1000 times (thus giving us 1000 PNM ensembles rather than one); each time, we find the 5th, 50th and 95th percentile of a certain crust property we are interested in. The values of the properties at these percentiles follow a normal distribution from which it is straightforward to extract 2-$\sigma$ errors on the percentiles.

\revision{We are sampling directly from a given input distribution of model parameters, rather than from the residuals of a fit of those parameters to empirical data. The bootstrap method can also be used to assess the errors on the latter; in that case, however, one must take care because the residuals - the differences between model calculations and data - are dependent on the model parameters \citep{Bertsch:2017wv} and this can result in underestimating the error. This effect is not present in our results.}

\subsubsection{The maximal information coefficient}

We want to be able to quantify the relationships between crust and nuclear observables. We will find that some of the most most notable relationships are non-linear, so we must move beyond measures of strictly linear correlations. We choose to employ the maximal information coefficient (MIC) \citep{Li:1990aa,Reshef:2011aa} - a measure of the mutual information content of pairs of variables.

Given a probability distribution over two variables $A$ and $B$, $P(A,B)$, and the marginal probability distributions $P(A)$ and $P(B)$, the mutual information correlation $I(A,B)$ is defined 
\begin{equation}
I(A,B) = \int P(A,B) \log\bigg(\frac{P(A,B)}{P(A)P(B)}\bigg) dA dB.
\end{equation}

\noindent This is a measure of the nonlinear correlation between variables $A$ and $B$ - the extend to which a measurement of $A$ contains information about the value of $B$, or the ``the degree of predictability'' in the relationship \citep{Li:1990aa}. This is most conveniently expressed by normalizing $I(A,B)$, which obtains the maximal information coefficient (MIC) which ranges from 0 (statistical independence) to 1 (noiseless relationship). The fact that it captures non-linear, non-monotonic relationships is an advantage over, for example, the often used Pearson correlation coefficient. The MIC tends to give lower power for many monotonic relationships compared to simpler correlation measures such as the Pearson coefficient \citep{Simon:2014aa}, and so can generally be considered a lower bound on the measure of a strength of a correlation.

%
%

\section{Results}

Before we start, let us note that although the crust core transition baryon density $n_{\rm cc}$ is ordinarily presented first in studies like this, and we follow suit, this is \emph{not} the quantity that determines the location of the crust-core boundary; that is the pressure and chemical potential (or equivalently energy density). $n_{\rm cc}$ is useful only if one uses the same EOS as was used to determine it to find the pressure and energy density at that density. Essentially, the pressure determines the mass coordinate of the crust-core boundary, and the chemical potential determines the radial coordinate.


\subsection{The crust-core boundary}

Let us first demonstrate how the thermodynamic and dynamic spinodal methods compare to the full compressible liquid drop model when it comes to calculating the crust-core transition properties. \revision{This will allow us to assess the model uncertainty inherent in the various descriptions of the symmetry energy.}

In Figs~9 and 10 we show histograms for the crust core transition density $n_{\rm cc}$, pressure $P_{\rm cc}$ and chemical potential $\mu_{\rm cc}$ for our uniform prior distribution (figure~9) and our PNM prior distribution (figure~10). Above the histograms we show points indicating the position of the 2.5th, 50th and 97.5th percentiles of the distributions. In addition, we indicate the sampling error with 2-$\sigma$ error bars on those points obtained using the bootstrap method.

\begin{figure}[!t]
    \centering
    \includegraphics[width=0.33\linewidth]{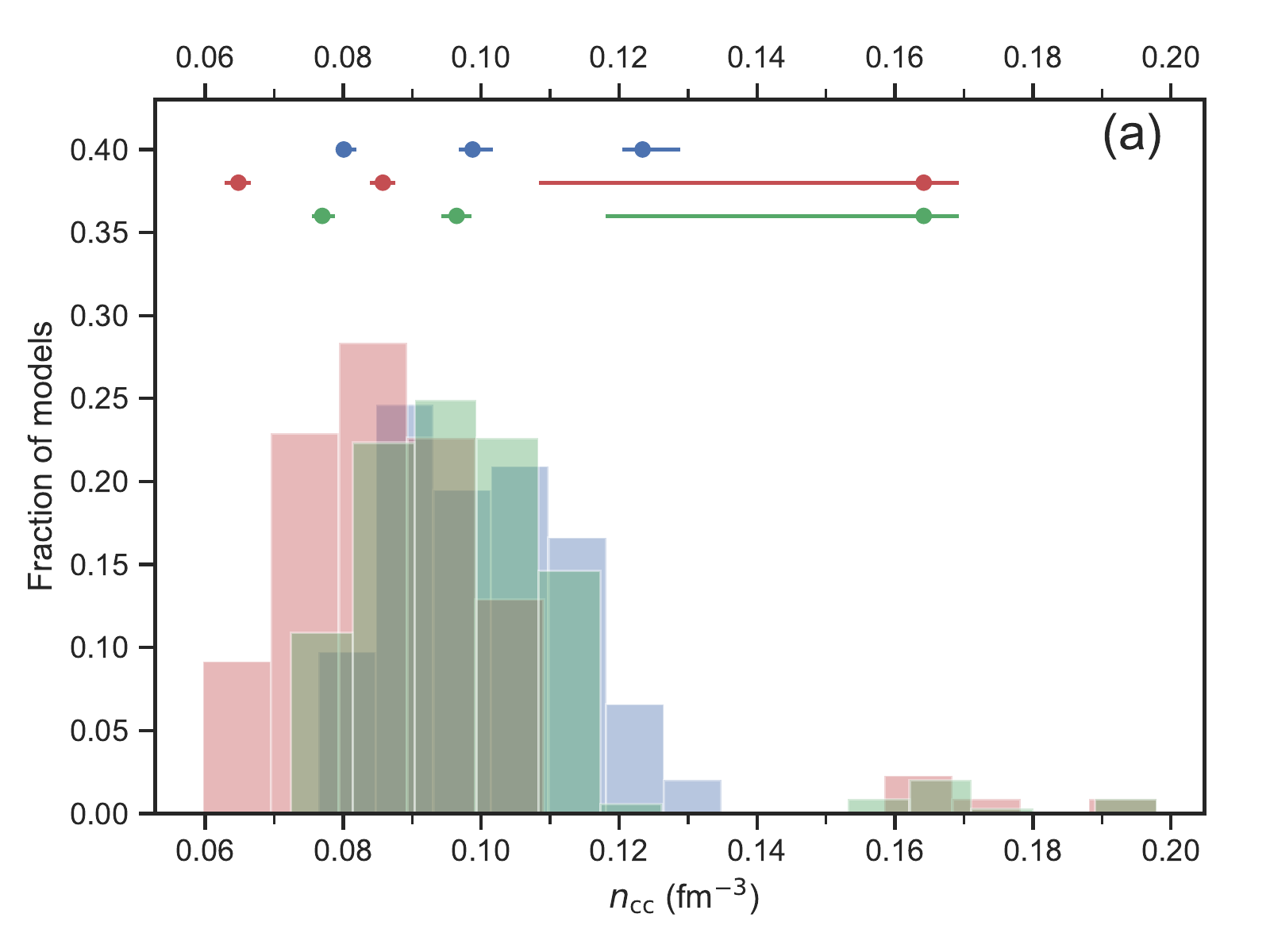}\includegraphics[width=0.33\linewidth]{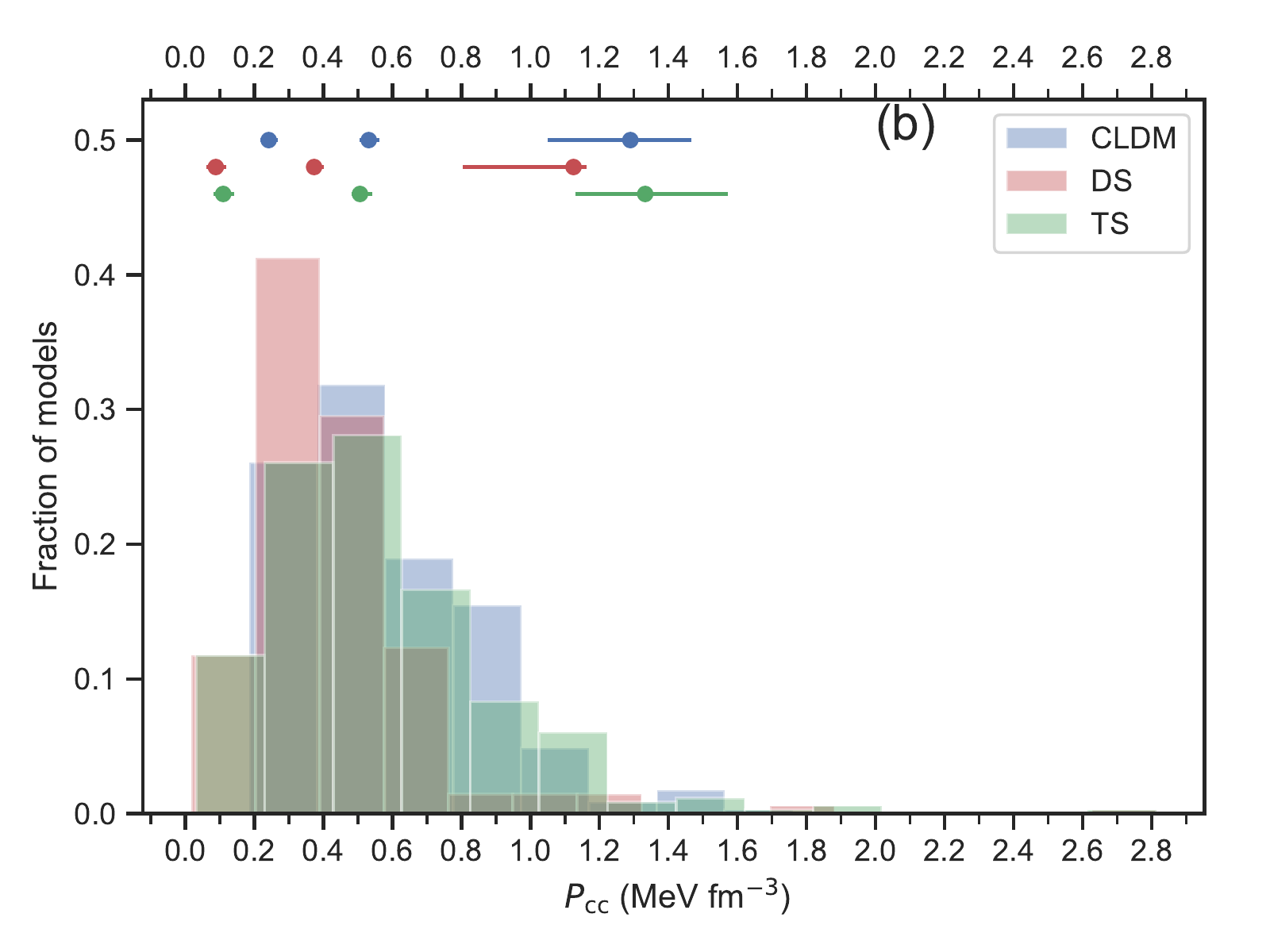}\includegraphics[width=0.33\linewidth]{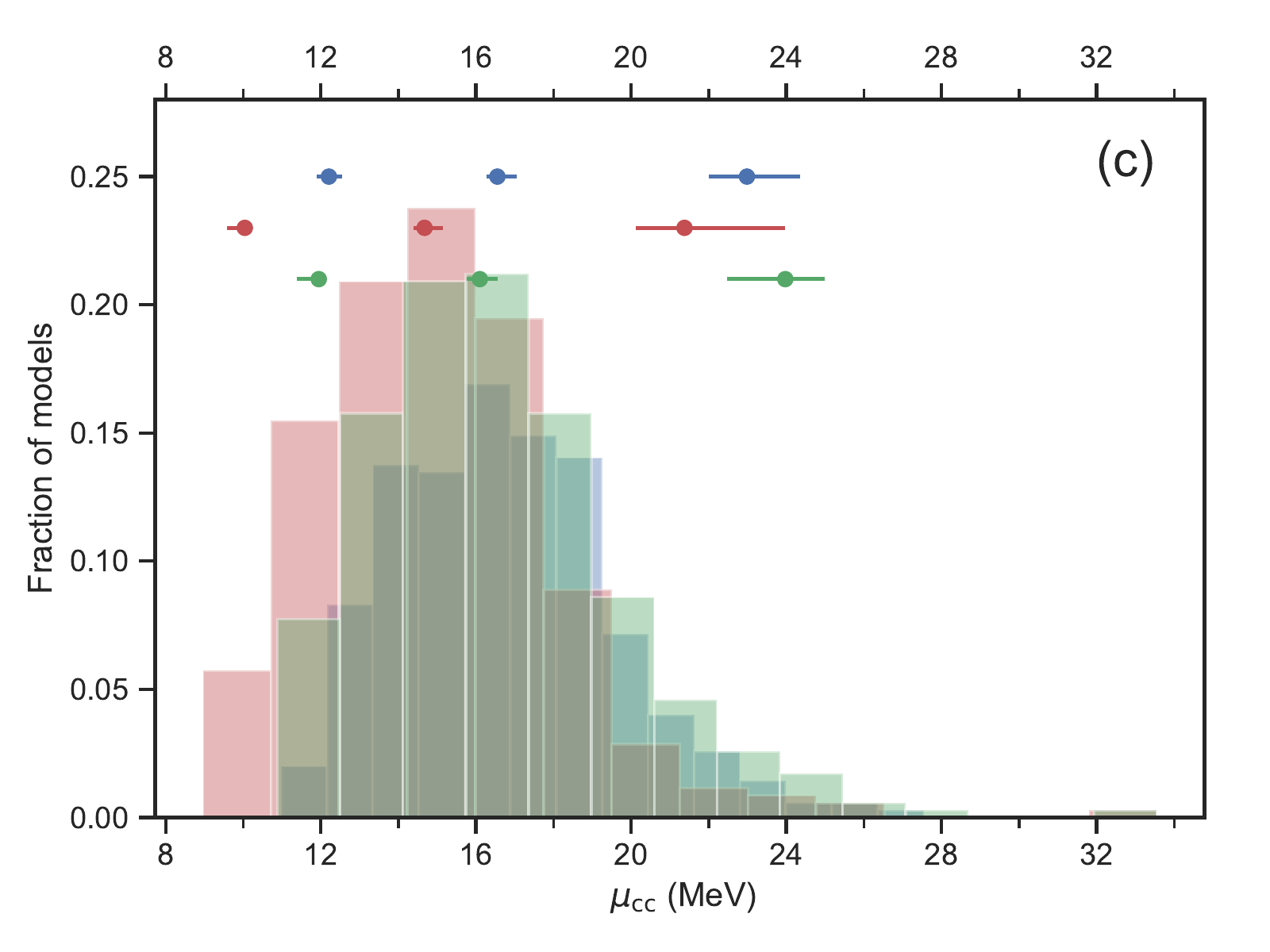}\\
    \caption{Histrograms comparing three ways of locating the crust-core boundary: the compressible liquid drop model (CLDM, blue bars), the dynamic spinodal (DS, red bars) and the thermodynamic spinodal (TS, green bars). For each transition model, we show their predicted values of the crust-core transition density ($n_{\rm cc}$) (a), pressure ($P_{\rm cc}$) (b), and chemical potential ($\mu_{\rm cc}$) (c), for the 558  Skyrme models comprising our uniform prior distribution. Points indicating the 2.5th, 50th and 97.5th percentiles of the distributions are indicated at the top of the plots with their associated 2-$\sigma$ sampling error bars estimated using the bootstrap method.  The compressible liquid drop model and thermodynamic spinodal are in good agreement with each other, whereas the dynamic spinodal underestimates the crust-core transition properties.}
    \label{fig:9}
\end{figure}

\begin{figure}[!t]
    \centering
    \includegraphics[width=0.33\linewidth]{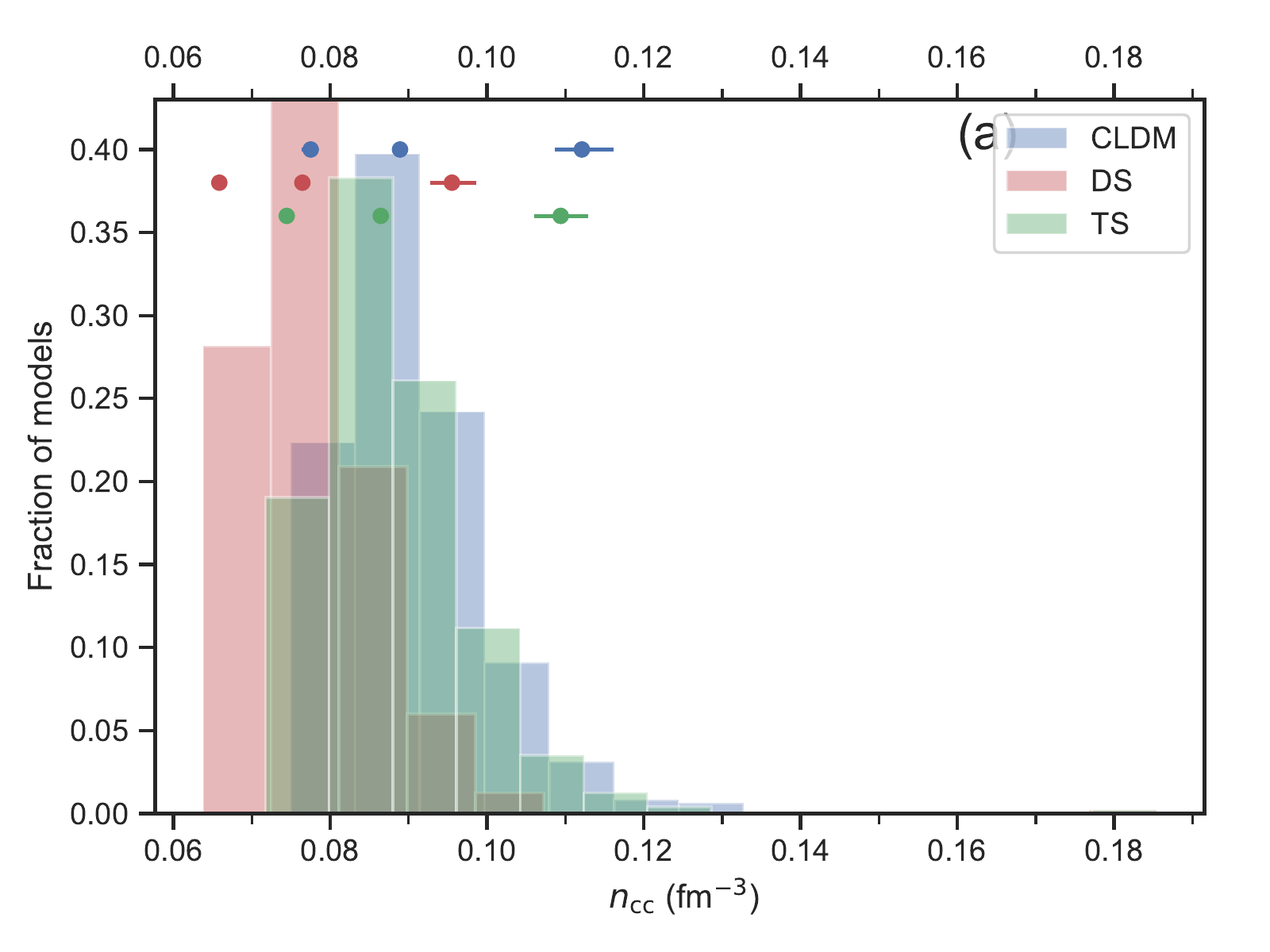}\includegraphics[width=0.33\linewidth]{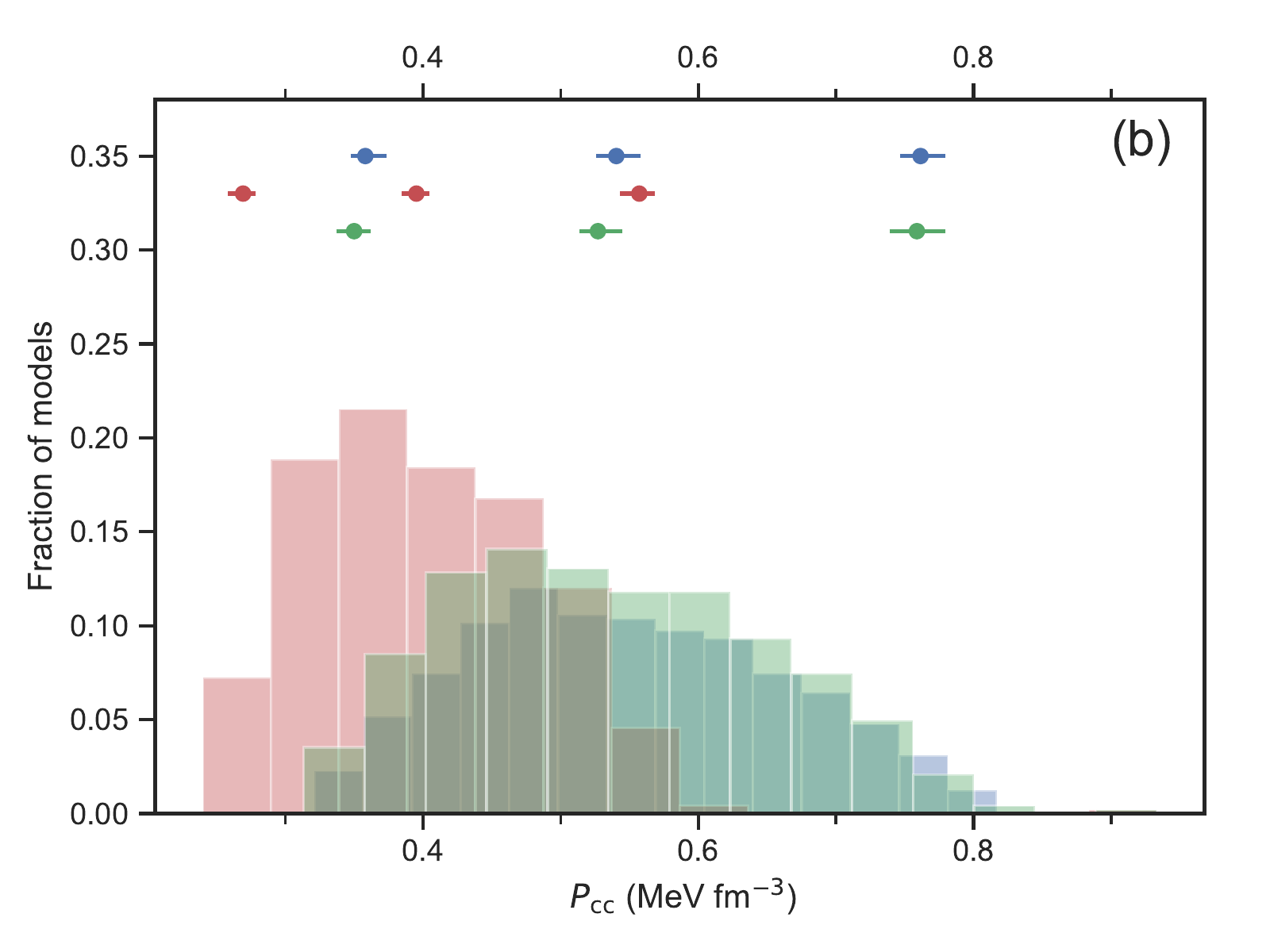}\includegraphics[width=0.33\linewidth]{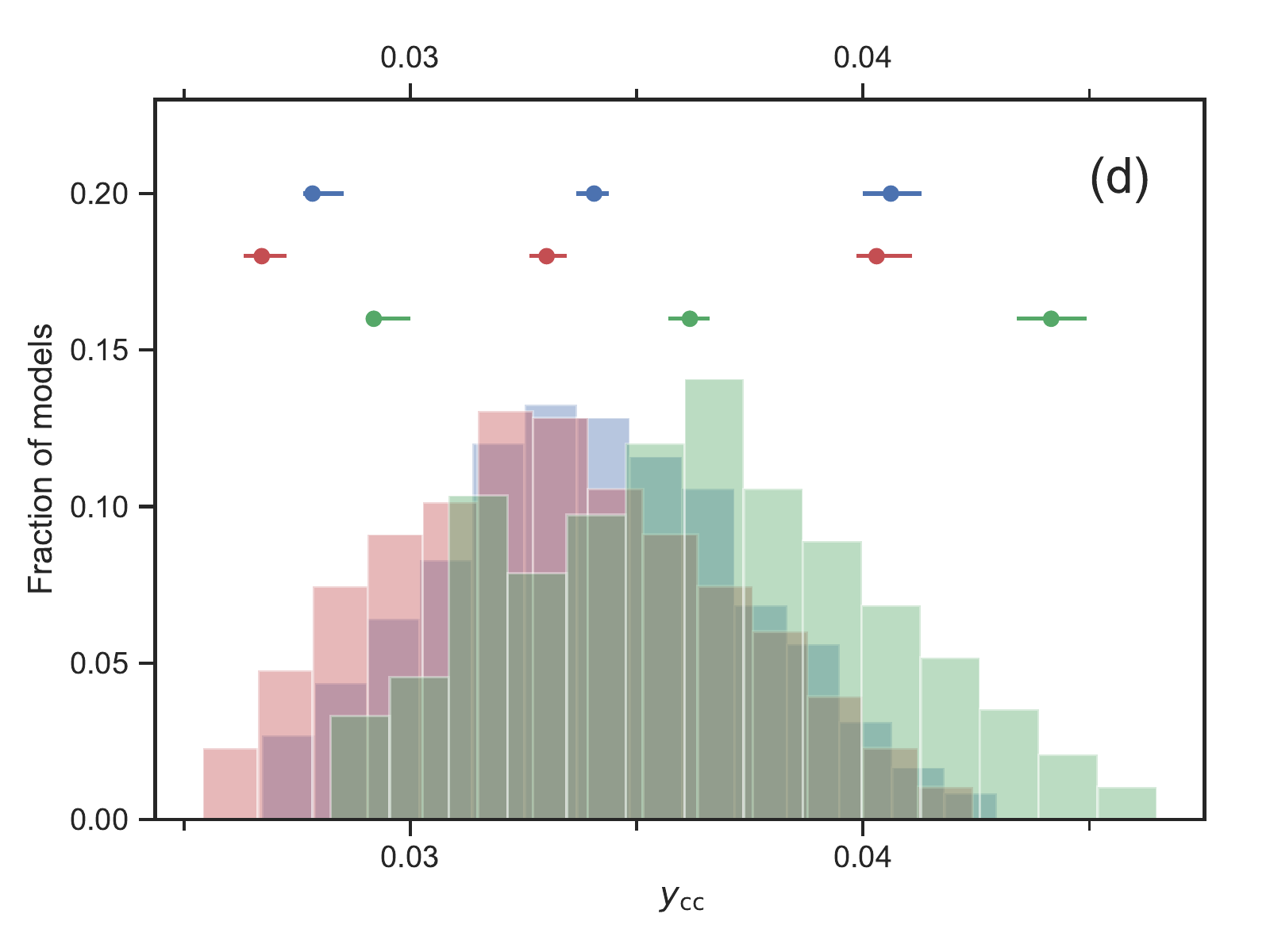}\\
    \caption{Same as Fig~9, but for the 658 Skyrme models comprising our PNM prior distribution.}
    \label{fig:10}
\end{figure}

The dynamic and thermodynamic spinodal distributions of the crust-core transition density $n_{\rm cc}$ are bimodal, with a small peak 0.16-0.2 fm$^{-3}$. This second mode in the distribution is due to the small number of PNM EOSs which give a local minimum in the pressure which leads to a vanishing compressibility of beta-equilibrium matter in the vicinity of that density range. The 97.5th percentiles of both distributions are consistent with each other, albeit with relatively large sampling errors because of the small number of points that contribute to the second mode of the distribution. However, for the bulk of the distribution the thermodynamic spinodal in consistent with the CLDM; the 2.5th and 97.5th percentiles are in close agreement, whereas the dynamic spinodal predicts crust-core transition densities about 0.01 fm$^{-3}$ below. Similarly, the thermodynamic spinodal and CLDM predictions for the crust-core transition pressure $P_{\rm cc}$ and chemical potential $\mu_{\rm cc}$ also agree, with the dynamic spinodal predicting a systematically smaller pressure by $\approx$ 0.1 MeV fm$^{-3}$ and chemical potential by $\approx 1$ MeV.

\revision{The thermodynamic spinodal method neglects gradient terms which penalize clustered matter and so is expected to provide an upper bound on the transition density, pressure and chemical potential. The CLDM follows the thermodynamic spinodal despite including surface terms. This is likely a result of the value of the $p$ parameters in the surface tension which governs the surface energy at high isospin asymmetry. We found it to be large ($p=3.8$) which corresponds to a smaller surface energy at high isospin asymmetry. On the other hand, the dynamic spinodal contains gradient terms whose magnitude is determined directly from the Skyrme models. Note that because we do not refit the Skyrme parameters, the gradient terms do not change with $J$, $L$ and $K_{\rm sym}$. There is still a dependence of the surface energy on bulk nuclear parameters in the dynamic spinodal model via the derivatives of the chemical potentials in equation~16. In the CLDM, the surface energy is correlated directly with $J$ through the parameter $c$, and indirectly on bulk nuclear parameters through the dependence on the proton fraction at beta-equilibrium. The dynamic spinodal generally leads to a lower bound on the transition density, pressure and chemical potential.}

\revision{These different surface energy treatments illustrate an important model uncertainty. The CLDM uses a surface energy form adapted to the neutron rich environment and fit to calculations of inner crust nuclei, but is not derived consistently from the EDF used for the bulk nuclear matter that also determines the symmetry energy. The dynamic spinodal contains density gradient terms from the EDF consistently but those terms are not refit to finite nuclear properties. How important that is for the crust-core transition, given the physical difference between the sharp finite nuclear surface and the diffuse surface with a neutron gas deep in the crust, should be further explored.}

\revision{In the light of these three methods, let us also discuss how we should compare the neutron skin calculations. The gradient terms are no longer consistent with the isovector parameters varied to give the sets of $J$,$L$ and $K_{\rm sym}$ values. We have discussed why this should not appreciably affect the relationship between the neutron skins and $J$,$L$ and $K_{\rm sym}$. The dynamic spinodal uses the same gradient terms, and while the neutron skins are insensitive to the gradient terms, the crust-core transition might not be \citep{Carreau:2019aa,Antic:2019tk}, and so a refit may alter the correlations between neutron skins and crust-core transition for the dynamic spinodal results, so the inconsistency could have an impact here. The thermodynamic spinodal uses only the bulk parameters and can also be compared consistently with the neutron skin calculations given the lack of sensitivity to gradient parameters. The CLDM uses a surface energy that is not consistently obtained from the gradient terms of the EDF.}

\revision{Let us therefore examine whether there is a significant difference in the relationships obtained between crust properties, symmetry energy parameters and neutron skins for each of the three methods used in this work.}

\begin{figure}[t!]
    \centering
    \includegraphics[width=0.95\linewidth]{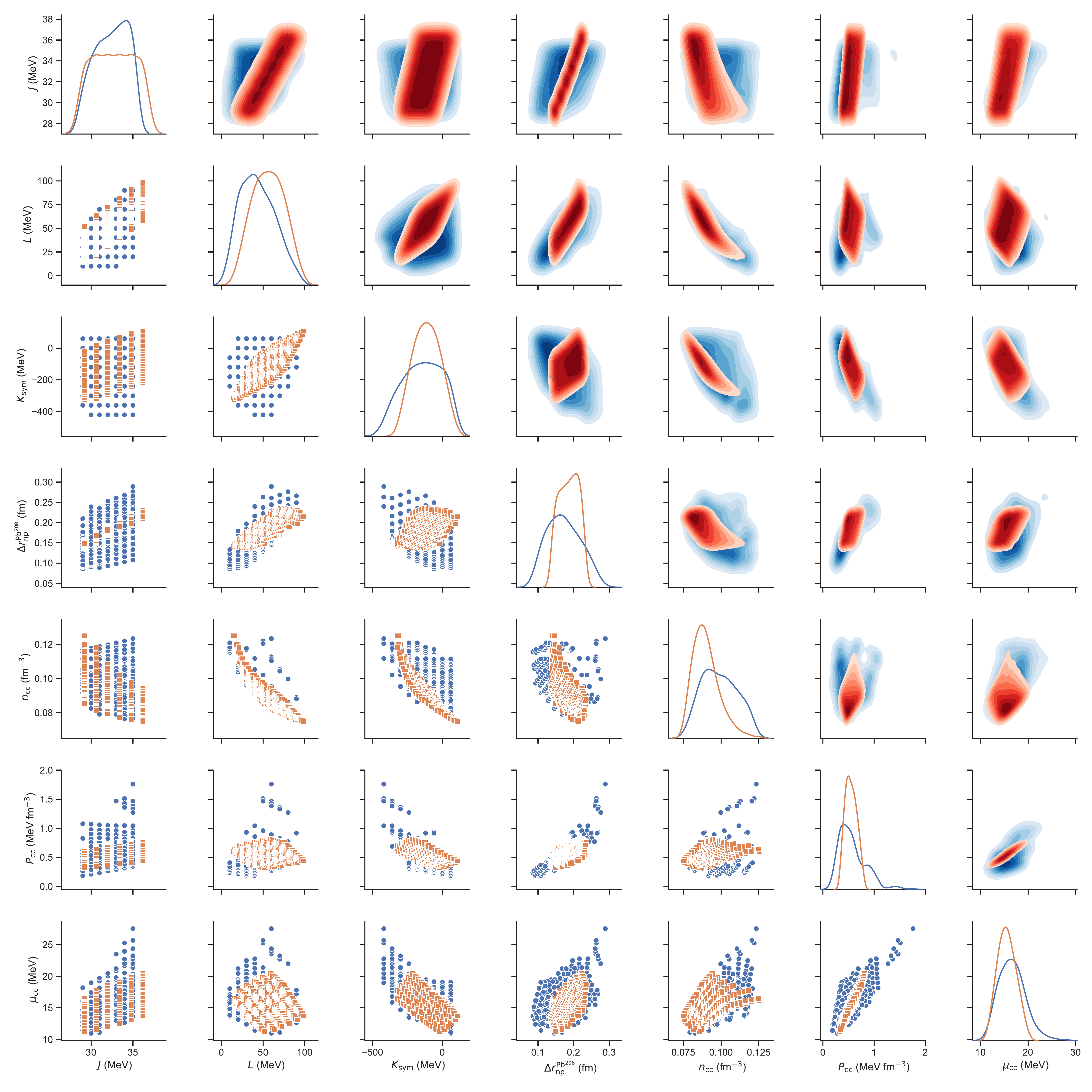}
    \caption{Plots of the relationships between symmetry energy parameters $J$, $L$ and $K$, the neutron skin of lead $\Delta r_{\rm np}^{Pb^{208}}$ and crust-core transition densities $n_{\rm cc}$, pressures $P_{\rm cc}$, and chemical potentials $\mu_{\rm cc}$ calculated using the CLDM for uniform (blue) and pure neutron matter (red) priors. Two different versions of the plots are displayed; below the diagonal are scatter plots where each point is a single model; square points show the PNM priors and circles shows the uniform priors. Above the diagonal are density plots that reveal more clearly for which regions of parameter space are the models are more concentrated.}
    \label{fig:11}
\vspace{11pt}
\vspace{11pt}
\vspace{11pt}
\end{figure}

\begin{figure}[t!]
    \centering
    \includegraphics[width=0.95\linewidth]{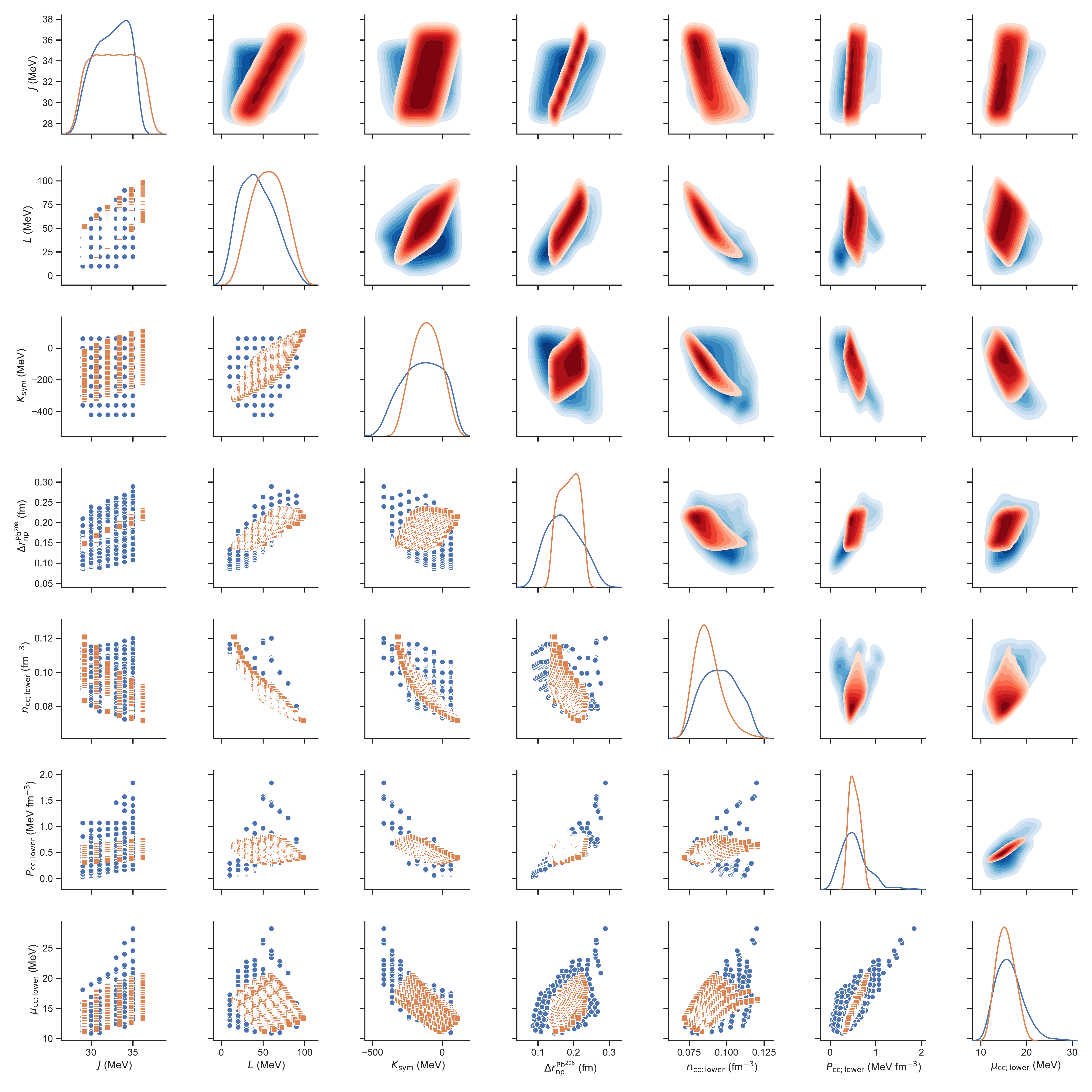}
    \caption{Plots of the relationships between symmetry energy parameters $J$, $L$ and $K$, the neutron skin of lead $\Delta r_{\rm np}^{Pb^{208}}$ and crust-core transition densities $n_{\rm cc}$, pressures $P_{\rm cc}$, and chemical potentials $\mu_{\rm cc}$ calculated from the thermodynamic spinodal for uniform (blue) and pure neutron matter (red) priors. Two different versions of the plots are displayed; below the diagonal are scatter plots where each point is a single model; square points show the PNM priors and circles shows the uniform priors. Above the diagonal are density plots that reveal more clearly for which regions of parameter space are the models are more concentrated.}
    \label{fig:12}
\vspace{11pt}
\vspace{11pt}
\vspace{11pt}
\end{figure}

\begin{figure}[t!]
    \centering
    \includegraphics[width=0.95\linewidth]{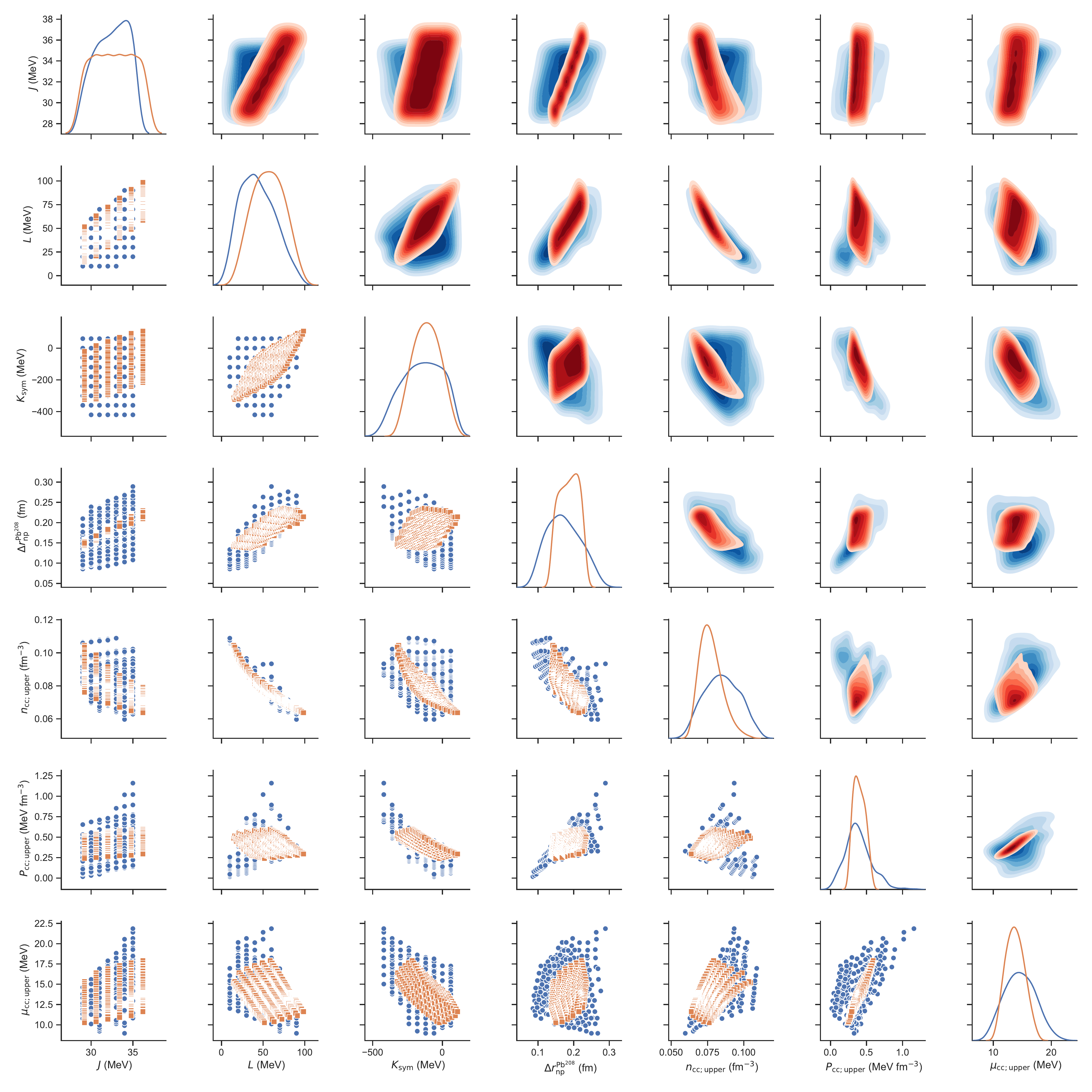}
    \caption{Plots of the relationships between symmetry energy parameters $J$, $L$ and $K$, the neutron skin of lead $\Delta r_{\rm np}^{Pb^{208}}$ and crust-core transition densities $n_{\rm cc}$, pressures $P_{\rm cc}$, and chemical potentials $\mu_{\rm cc}$ calculated from the dynamic spinodal for uniform (blue) and pure neutron matter (red) priors. Two different versions of the plots are displayed; below the diagonal are scatter plots where each point is a single model; square points show the PNM priors and circles shows the uniform priors. Above the diagonal are density plots that reveal more clearly for which regions of parameter space are the models are more concentrated.}
    \label{fig:13}
\vspace{11pt}
\vspace{11pt}
\vspace{11pt}
\end{figure}

\revision{In order to compare the effect of the different surface energy treatments, in Figures~9-11 we plot the relationships between the symmetry energy parameters, the neutron skin of $^{208}$Pb and the crust-core transition properties as calculated using the CLDM (Figure 9), the thermodynamic spinodal (Figure 10) and the dynamic spinodal (Figure 11). In each plot, the bottom-left corner shows scatter plots with a point for each model while the top right shows density plots to reveal the higher-probability regions. Uniform priors are shown in blue and PNM priors in red. We shall examine the relationships between the quantities in more detail in the next section, but here we want to contrast the relationships between pairs of quantities in the three figures. Qualitatively, we see that the relationships appear very similar; it looks like the correlations between symmetry energy, neutron skin and crust transition properties are not affected significantly by the different treatments of the surface; the main effect is that the dynamic spinodal systematically predicts a transition density around 0.01fm$^{-3}$ lower than the other two models.}

\revision{In order to quantitatively analyze the extent to which the different ways the surface energy treatment affects correlations between nuclear matter parameters, neutron skins and crust transition properties, in Figure~12 we show the maximal information coefficients characterizing the correlations between the symmetry energy parameters at saturation density $J$, $L$, $K_{\rm sym}$, the symmetry energy and its slope at 0.1fm$^{-3}$ and the crust-core transition density, pressure and chemical potential calculated using the CLDM (the subscript L), the thermodynamic spinodal (the subscript T) and the dynamic spinodal (the subscript D). The top chart gives the results from the uniform priors and the bottom for the PNM priors.}

\revision{The MIC shows quantitatively that the three methods of calculation give similar correlations. In a later section we will discuss the specific correlations in detail. Here, as representative examples consider the correlations between $L$, $K_{\rm sym}$ (0.1fm$^{-3}$) and $\Delta r_{\rm np}^{Pb^{208}}$. For uniform priors $L$ correlates strongly with the crust-core transition density and more weakly with the transition pressure. With PNM priors the correlation with the transition pressure disappears. These correlations are very similar for all three methods. $K_{\rm sym}$ (0.1fm$^{-3}$) correlates strongly with the transition density - especially when calculated using the CLDM and thermodynamic spinodal methods - for both uniform PNM priors. The neutron skin of lead shows some correlation with the transition pressure for uniform priors, albeit weaker using the dynamic spinodal method. There is little correlation between the neutron skin and transition properties for PNM priors independent of method used.}

\revision{The offset between the dynamic spinodal method, and the CLDM/thermodynamic spinodal results, is around 0.01 fm$^{-3}$ lower in density. We also note here that pushing $p$ to small values can lower the transition density by up to 0.02fm$^{-3}$, and indication of a conservative systematic model uncertainty. It has been previously found that the $c$ and $p$ surface parameters of the CLDM make little difference to the transition densities and pressures between spherical and pasta nuclei \citep{Newton:2013sp}. Therefore the dynamic spinodal would predict significantly less pasta. With all this in mind, in what follows we will use the CLDM model, as it also allows us to calculate the properties of nuclear pasta. The comparisons with the neutron skins should be read with the caveat that there is some inconsistency between the treatment of the surface which should be explored more thoroughly.}

\begin{figure}[!ht]
    \centering
    \includegraphics[width=0.7\linewidth]{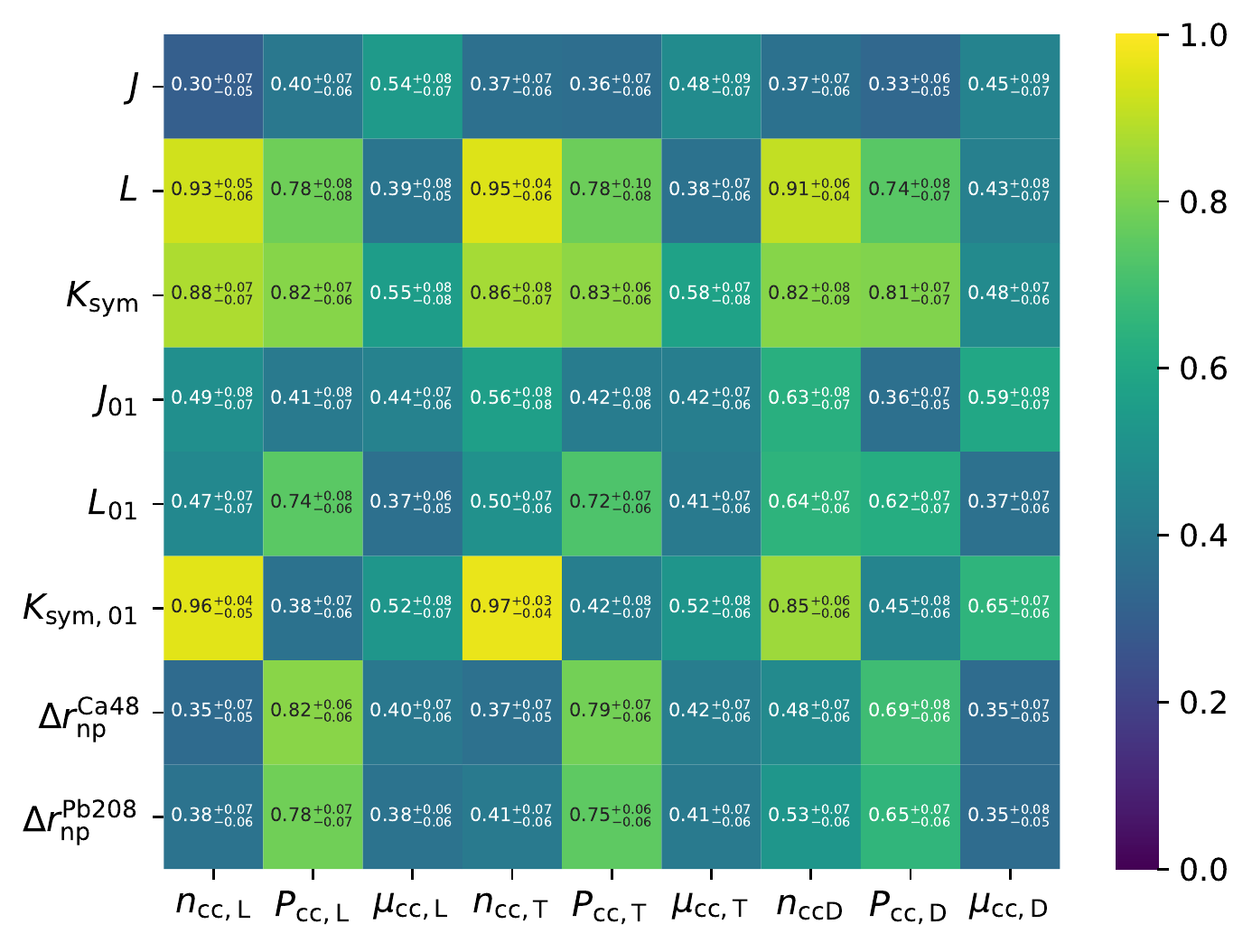}   \\ \includegraphics[width=0.7\linewidth]{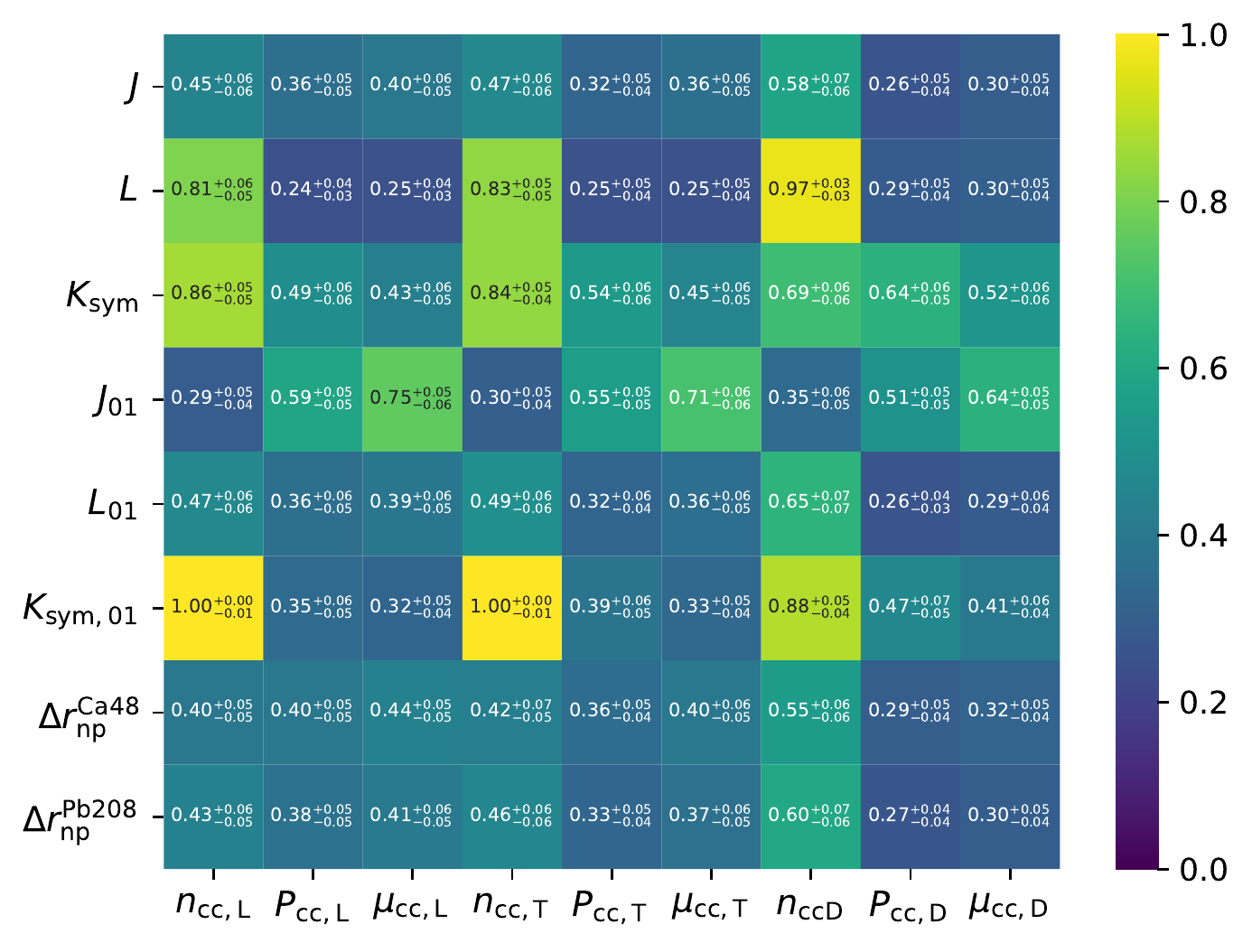}
    \caption{Maximal information coefficients between the crust-core transition properties using the CLDM (subscript L), thermodynamic spinodal (subscript T) and dynamic spinodal (subscript D) methods, and neutron skins and symmetry energy parameters for the uniform priors (top) and PNM priors (bottom). Using the bootstrap method, we show 2-$\sigma$ sampling uncertainties on each MIC.}
    \label{fig:14}
\vspace{11pt}
\vspace{11pt}
\vspace{11pt}
\end{figure}


\subsection{Crust-core and Pasta Transition Properties for Uniform and PNM priors}

In this section we present the results for the crust-core and pasta transition properties; see section~4 for a complete table of numerical values.

The crust-core transition properties for the uniform and PNM prior distributions are displayed in Figure~15. We show the distributions for the crust-core transition density $n_{\rm cc}$, pressure $P_{\rm cc}$, chemical potential $\mu_{\rm cc}$ and proton fraction $y_{\rm cc}$. The corresponding values of these quantities at the boundary between spherical nuclei and nuclear pasta are shown in Figure~16.    

It is notable that the \textit{range} of the crust-core transition density is very similar for both distributions: 0.096$^{+0.032}_{-0.020}$ fm$^{-3}$, and  0.092$^{+0.036}_{-0.015}$ fm$^{-3}$ respectively. However, there is a higher probability of a large ($>$0.1 fm$^{-3}$) crust-core transition densities for the uniform prior distribution. This illustrates the importance of considering the full probability distribution and not just the range. The ranges of the pasta transition density are, in contrast, quite different, although the median values are in agreement within sampling error. However, let us emphasize that the crust-core transition density is not the primary quantity that determines the location of the crust-core boundary.

\begin{figure}[!t]
    \centering
    \includegraphics[width=0.25\linewidth]{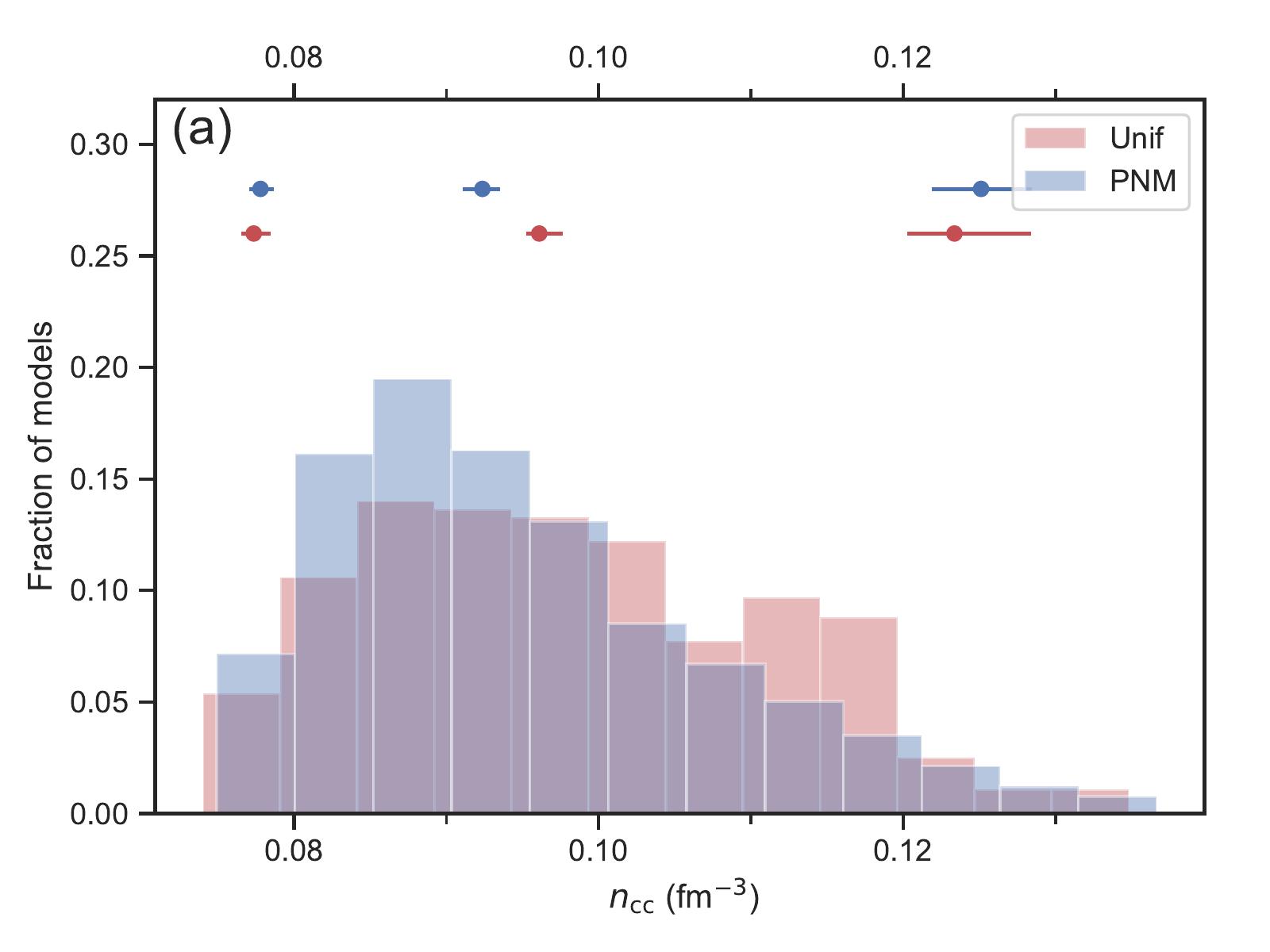}\includegraphics[width=0.25\linewidth]{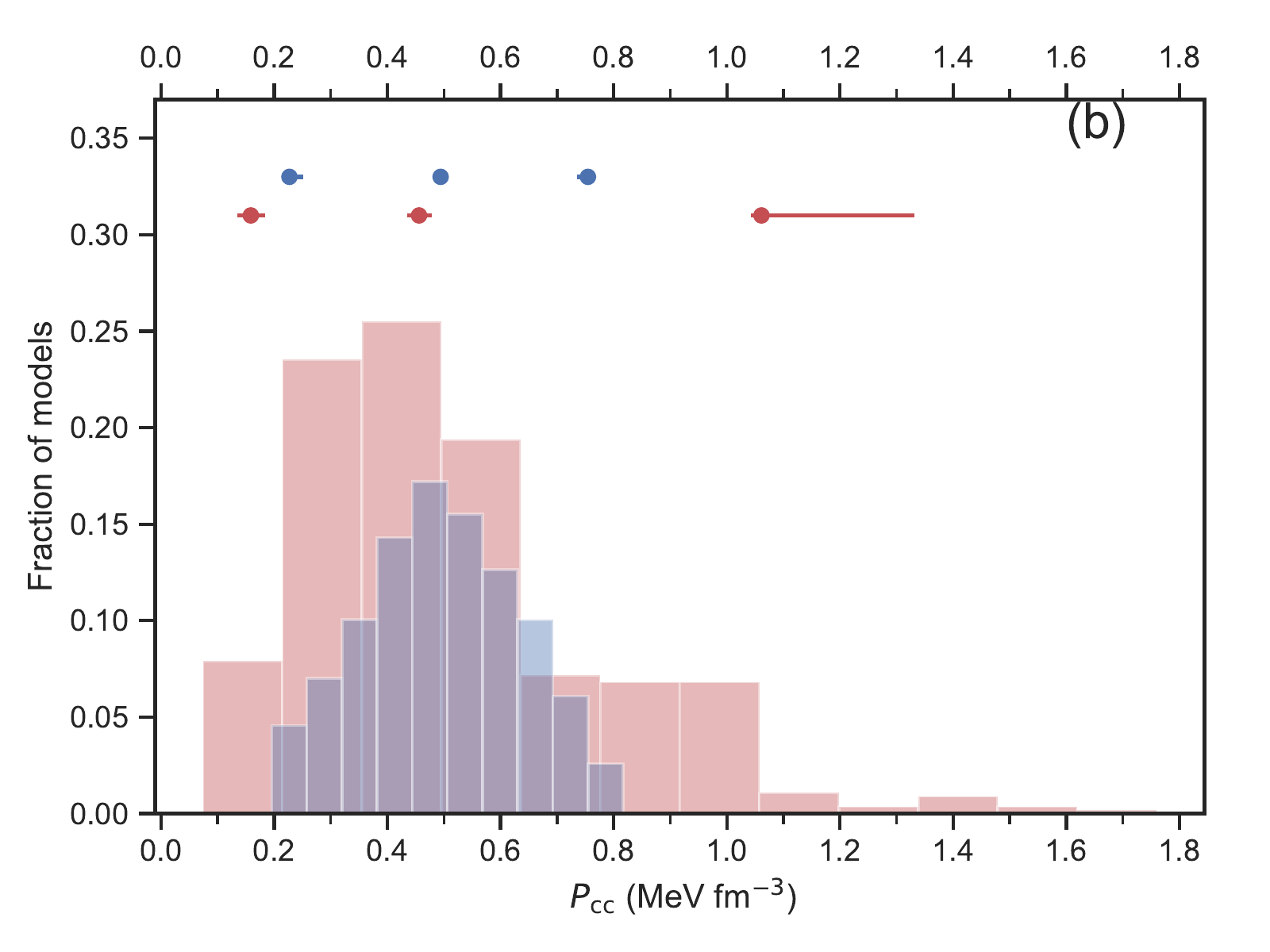}\includegraphics[width=0.25\linewidth]{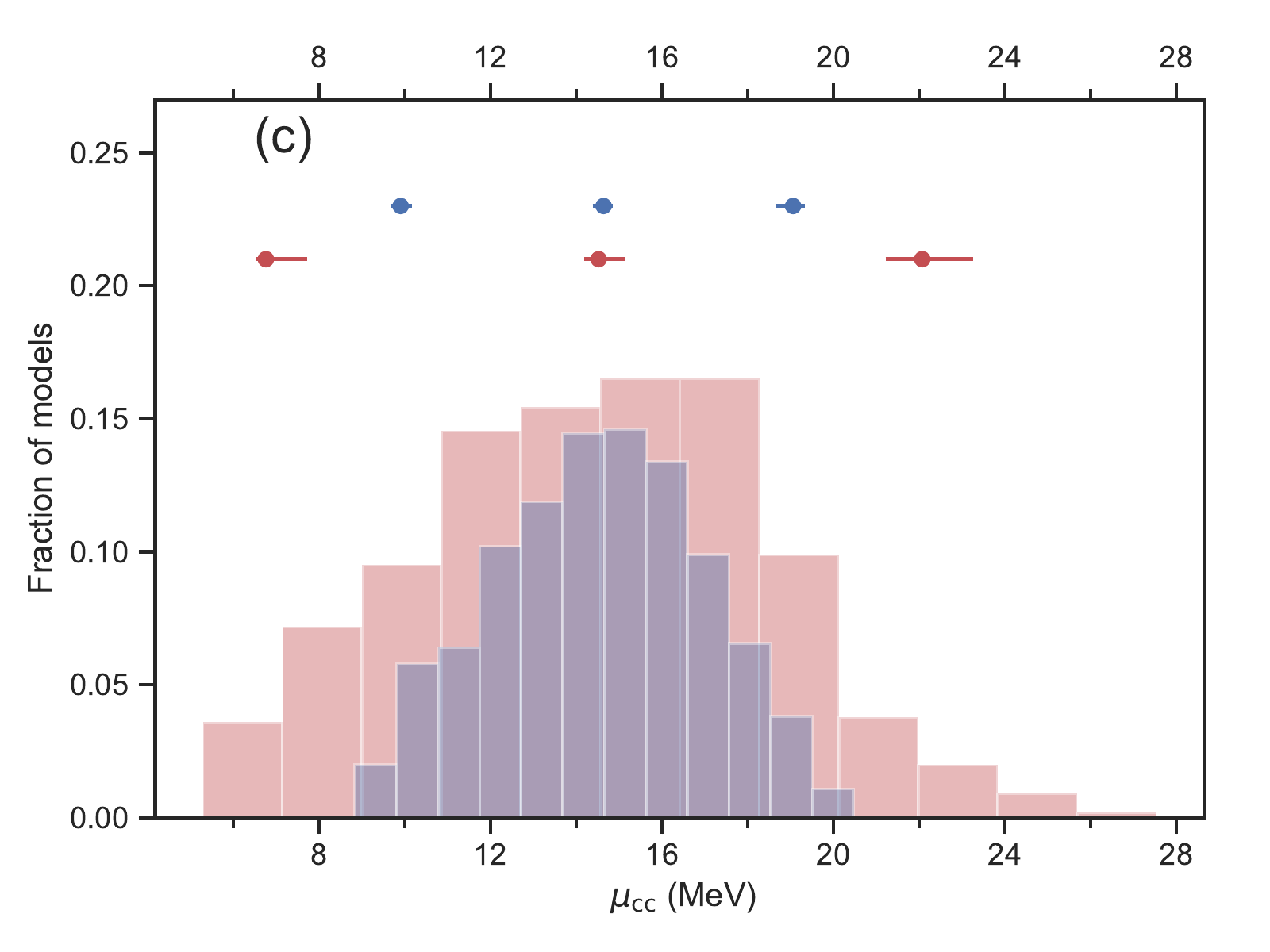}\includegraphics[width=0.25\linewidth]{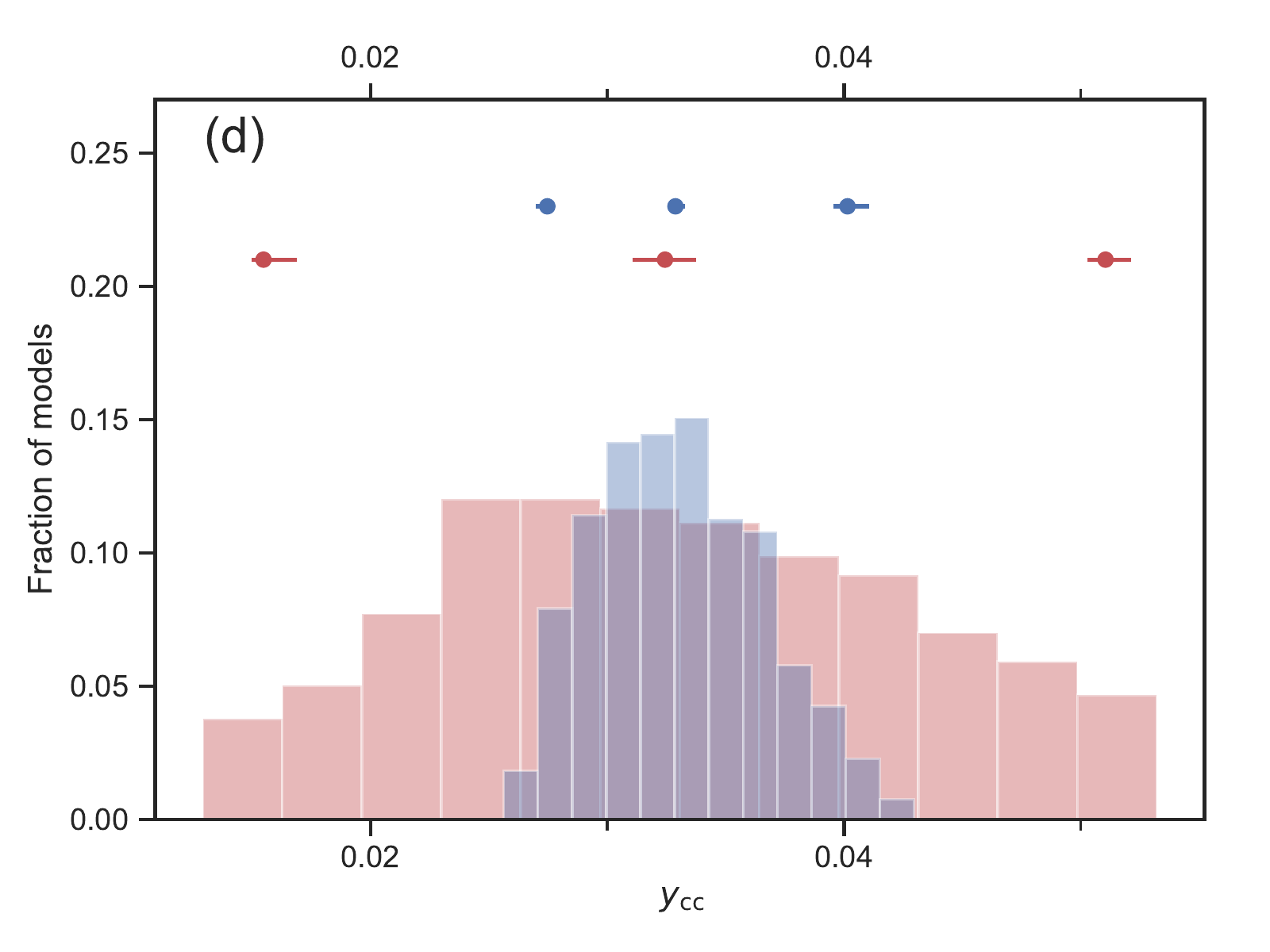}
    \caption{Crust-core transition density (a), pressure (b), chemical potential (c) and proton fraction (d) distributions from our uniform priors (red bars) and PNM priors (blue bars). Points indicating the 2.5th, 50th and 97.5th percentiles of the distributions are indicated at the top of the plots with their associated 2-$\sigma$ sampling error bars estimated using the bootstrap method.}
    \label{fig:15}
\end{figure}

\begin{figure}[!t]
    \centering
    \includegraphics[width=0.25\linewidth]{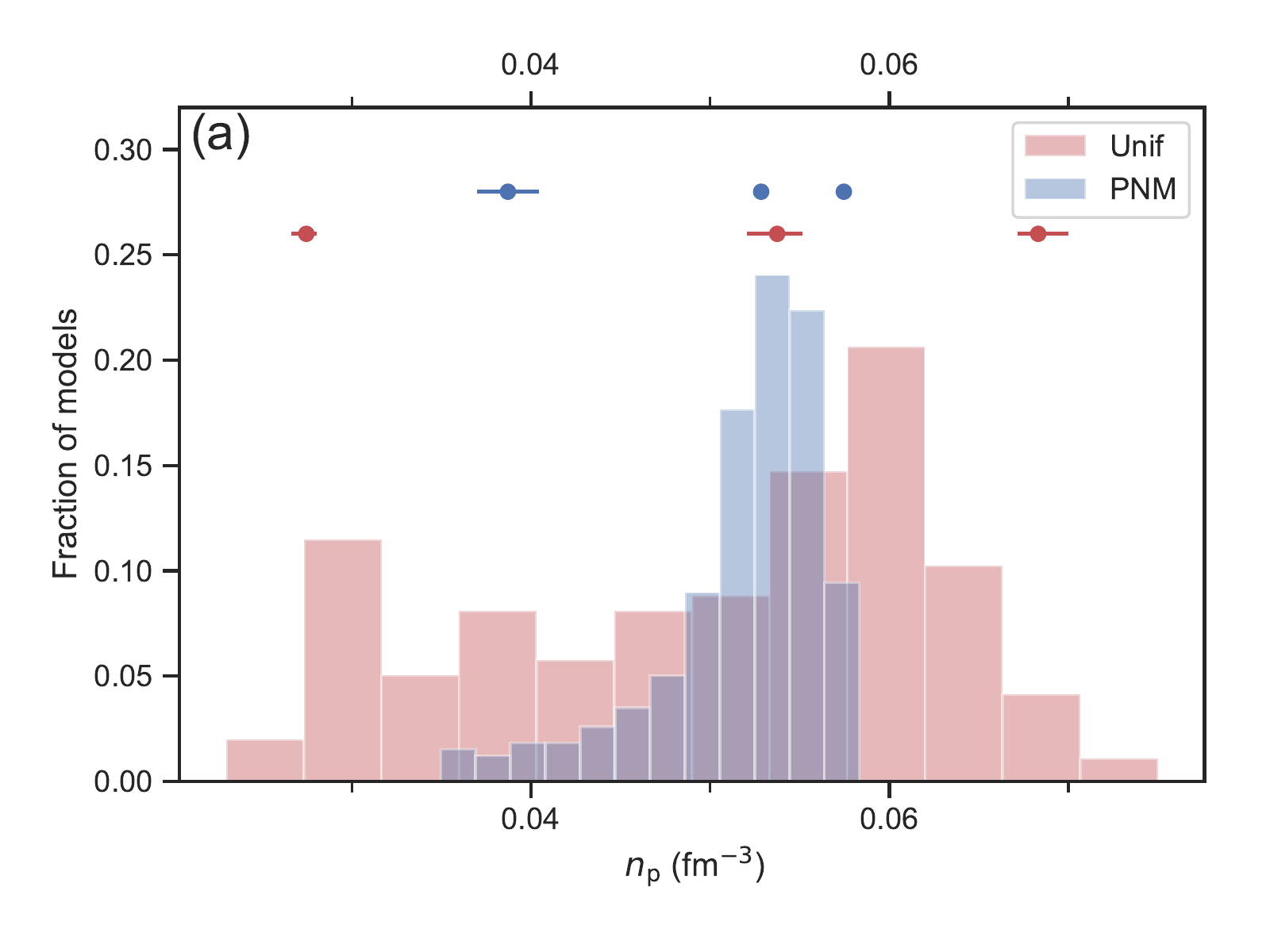}\includegraphics[width=0.25\linewidth]{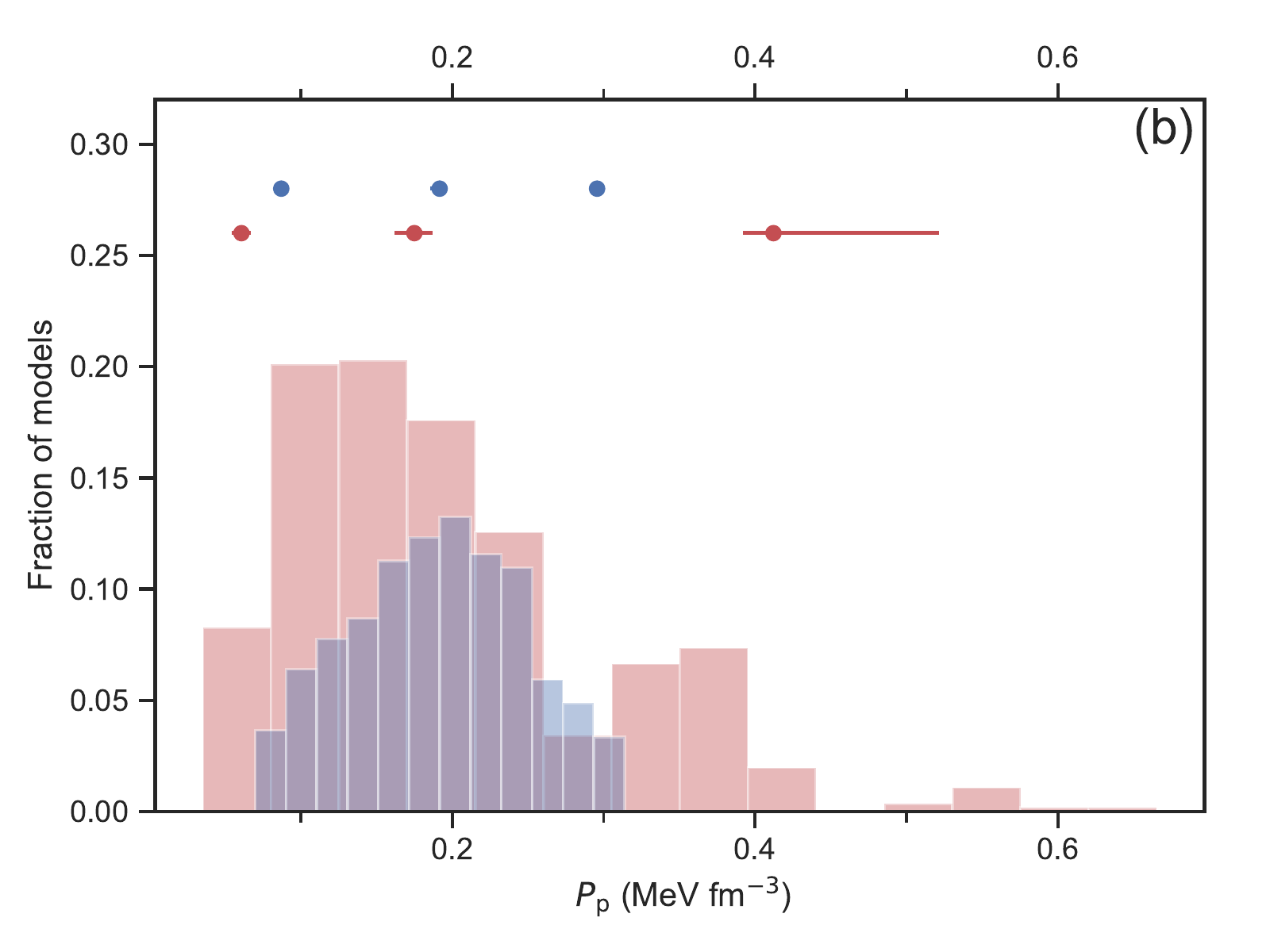}\includegraphics[width=0.25\linewidth]{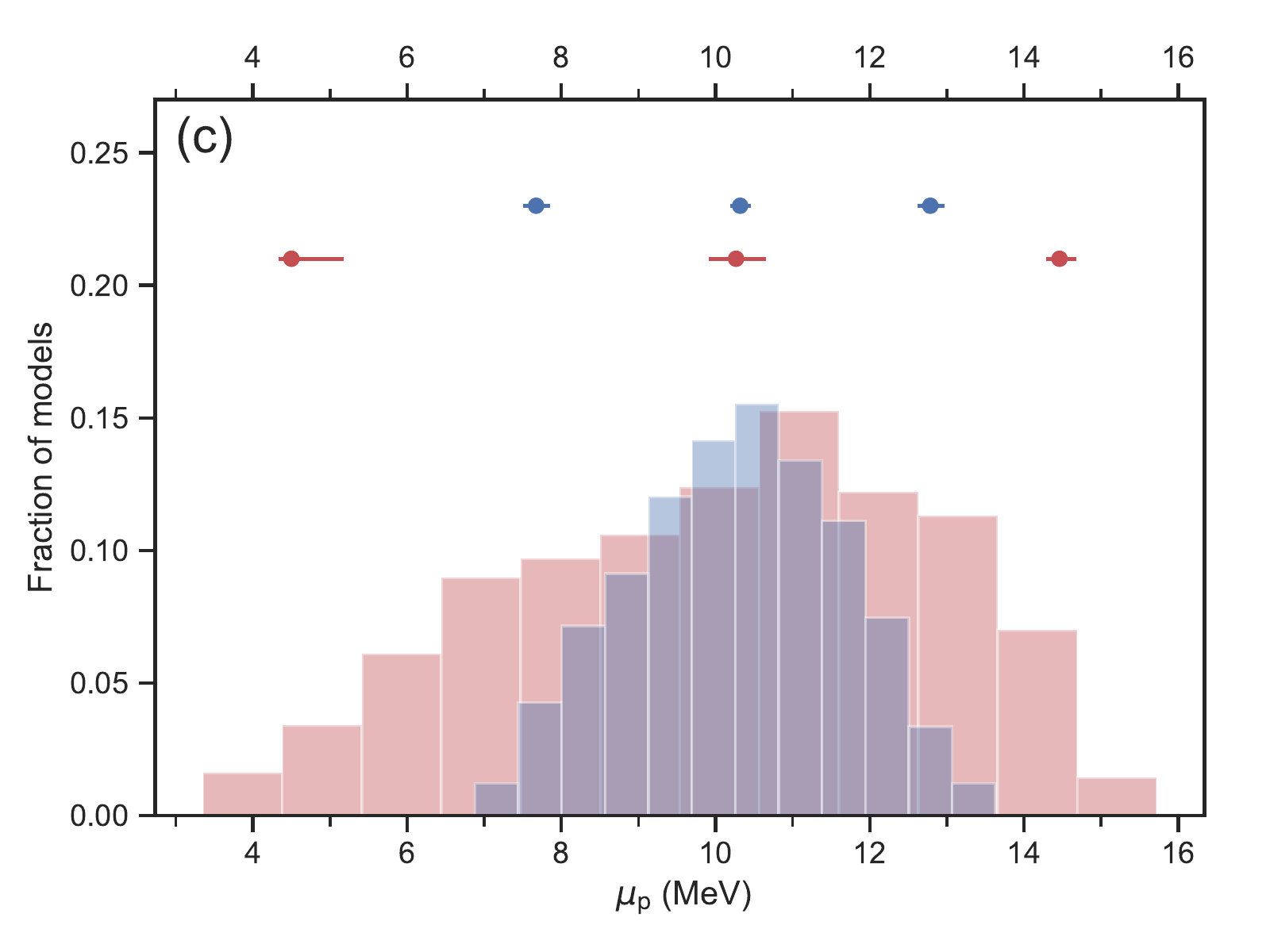}\includegraphics[width=0.25\linewidth]{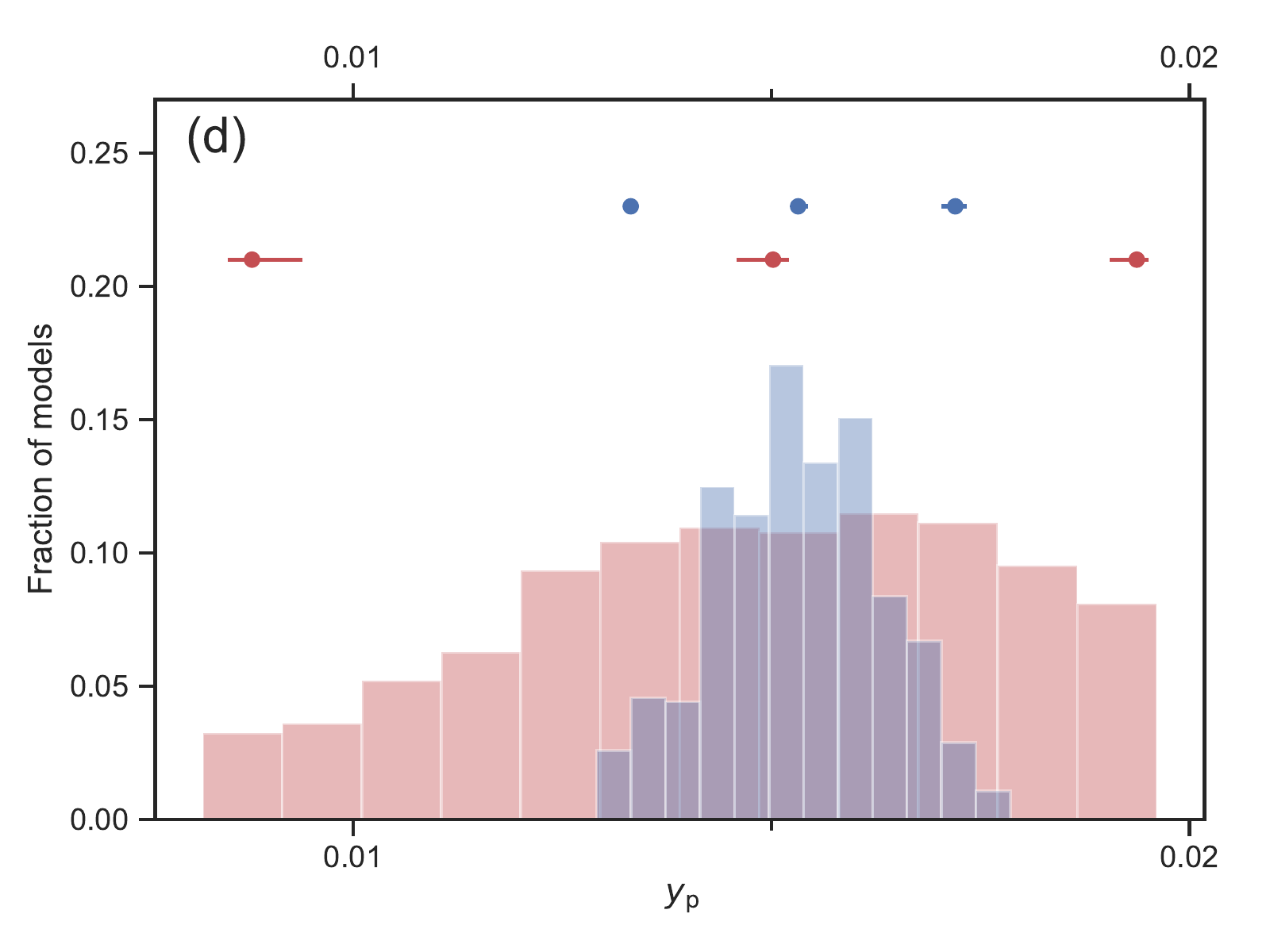}
    \caption{Spherical nuclei-pasta transition density (a), pressure (b), chemical potential (c) and proton fraction (d) distributions from our uniform priors (red bars) and PNM priors (blue bars). Points indicating the 2.5th, 50th and 97.5th percentiles of the distributions are indicated at the top of the plots with their associated 2-$\sigma$ sampling error bars estimated using the bootstrap method.}
    \label{fig:16}
\end{figure}

\begin{figure}[!t]
    \centering
    \includegraphics[width=0.5\linewidth]{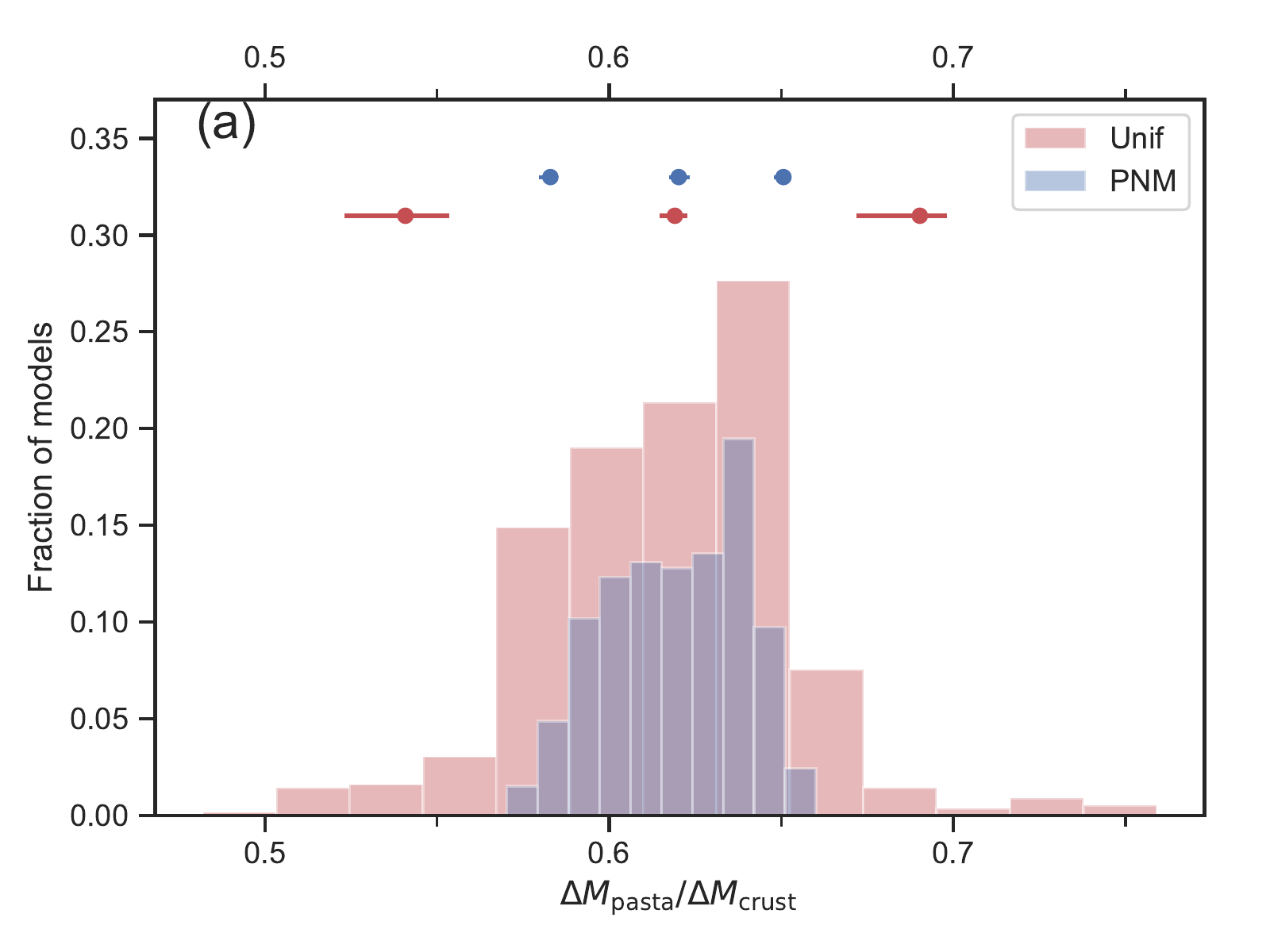}\includegraphics[width=0.5\linewidth]{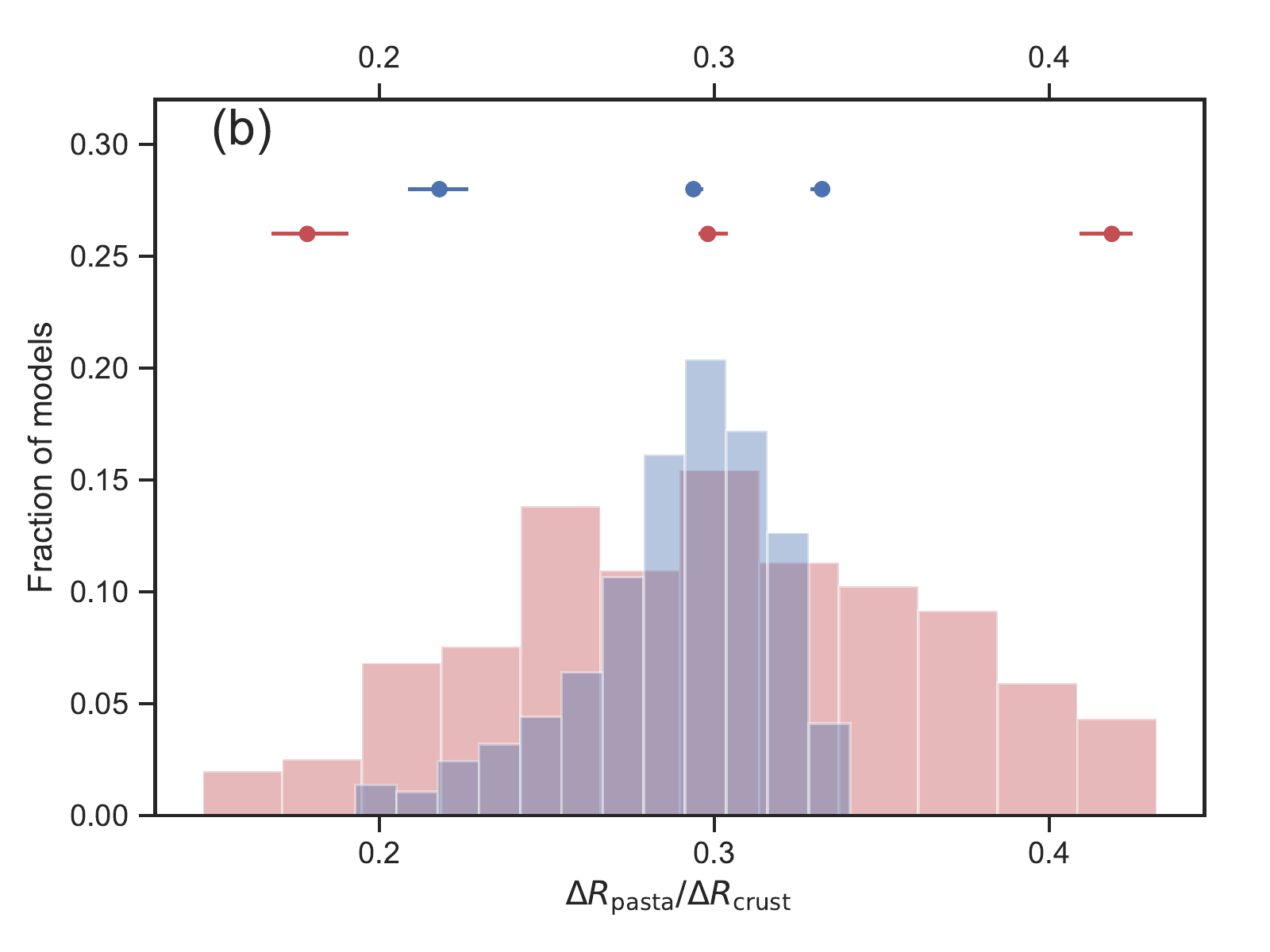}\\
    \caption{Mass (a) and radius (b) fraction of the pasta layers for the uniform prior distribution (red bars) and the PNM prior distribution (blue bars). Points indicating the 2.5th, 50th and 97.5th percentiles of the distributions are indicated at the top of the plots with 2-$\sigma$ sampling error bars.}
    \label{fig:17}
\end{figure}

The two quantities that play the biggest role in determining the location crust-core boundary and the pasta layer are the pressure and chemical potential. The median values for each of the two ensembles are in good agreement, with the uniform prior distribution, with its wider range of symmetry energy parameters, leading to a wider range of values for both quantities. At the crust-core boundary, the pressure is just below $\approx$ 0.5 MeV fm$^{-3}$ and the chemical potential is $\approx$ 14.5 MeV, with 95\% ranges for $P_{\rm cc}$ of widths $\approx$ 1.2 and $\approx$ 0.5 MeV fm$^{-3}$ for uniform and PNM priors respectively, and 95\% ranges for $\mu_{\rm cc}$ of widths $\approx$ 15 and $\approx$ 10 MeV for uniform and PNM priors respectively. At the pasta transition, the median values for pressure and chemical potential are $\approx$ 0.2 MeV fm$^{-3}$ and $\approx$ 10 MeV fm$^{-3}$ respectively; with 95\% ranges for $P_{\rm cc}$ of widths $\approx$ 0.5 and $\approx$ 0.2 MeV fm$^{-3}$ for uniform and PNM priors respectively, and 95\% ranges for $\mu_{\rm cc}$ of widths $\approx$ 10 and $\approx$ 5 MeV for uniform and PNM priors respectively. 

Finally, the proton fraction at the crust-core and pasta boundaries have median values of $\approx$0.03 and 0.015 respectively. The crust-core proton fraction has 95\% ranges with widths of 0.04 and 0.015 for uniform and PNM priors respectively, and proton fraction at the pasta boundary has 95\% ranges have widths of 0.01 and 0.005 for uniform and PNM priors respectively,

Our uniform prior distribution is similar to the prior distribution of \citet{Carreau:2019aa} and our PNM prior distribution is similar to the ``LD'' distribution in the same study. Comparing our distributions, the most noticeable difference is that the crust-core transition density and pressure distributions extend to much lower densities in \citet{Carreau:2019aa}. Our lower bound on the crust-core transition density and pressure is 0.075fm$^{-3}$ and 0.1 MeV fm$^{-3}$ compared to 0.01fm$^{-3}$ and below 0 MeV fm$^{-3}$. We attribute the differences to our filtering out of those crust models that do not give stable inner crust; this rules out the highest value of $L$ which lead to much smaller values of the symmetry energy and pressure at crust densities, and smaller transition densities and pressures. The other factor to take into account is that \citet{Carreau:2019aa} vary $p$ between 2.5 and 3.5 for their prior and ``LD'' distributions, whereas we find a higher value of $p$=3.8 is required to be consistent with microscopic calculations of nuclei in neutron rich matter. A higher value of $p$ leads to a smaller value of the surface tension, and thus allows inhomogeneous matter to be energetically favorable to higher densities, leading to a deeper crust-core transition. These are the main differences that account for the average values of $n_{\rm cc}$ reported by \citet{Carreau:2019aa} being around 0.02 fm$^{-3}$ smaller, and the average value of $P_{\rm cc}$ are around 0.15 MeV fm$^{-3}$ times smaller. The widths of the distributions are similar. It is important to note that their ranges include independent variation of the 4th parameter in the density expansion of the symmetry energy expansion, $Q_{\rm sym}$. Our values of $Q_{\rm sym}$ is determined uniquely by our chosen values of $J$, $L$, and $K_{\rm sym}$, which is a model dependence. 

The relative size of the pasta layer by mass and radius is calculated using the chemical potential at the crust-core and pasta transition \citep{Lorenz:1993aa,Lattimer:2007aa},

\begin{equation}
    \frac{\Delta R_{\rm pasta}}{\Delta R_{\rm crust}} =  \frac{(\mu_{\rm cc} -\mu_{\rm pasta})}{\mu_{\rm cc}}, \hspace{2cm} \frac{\Delta M_{\rm pasta}}{\Delta M_{\rm crust}} = \frac{(P_{\rm cc} - P_{\rm pasta})}{P_{\rm cc}},
\end{equation}

\noindent is shown in figure~17. It is also useful to note that the relative column depths of the pasta and crust-core boundaries is equal to the relative mass of the pasta layer. It is a robust prediction of our models that pasta accounts for more than 50\% of the mass and 15\% of the thickness of the crust, with median values of 62\% and 30\% by mass and thickness for both of our distribution. 95\% of models fall within the range of 50\% to 80\% of the mass of the crust and 15\% and 45\% of the thickness of the crust for uniform priors and 57\% and 66\% of the mass of the crust and 20\% and 34\% of the thickness of the crust for PNM priors.


\subsection{Relationships between nuclear and neutron star crust parameters}

\begin{figure}[t!]
    \centering
    \includegraphics[width=0.95\linewidth]{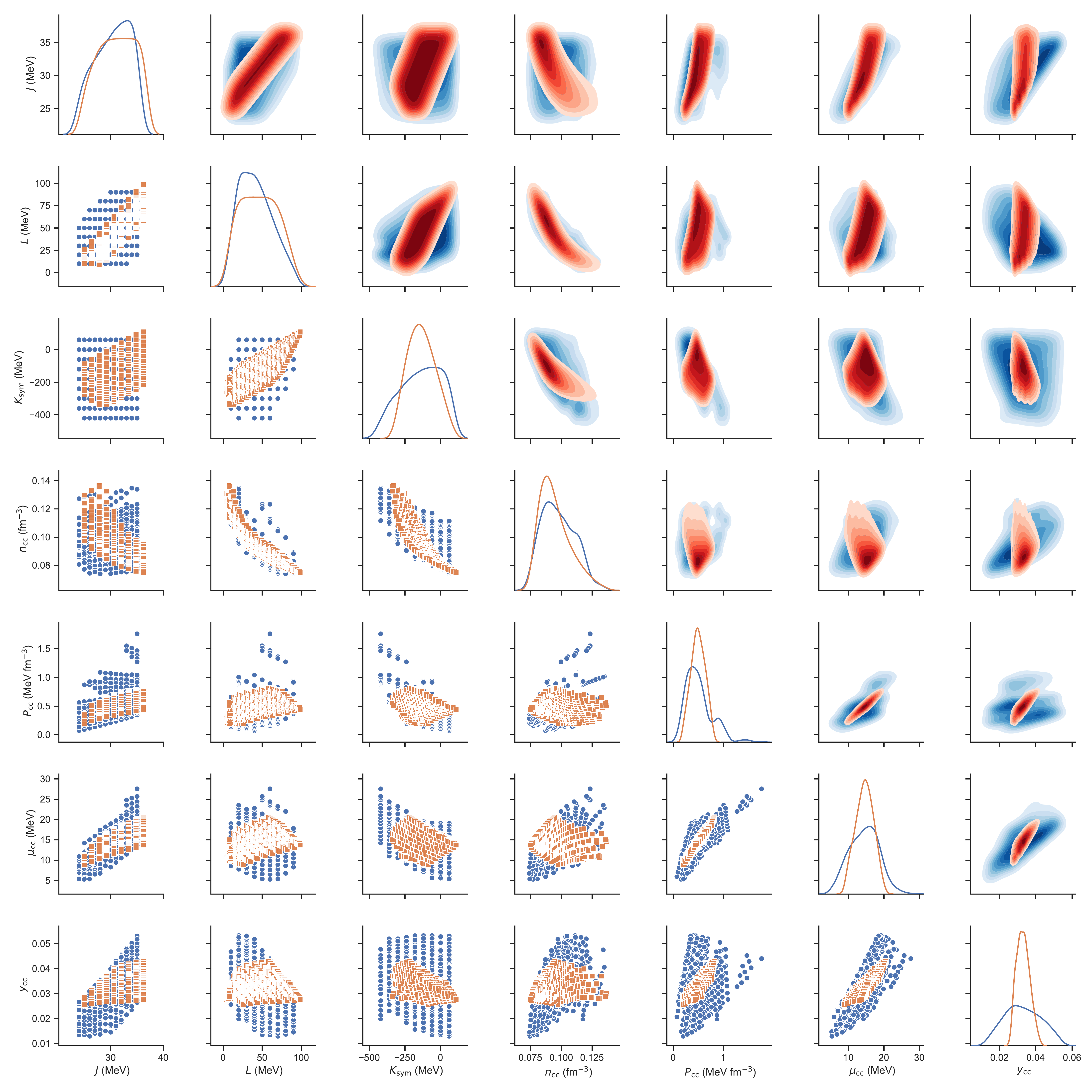}
    \caption{Plots of the relationships between symmetry energy parameters $J$, $L$ and $K_{\rm sym}$ and crust-core transition densities $n_{\rm cc}$, pressures $P_{\rm cc}$, chemical potentials $\mu_{\rm cc}$ and proton fractions $y_{\rm cc}$  for uniform (blue) and pure neutron matter (red) priors. Two different versions of the plots are displayed; below the diagonal are scatter plots where each point is a single model; square points show the PNM priors and circles shows the uniform priors. Above the diagonal are density plots that reveal more clearly for which regions of parameter space are the models are more concentrated.}
    \label{fig:18}
\vspace{11pt}
\vspace{11pt}
\vspace{11pt}
\end{figure}

\begin{figure}[t!]
    \centering
    \includegraphics[width=0.95\linewidth]{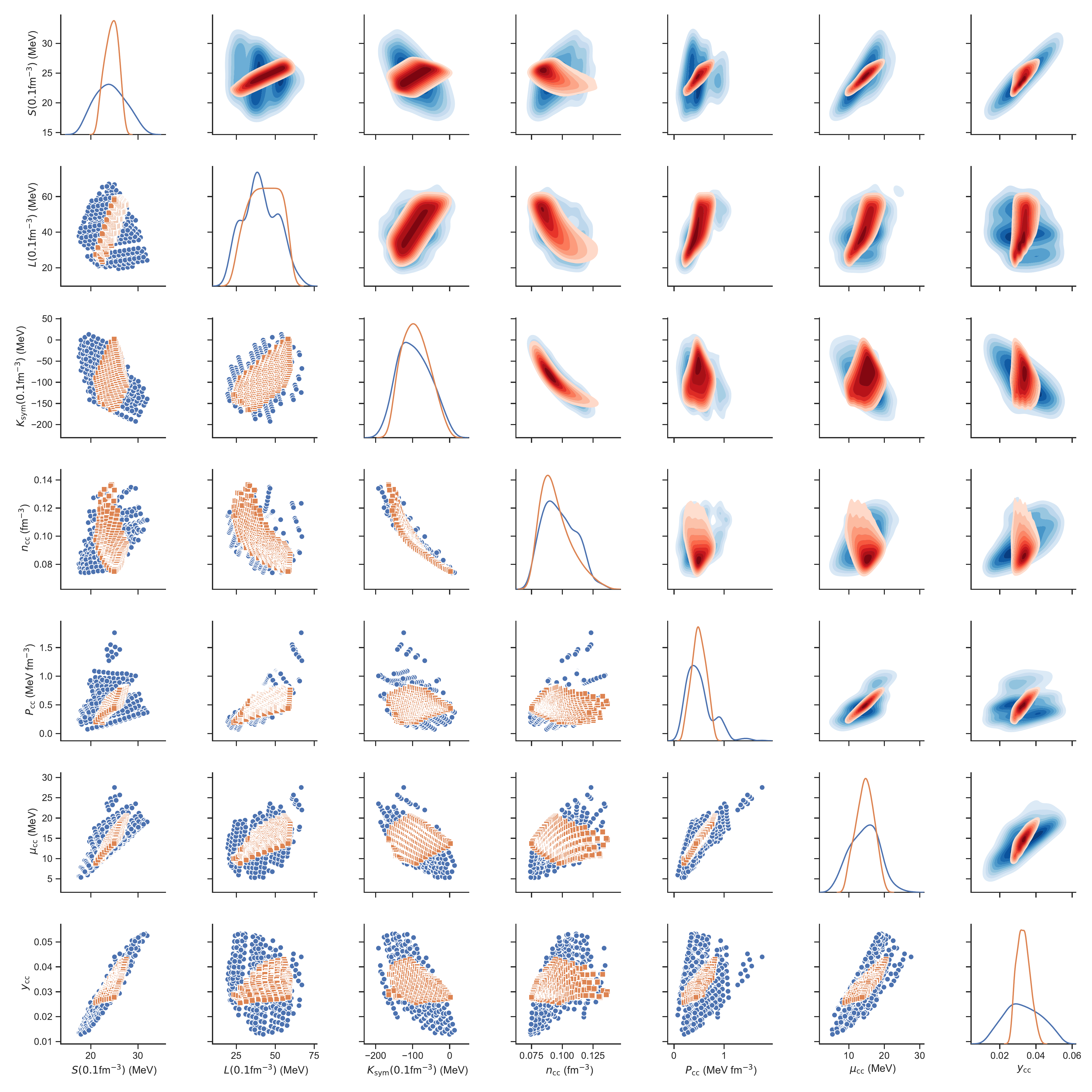}
    \caption{As in Fig~18, but now the symmetry energy parameters are calculated at $0.1$fm$^{-1}$.}
    \label{fig:19}
\vspace{11pt}
\vspace{11pt}
\vspace{11pt}
\end{figure}

Identifying correlations between crust observables and terrestrial nuclear observables helps us use experiments and observations in one domain to obtain meaningful information in the other. A number of correlations between crust properties and the nuclear symmetry energy and neutron skins have been identified in the past; however, these are often reflections of correlations between model parameters resulting from our choice of models. In this section we examine the relationships between the model parameters (the symmetry energy) and a number of nuclear and neutron star crust observables. We will highlight notable relationships ``by eye''; in the following section, we quantify the strength of the correlations seen here.

In Figures~18-26 we plot the relationships between the symmetry energy parameters, nuclear observables in the form of the neutron skins of $^{48}$Ca and $^{208}$Pb, and the crust-core and pasta transition properties. In each plot, the bottom-left corner shows scatter plots with a point for each model while the top right shows density plots to reveal the higher-probability regions.

In Figure~18 we plot the relationships between the symmetry energy parameters and the crust-core transition properties. One of the first correlations between symmetry energy and crust-core transition properties that was discovered was the negative correlation between $L$ and $n_{\rm cc}$. This is evident in our ensembles of models, but as has been noted in subsequent investigations that correlation is not as strong as once suspected \citep{Ducoin:2011aa}. For both sets of models, the correlation is strong but with a width of $\sim$0.02fm$^{-3}$; this is particular evident in the density plot. We also see a slightly negative correlation between $n_{\rm cc}$ and $K_{\rm sym}$; this correlation was also examined in \citet{Ducoin:2011aa, Zhang:2018aa}. 

Let us turn to the most important crust-core transition properties for astrophysical modeling purposes, the crust-core transition pressure and chemical potential. A strong correlation exists between these two parameters, especially when we restrict ourselves to the subset of models that give good descriptions of the PNM EoS.

No other notable relationships appear. Particularly there is no strong correlation between $P_{\rm cc}$, which determines the relative mass of the crust, with any of the symmetry energy parameters.

\begin{figure}[t!]
    \centering
    \includegraphics[width=0.95\linewidth]{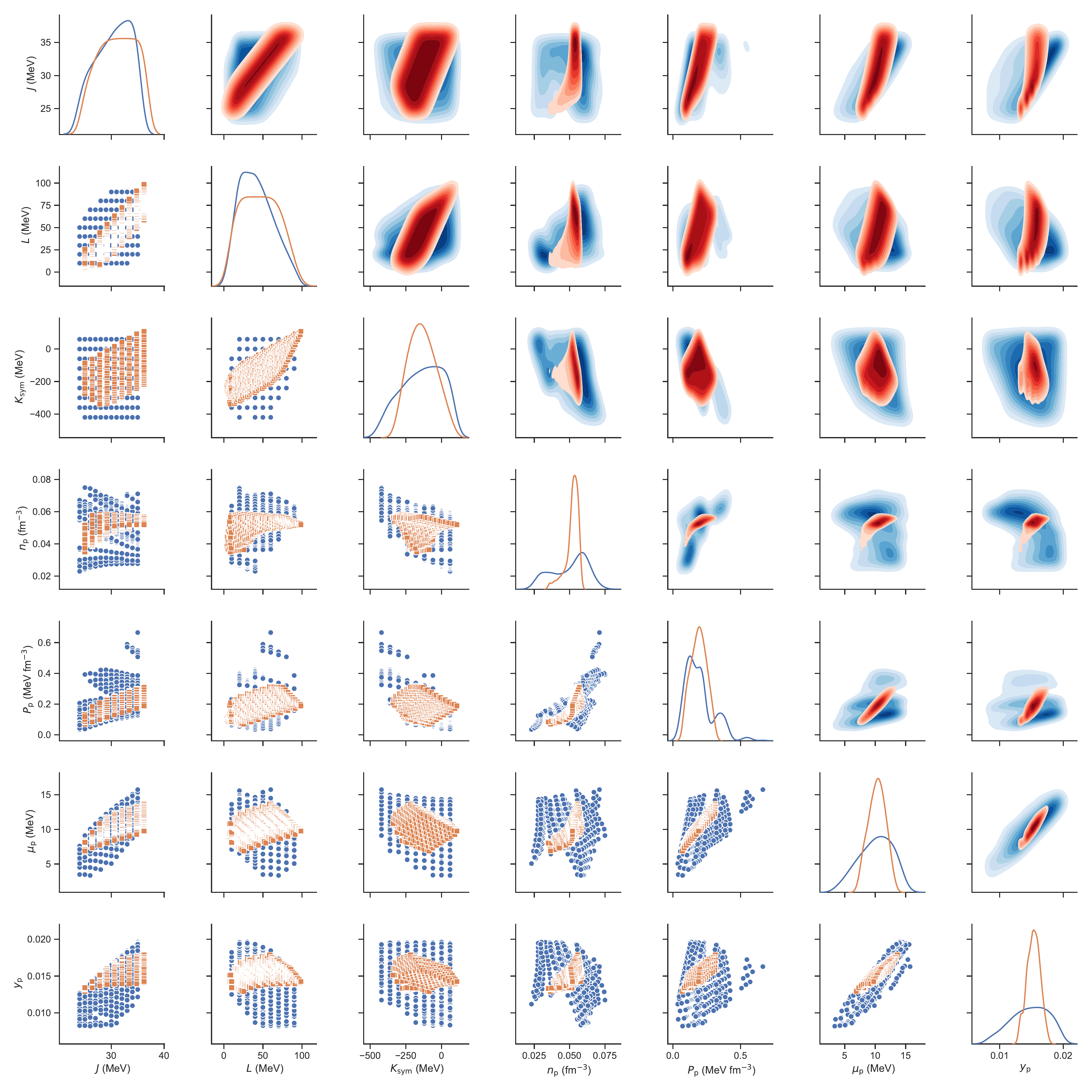}
    \caption{Plots of the relationships between symmetry energy parameters $J$, $L$ and $K_{\rm sym}$ and spherical nuclei-pasta transition densities $n_{\rm p}$, pressures $P_{\rm p}$, chemical potentials $\mu_{\rm p}$ and proton fractions $y_{\rm p}$  for uniform (blue) and pure neutron matter (red) priors. Two different versions of the plots are displayed; below the diagonal are scatter plots where each point is a single model; square points show the PNM priors and circles shows the uniform priors. Above the diagonal are density plots that reveal more clearly for which regions of parameter space are the models are more concentrated.}
    \label{fig:20}
\vspace{11pt}
\vspace{11pt}
\vspace{11pt}
\end{figure}

\begin{figure}[t!]
    \centering
    \includegraphics[width=0.95\linewidth]{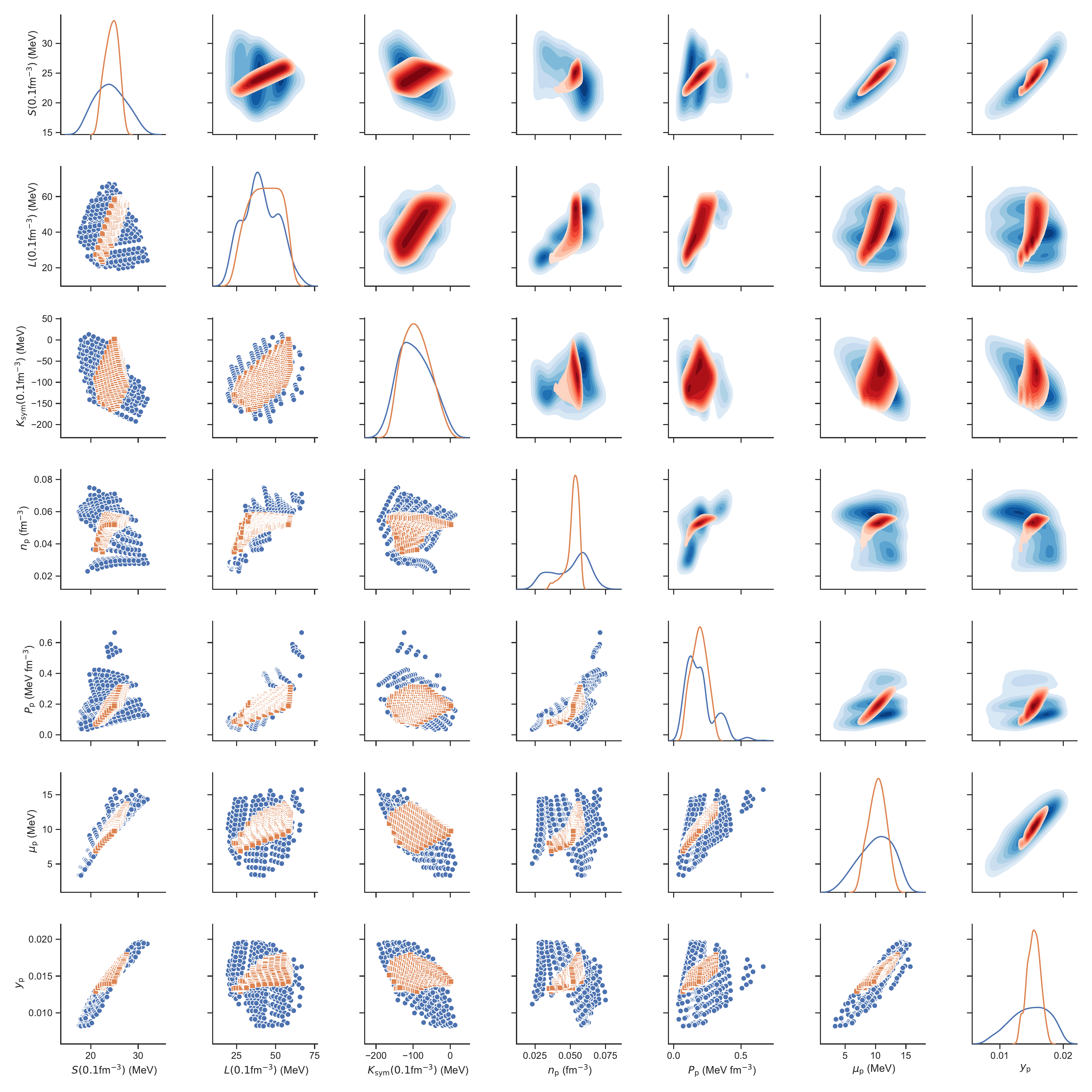}
    \caption{As in Fig~20, but now the symmetry energy parameters are calculated at $0.1$fm$^{-1}$.}
    \label{fig:21}
\vspace{11pt}
\vspace{11pt}
\vspace{11pt}
\end{figure}

The symmetry energy parameters here are the standard density expansion coefficients at saturation density - a density almost twice the median crust-core transition density predicted by our ensembles of crust models, so perhaps it is not surprising that there is no stronger correlations in evidence. In Fig~19 we therefore plot the symmetry energy parameters calculated at $0.1$fm$^{-1}$. A number of particularly strong correlations stand out now.

Firstly, the crust core transition density correlates very strongly with the curvature of the symmetry energy at sub-saturation density. This is not very surprising because of the role the curvature symmetry energy plays in determining the stability of uniform nuclear matter \citep{Kubis:2007aa,Lattimer:2007aa,Zhang:2018aa}.

Secondly, the crust-core transition proton fraction correlates strongly with the symmetry energy at sub-saturation density $S$($0.1$fm$^{-1}$), again not surprisingly given the symmetry energy's role in determining the proton fraction of uniform matter. 

Finally, the crust-core transition chemical potential, which locates the depth of the crust-core boundary, and the sub-saturation symmetry energy are correlated.

\begin{figure}[t!]
    \centering
    \includegraphics[width=0.95\linewidth]{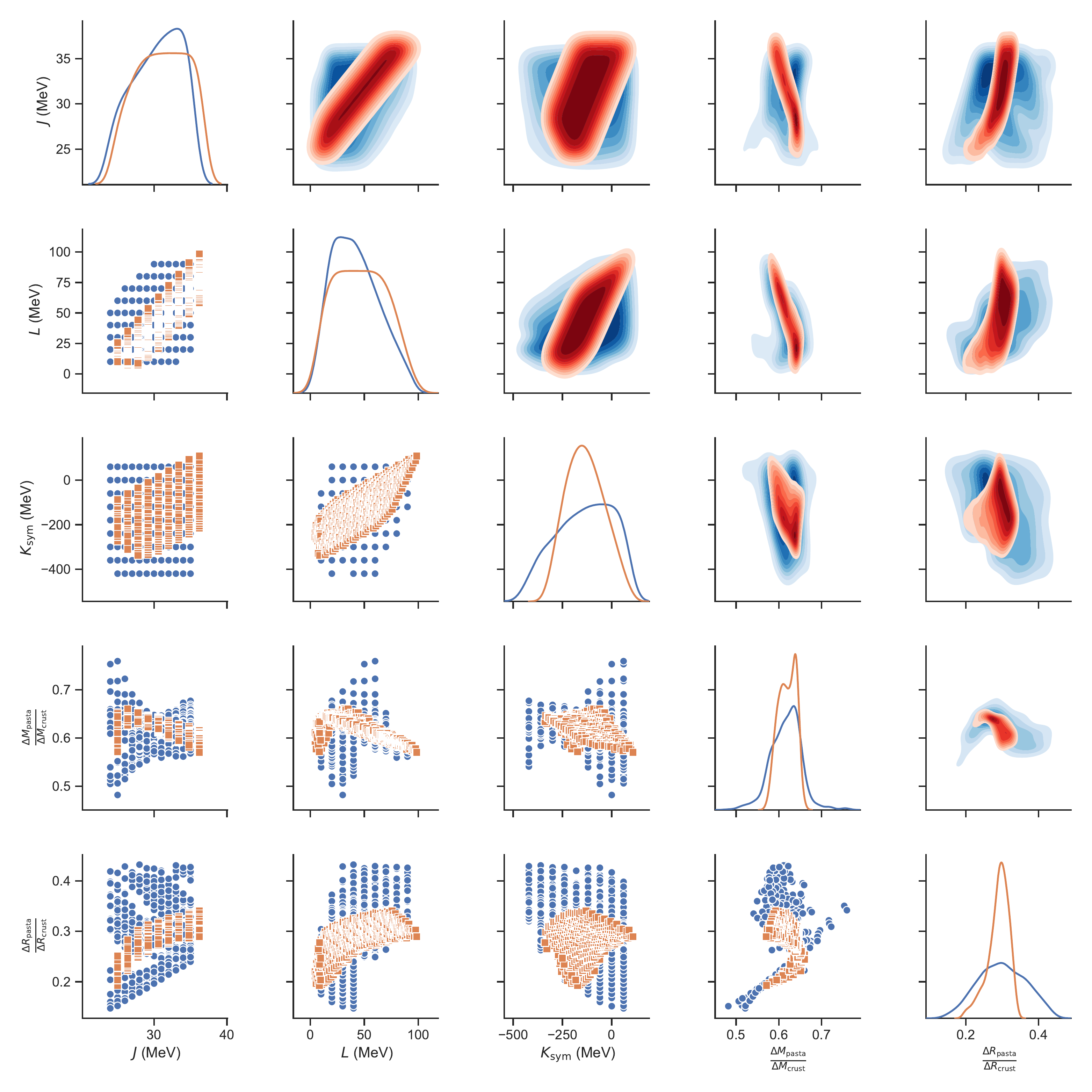}
    \caption{The symmetry energy parameters $J$, $L$ and $K_{\rm sym}$ plotted against the fractional mass $\Delta M_{\rm p}/\Delta M_{\rm c}$ and thickness $\Delta R_{\rm p}/\Delta R_{\rm c}$ of the crust occupied by pasta. As in Figs 14-17, uniform priors are blue and pure neutron matter priors are red. Two different versions of the plots are displayed; below the diagonal are scatter plots where each point is a single model; square points show the PNM priors and circles shows the uniform priors. Above the diagonal are density plots that reveal more clearly for which regions of parameter space are the models are more concentrated.}
    \label{fig:22}
\vspace{11pt}
\vspace{11pt}
\vspace{11pt}
\end{figure}

\begin{figure}[t!]
    \centering
    \includegraphics[width=0.95\linewidth]{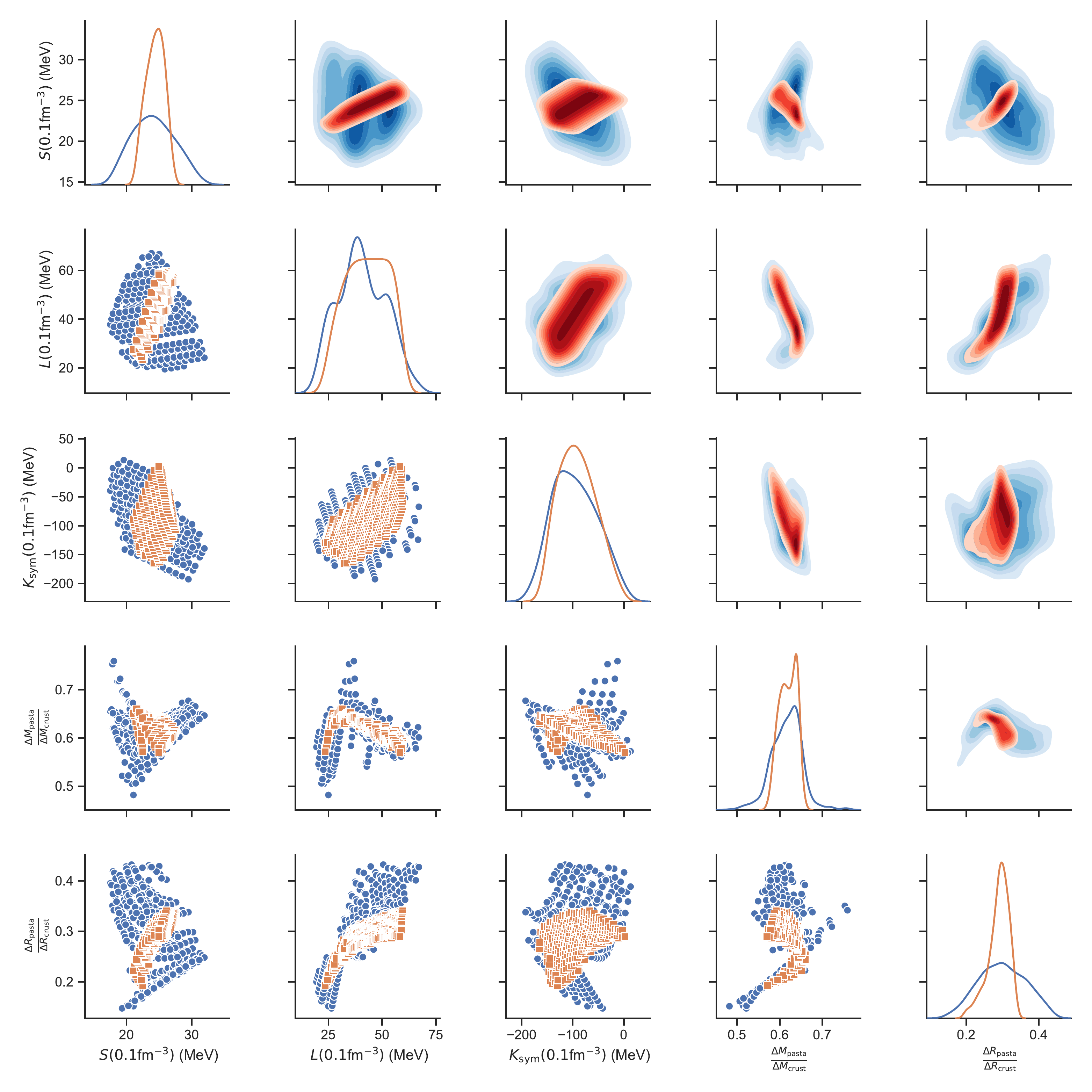}
    \caption{As in Fig~22, but now the symmetry energy parameters are calculated at $0.1$fm$^{-1}$.}
    \label{fig:23}
\vspace{11pt}
\vspace{11pt}
\vspace{11pt}
\end{figure}

In Figure~20 we turn our attention to the relationships between the pasta transition properties and the symmetry energy parameters at saturation density. We see that lower $L$ and $J$ tend to give an increased probability of the pasta phases appearing at lower density and pressure; the latter is consistent with previous studies \citep{Oyamatsu:2007aa,Bao:2015aa}. The spread of possible pasta transition densities decreases significantly with increasingly $L$: For $L<$50MeV, a large range of densities at which pasta first appears is predicted, from 0.02 to 0.08 fm$^{-3}$ for the uniform priors. At $L\approx100$ MeV, this has shrunk to a range of 0.05 to 0.07 fm$^{-3}$ for the uniform priors, and close to 0.05fm$^{-3}$ for PNM priors. 

Moving to the sub-saturation symmetry energy parameters to search for stronger correlations in Figure~21, we immediately see strong correlations standing out between the pasta transition chemical potential $\mu_{\rm p}$ (which determines the location of the top layer of pasta) and proton fraction $y_{\rm p}$ with $S(0.1$fm$^{-3})$. This is of particular note because of the relationship between $S(0.1$fm$^{-3})$ and the neutron skins of neutron rich nuclei, global nuclear mass fits and with giant dipole resonances \citep{Trippa:2008aa,Lattimer:2013aa}. A weaker correlation is seen between the pasta transition pressure - which locates the pasta transition boundary by mass - and the sub-saturation slope $L(0.1$fm$^{-3})$ of the symmetry energy.

Next up are the relationships involving the relative mass and thickness of pasta in the crust. In Fig~22 we show them plotted against the symmetry energy parameters at saturation density, and in Fig~23 we show them plotted against the symmetry energy parameters at sub-saturation density. Some distinct correlations emerge, particularly with sub-saturation symmetry parameters, but their structure is more complex. There is a strong correlation between the relative mass of pasta and the slope of the symmetry energy at sub-saturation density, but it is non-monotonic: at low $L(0.1$fm$^{-3})$ the mass of pasta increases rapidly with $L(0.1$fm$^{-3})$, reaching a maximum for values of $L(0.1$fm$^{-3})$ around 30-40 MeV before decreasing with $L(0.1$fm$^{-3})$. The slope of the symmetry energy at sub-saturation density also correlates with the relative thickness of the pasta layer, increasing monotonically with $L(0.1$fm$^{-3})$ although a clear break in the slop of the correlation appears to occur at around the same value as the peak in the value of the relative mass of pasta, 30-40 MeV. When we use the PNM priors we find strong correlations between the mass fraction of pasta and $J$ and $L$.

\begin{figure}[t!]
    \centering
    \includegraphics[width=0.95\linewidth]{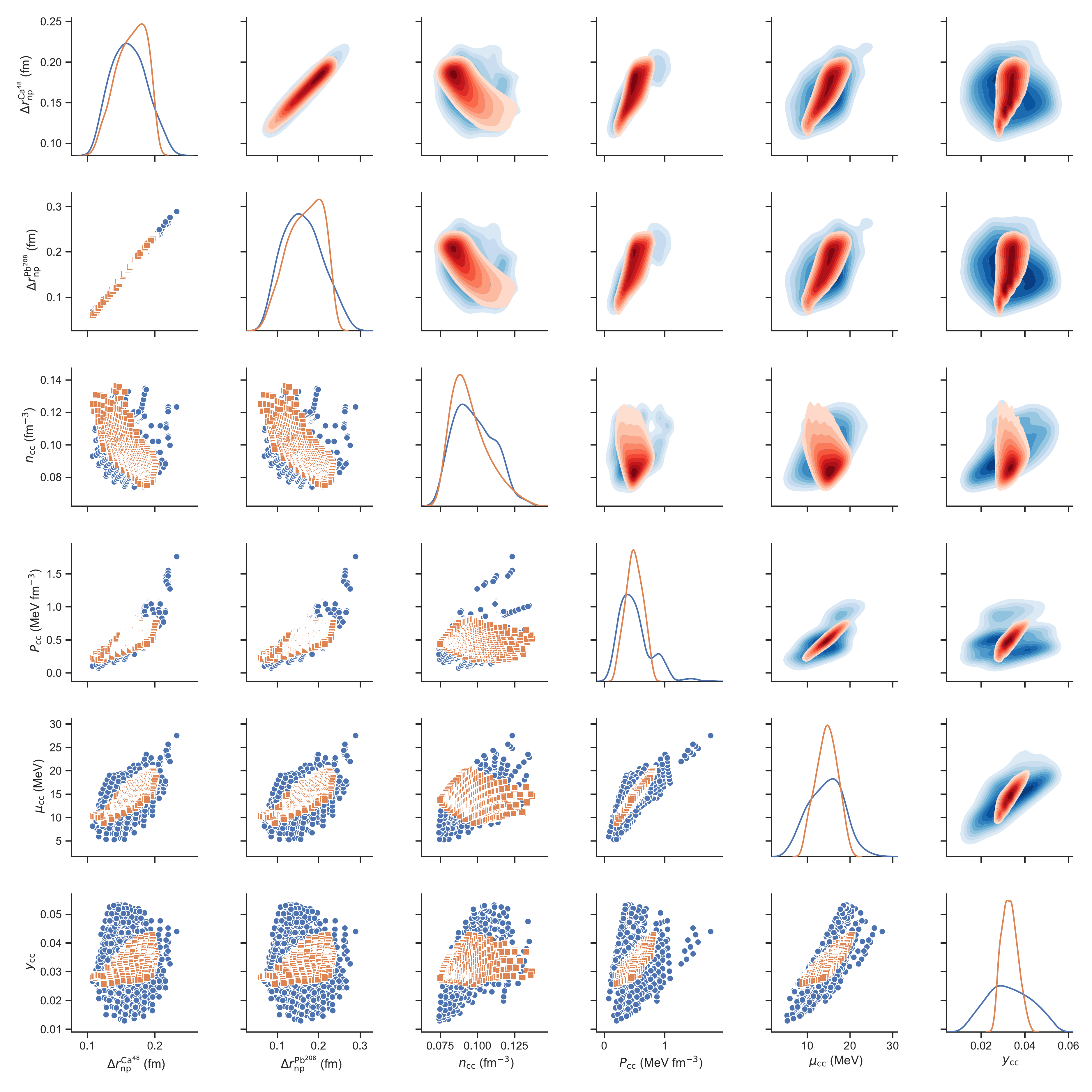}
    \caption{Plots of the relationships between the neutron skin thicknesses \textsuperscript{48}Ca
and \textsuperscript{208} Pb and crust-core transition densities $n_{\rm cc}$, pressures $P_{\rm cc}$, chemical potentials $\mu_{\rm cc}$ and proton fractions $y_{\rm cc}$  for uniform (blue) and pure neutron matter (red) priors. Two different versions of the plots are displayed; below the diagonal are scatter plots where each point is a single model; square points show the PNM priors and circles shows the uniform priors. Above the diagonal are density plots that reveal more clearly for which regions of parameter space are the models are more concentrated.}
    \label{fig:24}
\vspace{11pt}
\vspace{11pt}
\vspace{11pt}
\end{figure}

\begin{figure}[t!]
    \centering
    \includegraphics[width=0.95\linewidth]{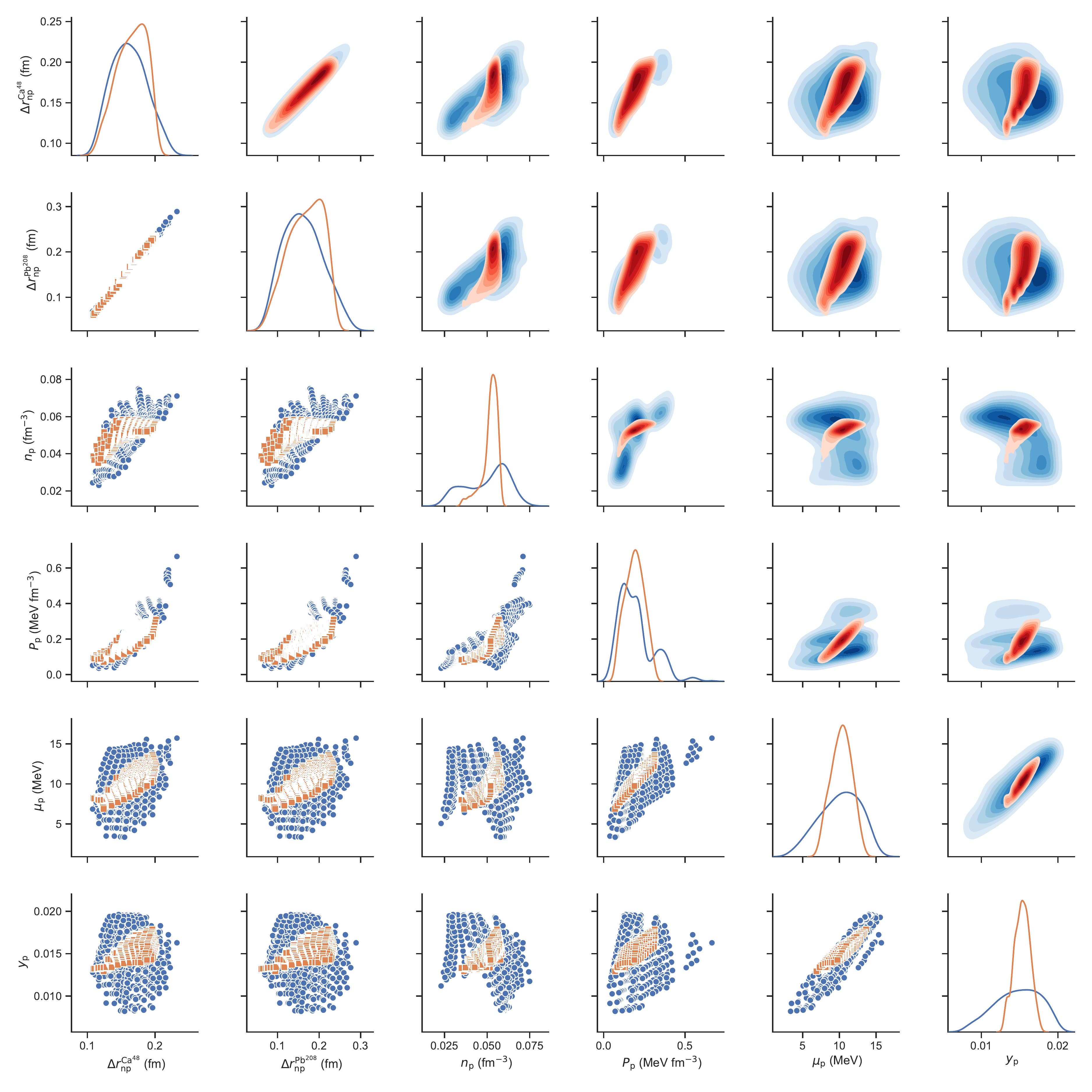}
    \caption{Plots of the relationships between the neutron skin thicknesses \textsuperscript{48}Ca
and \textsuperscript{208} Pb and crust-core transition densities $n_{\rm p}$, pressures $P_{\rm p}$, chemical potentials $\mu_{\rm p}$ and proton fractions $y_{\rm p}$  for uniform (blue) and pure neutron matter (red) priors. Two different versions of the plots are displayed; below the diagonal are scatter plots where each point is a single model; square points show the PNM priors and circles shows the uniform priors. Above the diagonal are density plots that reveal more clearly for which regions of parameter space are the models are more concentrated.}
    \label{fig:25}
\vspace{11pt}
\vspace{11pt}
\vspace{11pt}
\end{figure}

In figures~24 and~25 we show the neutron skin thicknesses - obtained using fully quantum Hartree-Fock calculations \citep{Newton:2020aa} with the same two ensembles of Skyrme models - plotted against their crust-core and pasta transition properties. The advantage of building our models out of Skyrme EDFs is the ability to calculate fully microscopically nuclear properties in a way that is consistent with the crust models. 

\begin{figure}[t!]
    \centering
    \includegraphics[width=1.0\linewidth]{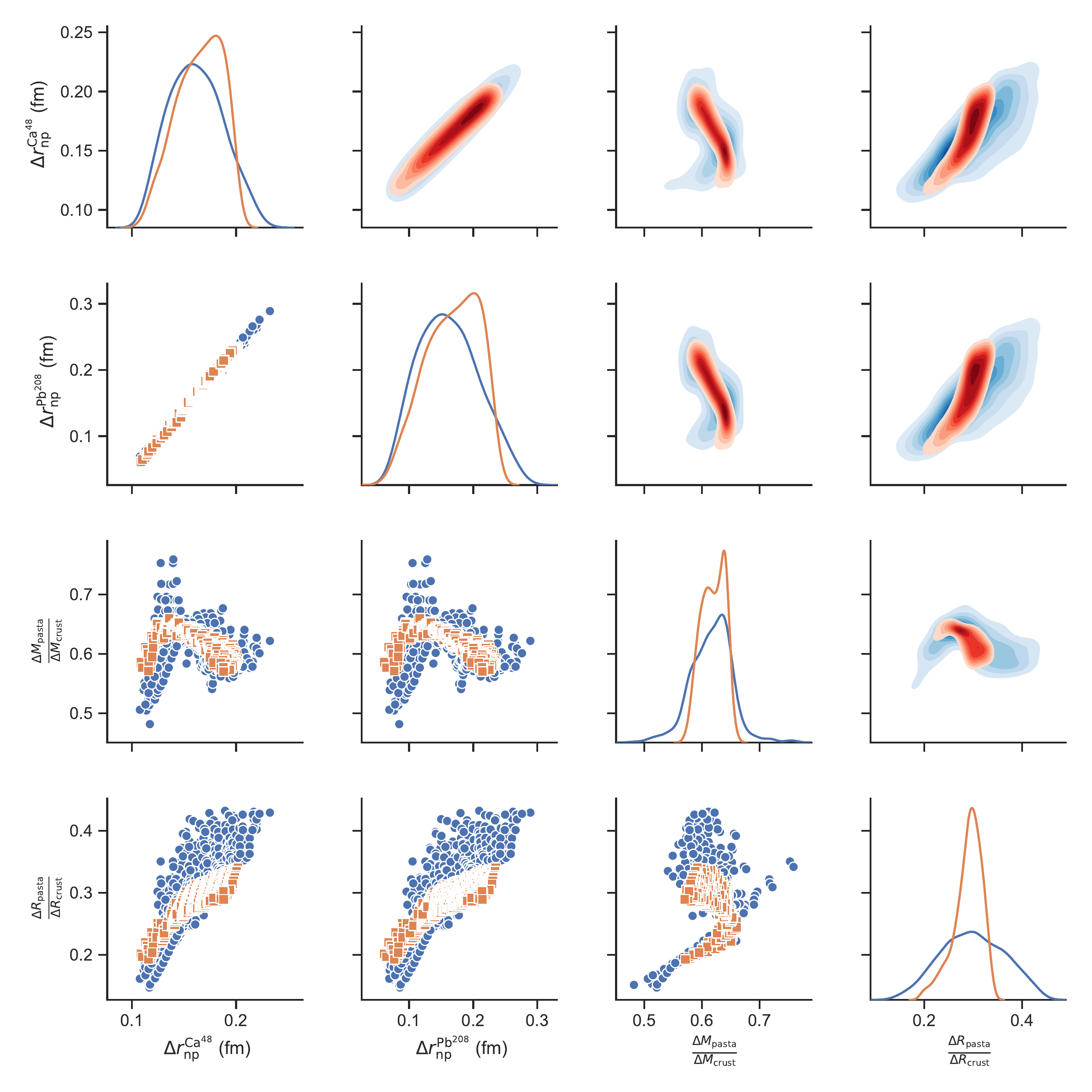}
    \caption{The neutron skin thicknesses \textsuperscript{48}Ca
and \textsuperscript{208} Pb plotted against the fractional mass $\Delta M_{\rm p}/\Delta M_{\rm c}$ and thickness $\Delta R_{\rm p}/\Delta R_{\rm c}$ of the crust occupied by pasta. Uniform priors are blue and pure neutron matter priors are red.Two different versions of the plots are displayed; below the diagonal are scatter plots where each point is a single model; square points show the PNM priors and circles shows the uniform priors. Above the diagonal are density plots that reveal more clearly for which regions of parameter space are the models are more concentrated.}
    \label{fig:26}
\vspace{11pt}
\vspace{11pt}
\vspace{11pt}
\end{figure}

The relationship between the crust-core transition pressure and the neutron skin thickness of \textsuperscript{208}Pb has been a matter of much study. We see a positive correlation for both our ensembles between $P_{\rm cc}$ and the neutron skin thicknesses of both nuclei, albeit with significant spread of models, and a similar correlation between the pasta transition pressure and the neutron skins. A weaker negative correlation between $n_{\rm cc}$ and the neutron skins is also evident. The pasta and crust-core transition chemical potential shows little correlation, suggesting the masses of the crust and the pasta within are sensitive to the neutron skins of \textsuperscript{208}Pb and \textsuperscript{48}Ca more so than their thicknesses.

In Fig~26 we show the relationships between the neutron skin thicknesses of \textsuperscript{208}Pb and \textsuperscript{48}Ca and the relative mass and thickness of the crust. Both the mass and thickness of pasta are strongly correlated with the neutron skin thicknesses, although the mass of pasta exhibits a peak in the distribution very similar to its relationship with the slope of the symmetry energy at $L(0.1$fm$^{-3})$. The peak mass of pasta relative to the crust is at neutron skin thicknesses of \textsuperscript{208}Pb of around 0.13fm and neutron skin thicknesses of \textsuperscript{48}Ca of around 0.14fm. 



\begin{figure}[!ht]
    \centering
    \includegraphics[width=0.7\linewidth]{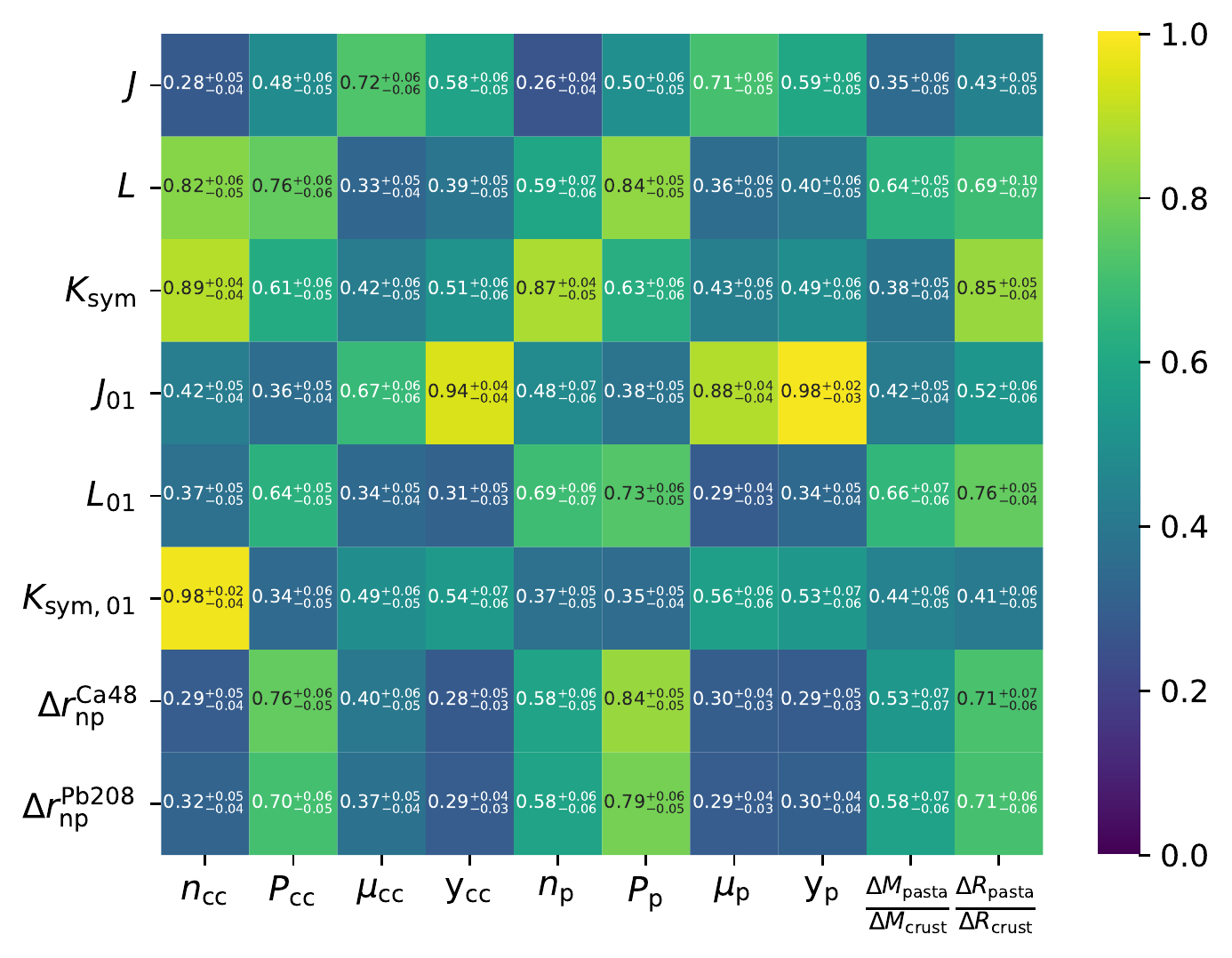}   \\ \includegraphics[width=0.7\linewidth]{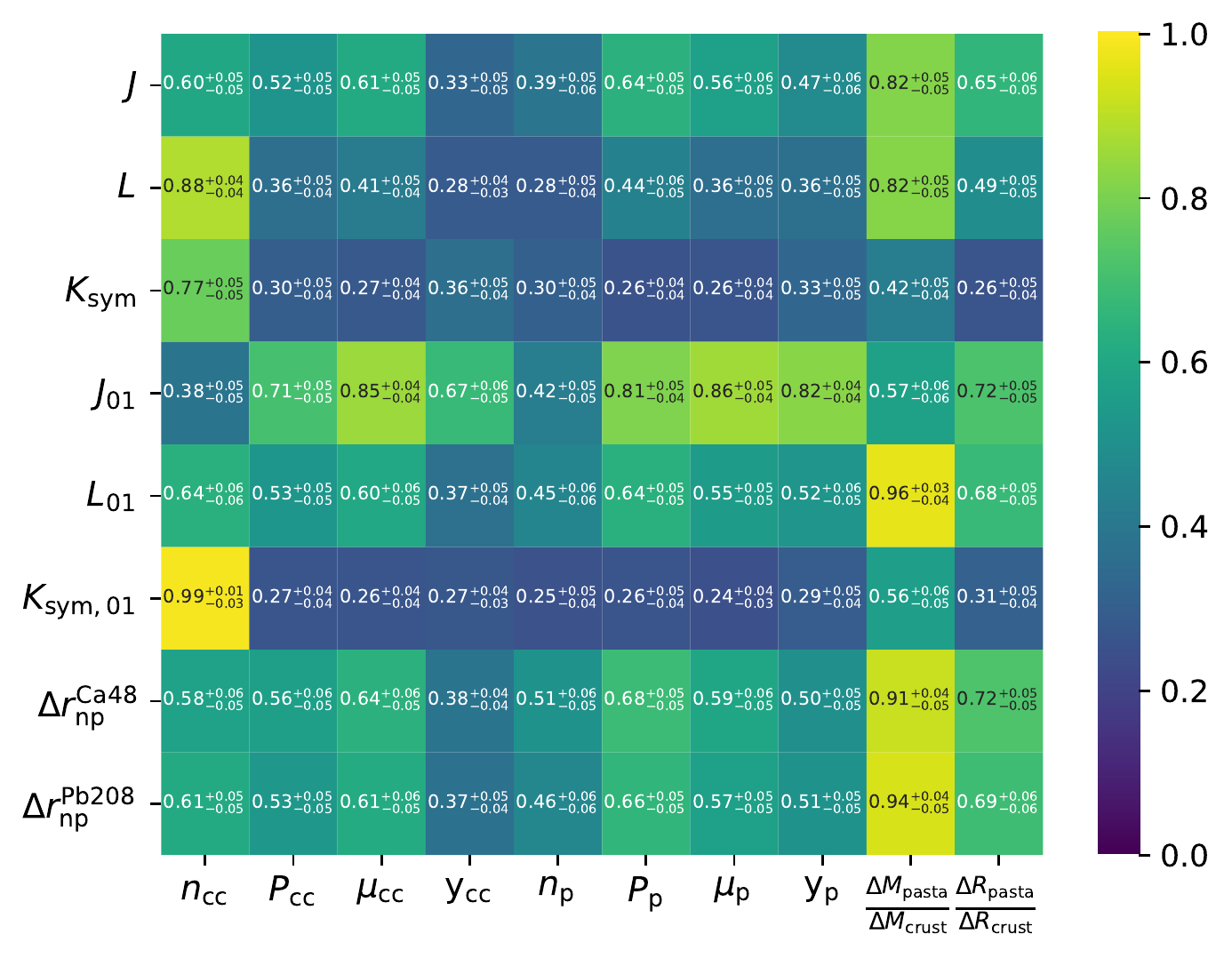}
    \caption{Maximal information coefficients between pairs of crust-core and pasta transition properties, neutron skins and symmetry energy parameters for the uniform priors (top) and PNM priors (bottom). Using the bootstrap method, we show 2-$\sigma$ sampling uncertainties on each MIC.}
    \label{fig:27}
\vspace{11pt}
\vspace{11pt}
\vspace{11pt}
\end{figure}

\noindent
\subsection{Quantifying correlations with the maximum information coefficient}

In the previous section we examined some relationships between crust-core and pasta transition properties, symmetry energy parameters and neutron skins ``by eye''. We now quantify the strength of these relationships. We calculate MICs between all pairs consisting of one nuclear matter parameter or neutron skin and one crust parameter.
Figure~27. We use the bootstrap method to estimate the sampling uncertainties on the MIC coefficients. The 2-$\sigma$ sampling uncertainties in the MIC are displayed on the plots; they are generally around $\pm$0.05.

For uniform priors, it can clearly be seen that the highest MIC coefficients occur for the pairs of variables ($K_{\rm sym}(0.1$fm$^{-3}$), $n_{\rm cc}$), ($J_{01} \equiv S(0.1$fm$^{-3}$),$y_{\rm p}$) and ($J_{01}$,$y_{\rm cc}$). Measures of the column depth and mass above the crust-core transition, and above the pasta layers - $P_{\rm p}$, $P_{\rm cc}$ - are most sensitive to neutron skins of \textsuperscript{48}Pb and \textsuperscript{208}Pb and the slope of the symmetry energy $L$. The depth of the crust down to the pasta layers and the crust-core transition - determined by $\mu_{\rm p}$ and $\mu_{\rm cc}$ is most sensitive to $J_{01}$ and $J$ respectively. There are no particularly strong correlations with the mass fraction of pasta in the crust, but the thickness of the pasta layers correlates strongly with $K_{\rm sym}$. Nuclear experimental measurements of the neutron skin thicknesses of \textsuperscript{48}Pb and \textsuperscript{208}Pb, fits to nuclear masses (sensitive to $J_{01}$) and measurements of giant resonances (sensitive to $J_{01}$ and $K_{\rm sym}$) can therefore help pin down the location of the crust-core boundary and pasta layers, and the relative thickness of the pasta layer.

For PNM priors, as well as the strong correlation between $K_{\rm sym}(0.1$fm$^{-3}$) and $n_{\rm cc}$, we see very strong correlations between the relative mass of pasta in the crust and the neutron skins of \textsuperscript{48}Pb and \textsuperscript{208}Pb and $L(0.1$fm$^{-3}$). Thus, when combined with information about the PNM EOS, measurements of neutron skins can give us information about the relative mass of pasta in the crust. The crust-core and pasta transition pressures and chemical potentials are sensitive to $J_{01}$, so once again, coupled with PNM EOS constraints, nuclear mass fits and measurements of giant dipole resonances can deliver information about the location of the crust-core boundary and the pasta layers.


\begin{figure}[t!]
    \centering
    \includegraphics[width=1.0\linewidth]{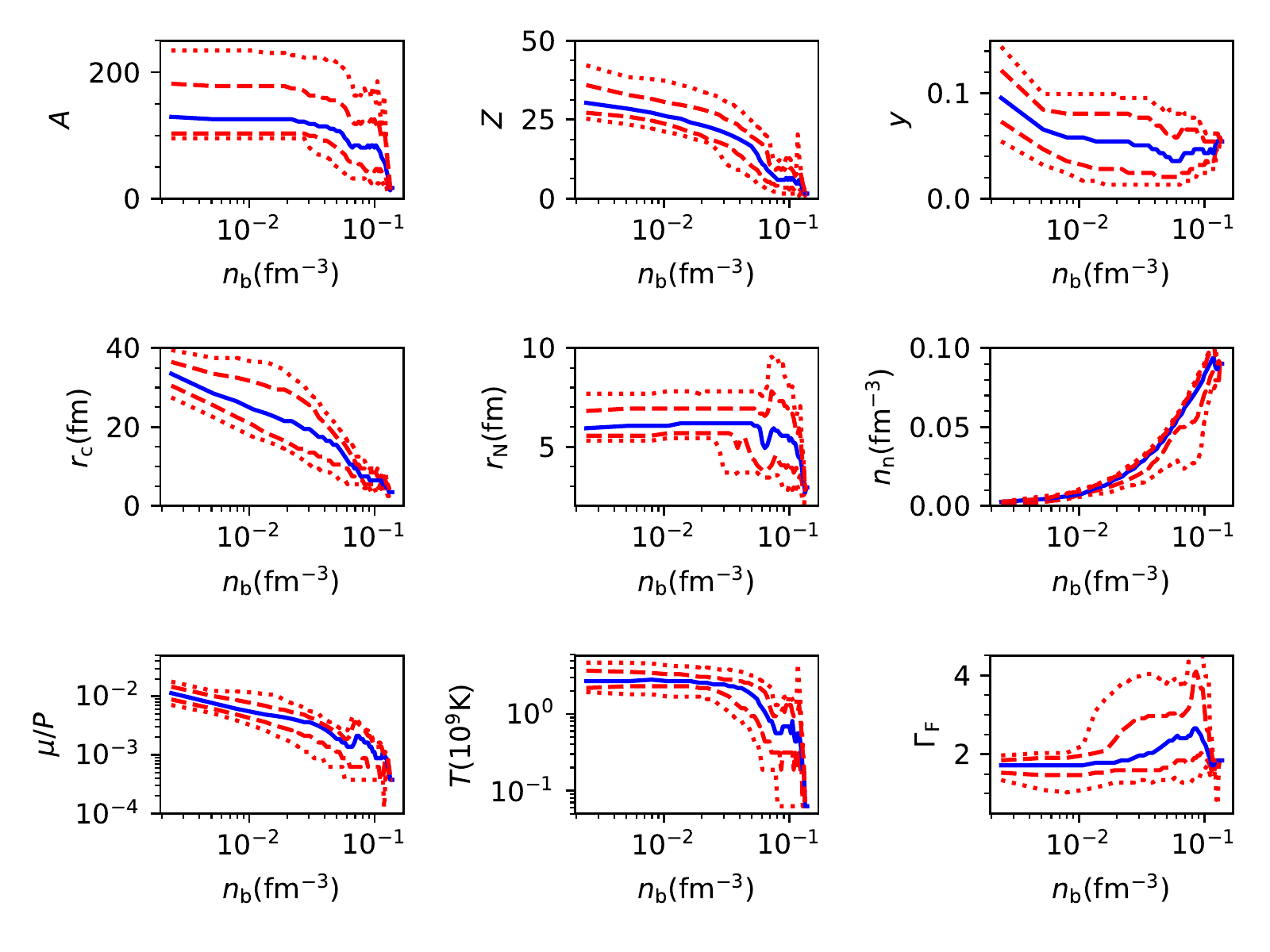}
    \caption{Crust composition for uniform priors. From top left to bottom right: the number of nucleons in the nucleus $A$, the number of protons in the nucleus $Z$, the global proton fraction $y$, the Wigner-Seitz cell radius $r_{\rm c}$, the nuclear or pasta radius $r_{\rm N}$, the density of the neutron gas $n_{\rm n}$, the shear modulus relative to the pressure $\mu/P$, the melting temperature $T$ in $10^{-2}$MeV and the frozen-composition adiabatic index $\gamma_{\rm f}$. The blue line shows the median value at a given density, with the dashed and dotted lines bounding the 68\% and 95\% ranges.}
    \label{fig:28}
\vspace{11pt}
\vspace{11pt}
\vspace{11pt}
\end{figure}

\begin{figure}[t!]
    \centering
    \includegraphics[width=1.0\linewidth]{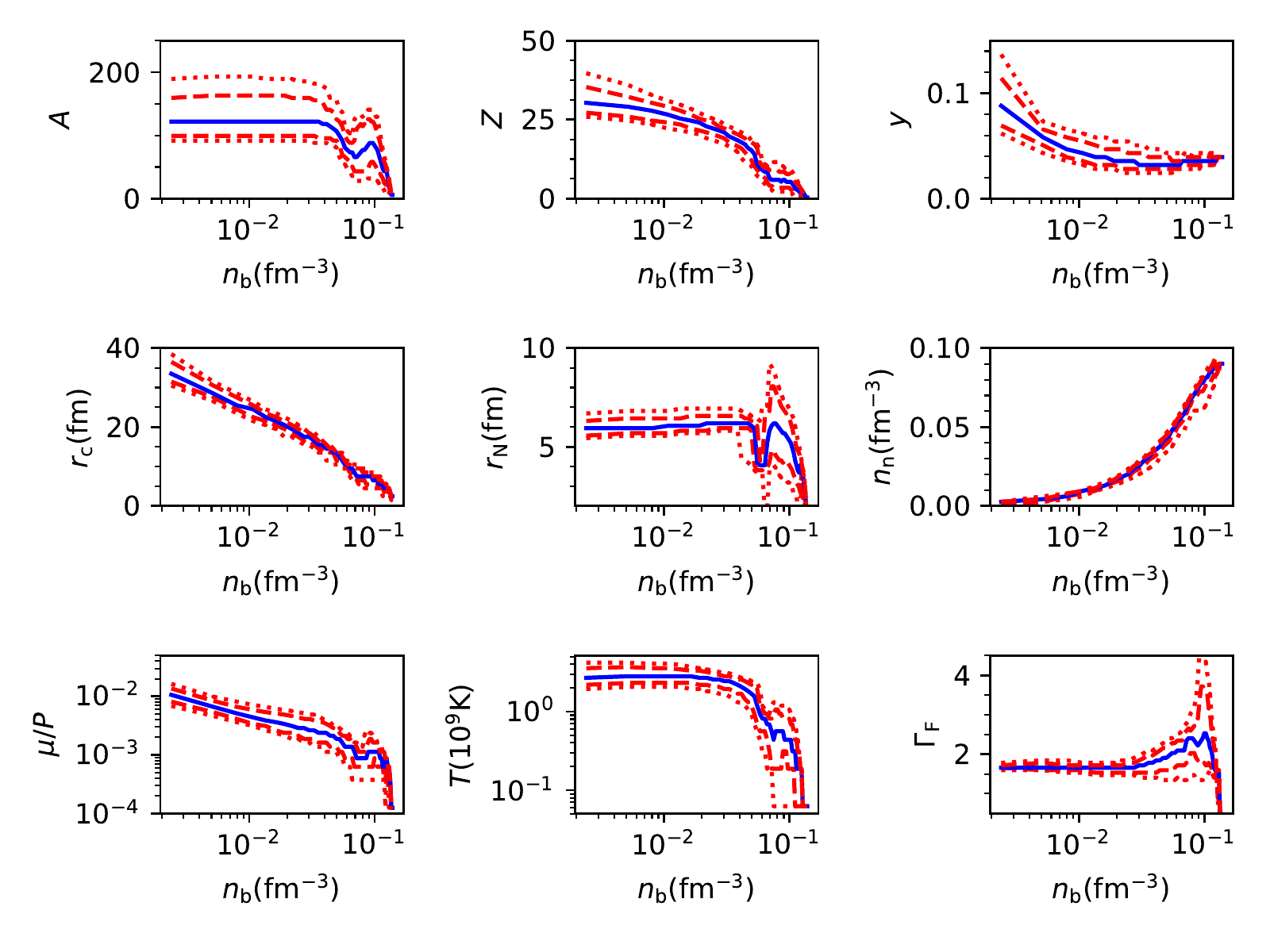}
    \caption{Same as Fig~28 but for the PNM priors ensemble.}
    \label{fig:29}
\vspace{11pt}
\vspace{11pt}
\vspace{11pt}
\end{figure}

\subsection{Inner crust composition}

We have been focusing on the location of the crust-core boundary and the transition between spherical and pasta nuclei, but of course the composition of the whole of the inner crust is an important ingredient for modeling diverse physics such as crust cooling, crustal oscillations, glitches and crustal magnetic fields.

In Figures~28 and~29 we plot 9 parameters associated with the composition of the crust for our uniform priors (Fig~28) and PNM priors (Fig~29) respectively. Alongside the mass number and atomic number of nuclei (or nuclear pasta structures) in the WS cell $A$ and $Z$, the average proton fraction $x$, the WS cell and nuclear sizes $r_{\rm c}$ and $r_{\rm N}$, the density of the neutron gas $n_{\rm n}$, we plot the shear modulus calculated according to the fit to molecular dynamics simulations \citep{Strohmayer:1991aa,Chugunov:2010aa}

\begin{equation} \label{eq:shear_mod}
	\mu = 0.1106 \left(\frac{4\pi}{3}\right)^{1/3} A^{-4/3} n_{\rm b}^{4/3} (1-X_{\rm n})^{4/3} (Ze)^2,
\end{equation}

\noindent the melting temperature of the crust \citep{Chamel:2008aa},

\begin{equation} \label{eq:melting_temp}
T_{\rm m} = \frac{(Ze)^2}{175 k_{\rm B} r_{\rm C}}
\end{equation}

\noindent and the frozen-composition adiabatic index

\begin{equation}
\Gamma_{\rm F} = \frac{n_{\rm b}}{P} \frac{dP}{dn_{\rm b}}\bigg|_{\rm constant \; composition} =  \frac{n_{\rm b}}{P} \bigg[\frac{dP_n}{dn_{\rm n}} +  x \frac{dP_e}{dn_{\rm e}}  \bigg]
\end{equation}

\noindent which is a useful ingredient in the calculation of crust oscillations \citep{Haensel:2002aa}. In the above equations, $X_{\rm n}$ is the density fraction of dripped neutrons $X_{\rm n} = (1-v)n_{\rm n}/n_{\rm b}$ where $v$ is the volume fraction of the neutron gas.

In the plots, the blue lines show the median values and the dashed and dotted red lines the bounds of the 68\% and 95\% credible ranges respectively. The difference in predictions for the composition between the two ensembles of crust models shows up in the higher density region where the increasingly dense neutron gas and isospin-asymmetric nuclei are influenced more by the symmetry energy. The range of Wigner-Seitz cell size $r_{\rm c}$ and the global proton fraction $x$ is significantly larger throughout the crust for the uniform priors. The lower bound of the neutron gas density and upper bound on the adiabatic index is significantly more constrained by PNM priors, but the two ensembles of models predict similar ranges for $A$, $Z$, $r_{\rm N}$, the shear modulus and melting temperature. In our previous extensive study of the CLDM, we presented absolute limits rather than statistical properties of predictions for many of these quantities. There, our baseline models were those consistent with unitary gas predictions for the low density PNM, and give results that fall within the ranges obtained here \citep{Newton:2013sp}. 

In \citet{Tews:2017aa}, using crust models based on EOSs that were consistent with chiral-EFT calculations of PNM, ranges for $r_{\rm c}$, $r_{\rm N}$ and $A$ were presented which are broadly consistent with our results. Note when comparing our results that \citet{Tews:2017aa} did not include nuclear pasta phases, which substantially alters, for example, the behavior of the mass number of nuclei (pasta structures with smaller mass number in the WS cell are preferred to large spherical nuclei near the crust-core transition.)



\section{Discussion and Conclusions}

We have presented two different ensembles of neutron star crust models designed for use in inferring astrophysical or nuclear model parameters from observational and experimental data, and conducted a rigorous statistical analysis of their properties. Both ensembles are parameterized by the first three density-expansion coefficients of the symmetry energy at saturation density, $J$, $L$ and $K_{\rm sym}$. The first ensemble takes as its input parameter set a uniform distribution over $J$, $L$ and $K_{\rm sym}$ (uniform priors). The second ensemble is a non-uniform distribution determined by theoretical considerations of the pure neutron matter EOS coupled with constraints from chiral-EFT calculations (PNM priors). The input distributions are used to create ensembles of extended Skyrme energy-density functional (EDF) parameterizations, which are used in both a compressible liquid drop model (CLDM) to calculate the crust composition, and mean field calculations of neutron skins.

One of our aims is to take steps towards characterizing all the ways modeling uncertainty arises and increasing the internal model consistency required to confidently infer crust constraints from neutron skin measurements and vice versa. We do this by using the same EDFs to calculate crust and finite nuclear properties and by fitting the surface parameters of the CLDM to 3D Hartree-Fock calculations of inner crust nuclei immersed in a neutron gas over a density range of 0.025-0.04fm$^{-3}$ and a range of proton fractions from 0.5 down to 0.1. We find in particular that the results are best fit with a value of the parameter that controls the magnitude of the surface tension in neutron rich matter, $p$, equal to 3.8, higher than found from fits to nuclear masses and corresponding to a smaller surface tension in extremely isospin-asymmetric matter. This value is consistent with another recent study \citep{Jose-Furtado:2020vb}. This still leaves us with a number of places where the model is inconsistent or uncertain beyond the symmetry energy parameters, which we summarize in the caveats below.

The ensembles are filtered to remove models that do not predict stable neutron star crusts due to the supporting pure neutron pressure becoming negative. Our uniform ensemble consists of 558 models and our PNM ensemble consists of 658. When we analyze our results, we use the bootstrap method to estimate the sampling errors.


In table~2 we summarize our predictions for the crust parameters that locate the onset of nuclear pasta and the crust-core boundary. The pressures at a given layer in the crust determine the column depth and mass of the crust down to that layer. The chemical potentials determine the depth of the layer. Given these quantities, the relative mass and thickness of the pasta layers can be calculated. These quantities are essential inputs into the modeling of astrophysical observables  such as crust cooling, pulsar glitches, crust oscillations and shattering, magnetic field evolution and crust-core coupling \citep{Newton:2014pb}.
 
We give the median values and the 95\% ranges (68\% ranges in parentheses). The median values of the parameters for the two ensembles are very similar. Since these are prior probability distributions, making use of minimal physical information other than very conservative bounds on the range of PNM EOS, the ranges predicted for crust properties are quite wide. For example, the PNM priors for $P_{\rm cc}$ have a 95\% range of 0.21-0.76 MeV fm$^{-3}$, the wide range indicative of the. The more localized ranges of 0.2-0.65 MeV fm$^{-3}$ \citep{Lattimer:2007aa}, 0.4-0.6 MeV fm$^{-3}$ \citep{Lattimer:2013aa} and 0.41-0.47 MeV fm$^{-3}$ \citep{Hebeler:2013aa}, which incorporate more information from nuclear experiment and pure neutron matter calculations, are nevertheless consistent with our ranges and have medians close to ours. 

Let us compare our results to the other recent study of probability distributions of crust models \citep{Carreau:2019aa}, who used very similar priors. Our 68\% range for uniform priors gives $P_{\rm cc}$=0.26-85 MeV fm$^{-3}$ compared to their 1-$\sigma$ range of -0.084 - 0.768 MeV fm$^{-3}$. Our PNM priors give a 68\% range of 0.33-0.66 MeV fm$^{-3}$, compared to their 1-$\sigma$ range of 0.24-0.48 MeV fm$^{-3}$ using a distribution constrained by PNM calculations. Two factors likely account for the generally higher range of crust-core transition pressures we obtain. Firstly, their range accommodates negative pressure, whereas our filtering of models that predict unstable crusts restricts us to positive pressures. Secondly, they restrict themselves to $p\leq3.5$, lower than our best fit value of 3.8. Lower $p$ corresponds to larger surface tension at low proton fractions, thus making is energetically favorable to transition to uniform matter at lower densities. Determining this surface parameter by fitting to properties of nuclei in neutron rich matter turns out to be very important.

We present the first predictions for the prior probability distributions of the nuclear pasta transition pressure, density, chemical potential, proton fraction and the relative thickness and mass of pasta. The 95\% ranges for relative the thickness of pasta is 17-43\% for uniform priors and 20-33\% for PNM priors, with median values around 30\%. The 95\% ranges for the relative mass of pasta is 52-70\% for uniform priors and 59-66\%, for PNM priors, with median values of 62\%. This is consistent with \citet{Lorenz:1993aa} who, with just one model, found 50\% of the mass was pasta. These values are larger than those obtained in \citep{Grill:2014vp} using relativistic mean field models in a Thomas-Fermi framework by a factor of around 3. They calculate the surface thickness exctly by integrating the TOV equations. It likely the discrepancy has its origins in the surface energy, highlighting the need to investigate the model uncertainties in the surface energy more. It is also consistent with the ranges of \citet{Newton:2013sp}. It is thus a robust prediction of our models that nuclear pasta accounts for over 50\% of the crust by mass, and one quarter to one third of the thickness of the crust. 

We put our ensembles of models to work examining the relationships between nuclear and neutron star parameters. Quantifying the strength of their correlations allows us to determine the nuclear observables that contain information about given crust properties and vice versa. As an example of a nuclear observable, we perform Hartree-Fock calculations of the neutron skins of \textsuperscript{48}Ca and \textsuperscript{208}Pb, using the same sets of Skyrme models as for the crust models \citep{Newton:2020aa}. We characterize the strength of the correlations using the maximal information coefficient (MIC) which accounts for non-linear, non-monotonic relationships.

The column depth (or equivalently the depth by mass co-ordinate) at which the pasta phases are first found and the column depth of the crust-core transition are determined by the pressures $P_{\rm p}$ and $P_{\rm cc}$. For uniform priors these are most strongly correlated with the slope of the symmetry energy $L$, and the neutron skins \textsuperscript{48}Pb and \textsuperscript{208}Pb. 

For the PNM priors - incorporating the information about pure neutron matter from chiral-EFT calculations - the pressures are most strongly correlated with symmetry energy at 0.1 fm$^{-3}$. This is closely related to the symmetry coefficient of finite nuclei in the droplet model, and can be constrained by mass fits \citep{Lattimer:2013aa} and giant dipole resonances \citep{Trippa:2008aa}.

\begin{table*}[!t]
\caption{\label{tab:table2} 95\% credible ranges (68\% ranges in parentheses) for parameters that locate the crust-core transition, the upper boundary of the pasta layers, and the relative amount of pasta in the crust.}
\begin{ruledtabular}
\setlength{\extrarowheight}{5pt}
\begin{tabular}{ccc}
 &Uniform priors&PNM priors \\
\hline

$n_{\rm cc}$ (fm$^{-3}$)& 0.096$\substack{+0.032 (0.018) \\ -0.020 (0.013)}$ & 0.092$\substack{+0.036 (0.018) \\ -0.015 (0.010)}$ \\ 

$P_{\rm cc}$ (MeV fm$^{-3}$) & 0.46$\substack{+0.88 (0.39) \\ -0.32 (0.20)}$ & 0.49$\substack{+0.27 (0.17) \\ -0.28 (0.16)}$ \\

$\mu_{\rm cc}$ (MeV)& 14.5$\substack{+8.7 (4.2) \\ -8.0 (5.0)}$ & 14.7$\substack{+4.7 (2.7) \\ -5.0 (3.1)}$ \\

$y_{\rm cc}$& 0.032$\substack{+0.020 (0.012) \\ -0.017 (0.011)}$ & 0.033$\substack{+0.008 (0.004) \\ -0.006 (0.004)}$ \\

$n_{\rm p}$ (fm$^{-3}$)& 0.054$\substack{+0.016 (0.009) \\ -0.027 (0.022)}$ & 0.053$\substack{+0.0049 (0.0032) \\ -0.015 (0.006)}$ \\

$P_{\rm p}$ (MeV fm$^{-3}$)& 0.17$\substack{+0.35 (0.16) \\ -0.121 (0.082)}$ & 0.19$\substack{+0.11 (0.07) \\ -0.11 (0.07)}$ \\

$\mu_{\rm p}$ (MeV)& 10.3$\substack{+4.4 (2.9) \\ -5.9 (3.7)}$ & 10.3$\substack{+2.7 (1.6) \\ -2.8 (1.7)}$ \\

$y_{\rm p}$& 0.015$\substack{+0.005(0.003) \\ -0.007 (0.004)}$ & 0.015$\substack{+0.002 (0.001) \\ -0.002 (0.001)}$ \\

$\Delta M_{\rm pasta}/\Delta M_{\rm crust}$& 0.62$\substack{+0.08 (0.03) \\ -0.10(0.04)}$ & 0.62$\substack{+0.03 (0.02) \\ -0.04 (0.03)}$ \\

$\Delta R_{\rm pasta}/\Delta R_{\rm crust}$& 0.30$\substack{+0.13 (0.08) \\ -0.013 (0.08)}$ & 0.29$\substack{+0.04 (0.02) \\ -0.09 (0.04)}$ \\

\end{tabular}
\end{ruledtabular}
\end{table*}

Taken together, $P_{\rm p}$ and $P_{\rm cc}$ determine the relative mass of the pasta layers, $\Delta M_{\rm p}/\Delta M_{\rm cc}$. Uniform priors show no strong correlation with any of the nuclear parameters, whereas PNM priors show strong correlations with neutron skins and the slope of the symmetry energy at 0.1 fm$^{-3}$. Thus determining the mass fraction of pasta requires combining information about the pure neutron matter EOS with, for example, nuclear experimental measurements of neutron skins.

The depth at which the pasta phases are first found and the depth of the crust-core transition are determined by the chemical potentials $\mu_{\rm p}$ and $\mu_{\rm cc}$. Uniform priors show strong correlations between $\mu_{\rm p}$ and $\mu_{\rm cc}$ and the symmetry energy at saturation density $J$ and at 0.1 fm$^{-3}$, $J_{01}$. The PNM ensemble shows strong correlation between $\mu_{\rm p}$ and $J$, and between $\mu_{\rm cc}$ and $J_{01}$. 

Taken together, $\mu_{\rm p}$ and $\mu_{\rm cc}$ determine the relative thickness of the pasta layers, $\Delta R_{\rm p}/\Delta R_{\rm cc}$. For both ensembles, this is most strongly correlated with \textsuperscript{48}Pb and \textsuperscript{208}Pb and the slope of the symmetry energy at 0.1 fm$^{-3}$. For the uniform priors, it is also correlated with the slope and curvature of the symmetry energy at saturation density $L$ and $K_{\rm sym}$; for the PNM priors, there is also a correlation with the symmetry energy at 0.1 fm$^{-3}$.

Finally, uniform priors show strong correlations between the symmetry energy at 0.1 fm$^{-3}$ and the proton fraction at the crust-core transition and the onset of pasta. PNM priors show a strong correlation between the symmetry energy at 0.1 fm$^{-3}$ and the proton fraction at the onset of pasta only.

We can conclude that, neutron skin measurements effectively provide information about the relative mass of pasta in the crust and the column depths of pasta and the crust-core boundary, while nuclear observables sensitive to the symmetry energy at 0.1 fm$^{-3}$ - giant dipole resonances and mass fits - provide information about the proton fractions and depths of the pasta and crust-core transition. When combined with information about the PNM EoS, they also constrain the relative thickness of the pasta layer of the crust. 

Some studies have shown a non-monotonic relationship between $P_{\rm cc}$ and the neutron skin of \textsuperscript{208}Pb for certain sequences of relativistic EDFs \citep{Fattoyev:2010aa, Pais:2016aa}. Our results confirm that this relationship is an expression of the path traced out in symmetry energy parameter space by the sequence of models used. We show that, in fact, for uniform priors (uniformly ranging over $L$, $J$ and $K_{\rm sym}$) there is a positive correlation between $P_{\rm cc}$ and the neutron skin of \textsuperscript{208}Pb. This correlation weakens for the PNM prior distribution.

Finally, although the crust-core transition density is of only secondary importance to the determination of crust properties, we point out that the strongest correlation involving $n_{\rm cc}$ is with $K_{sym}$ at half saturation density.

\subsection{Model uncertainties}

Here we discuss modeling uncertainties that arise because of parameter uncertainties or inconsistencies, and discuss the likely impact on our results. 

\revision{There are a number of inconsistencies in the modeling of the surface energy in our calculations that we must be aware of. Although we have argued that evidence points towards the effect of these model uncertainties being small, more work needs to be done to investigate and eliminate them. As part of our effort to assess these uncertainties, we compared our CLDM results with the more efficient, approximate method of calculating the crust-core transition depth by locating the nuclear matter spinodal. In agreement with many other studies, we find that the CLDM agrees with the thermodynamic spinodal method, while the dynamic spinodal predicts a transition systematically lower in density by 0.01fm$^{-3}$. We caution, however, that using this method for a particular nuclear matter model while ignoring the predictions for inner crust composition leads to including crust-core transition properties for models which do not predict a stable crust, which can unphysically skew the resulting distribution of crust properties. We now outline the main sources of uncertainty in our model.}

\begin{itemize}
    \item \revision{When adjusting the Skyrme parameters to systematically survey $J$, $L$ and $K_{\rm sym}$ space, we do not refit the remaining parameters, particularly those that control the gradient terms $t_1$,$t_2$,$x_1$,$x_2$. Such a refit would introduce extra correlation between the Skyrme parameters that might affect the correlations between, for example, the neutron skins and the bulk matter properties. There is some evidence that the neutron skins are relatively insensitive to the gradient terms \citep{Chen:2010aa,Zhang:2013uv}, and that refits do not qualitatively alter the correlations between neutron skins and crust-core transition properties \citep{Fattoyev:2010aa,Pais:2016aa}, but nonetheless we are in the process of conducting those refits and will assess their effect in an upcoming work. Our results also allow direct comparison with constraints extracted from neutron skin data using similar ensembles of Skyrme models \citep{Chen:2010aa,Newton:2020aa}.}
    
    \item \revision{The surface energy in the CLDM, although fit to EDF caclulations, has a functional form that is not derived from the Skyrme EDFs directly. Of the parameters that control the isospin dependence, $c$ and $p$, only $c$ depends on the symmetry energy. The rest of the dependence on the symmetry energy parameters comes from the functional dependence on the proton fraction at beta-equilibrium. Although this is consistent with our fits to 3DHF crust calculations, their computationally intensive nature means only three EDFs were used in the fit. This may affect the values and correlations we obtain, although the value of $p$ we obtain is consistent with another recent study \citep{Jose-Furtado:2020vb}. There is some evidence that this will not qualitatively alter the correlations we obtain: comparison of the CLDM crust-core transition properties with the spinodal methods - using different surface energy treatments - shows similar correlations between neutron skins, symmetry energy parameters and crust-core transition properties.}
    
    \item \revision{The dynamic spinodal method uses the EDF gradient terms, which are held constant as we do not refit the Skyrme parameters $t_1, x_1, t_2, x_2$. If the changes to the gradient terms induced by refitting the Skyrmes affect the dynamic spinodal transition density in a different way to the neutron skins, the correlations we obtain could change. There is evidence that this is indeed the case \citep{Antic:2019tk}. However microscopic calculations such as 3DHF appear to give results closer to the thermodynamic spinodal \citep{Pais:2014uj} like the CLDM. Dependence of the surface energy on bulk nuclear properties in the dynamic spinodal method does still enter through the derivatives of the chemical potentials in equation~16. We also find that the correlations do not qualitatively change comparing the thermodynamic and dynamic spinodal methods and the CLDM.}
    
    \item \revision{The dynamic spinodal predicts a crust-core transition at a lower depth (for example $n_{\rm cc}$ lower by $\sim 0.1$fm$^{-3}$). Decreasing $p$ has the same effect, lowering $n_{\rm cc}$ by up to $\sim 0.2$fm$^{-3}$ \citep{Newton:2013sp}. At the same time, changing $p$ (or $c$) has very little effect on the density at which pasta appears. Therefore the overall effect of these changes would be to substantially reduce the amount of pasta predicted \citep{Newton:2013sp}. However, the amount of pasta predicted by the CLDM with $p=3.8$ here is consistent with detailed 3D microscopic calculations of the pasta phases \citep{Newton:2021vk}. We also note that there is not necessarily a straightforward relationship between the EDF gradient terms determined from fits to nuclei and surface energy in very neutron rich matter where the surface is very diffuse and the external neutron density approaches the density inside nuclei.}
\end{itemize}

\revision{Although we argue that the uncertainties are unlikely to qualitatively change the relationships between neutron skin, symmetry energy and neutron star properties, this must be tested in the future. More work needs to be done to disentangle the relationships between neutron skins, the size of the gradient terms in the EDF, and the surface energy in very neutron rich matter near the crust-core boundary where the difference in density between neutron gas and nuclear cluster is small and the surface itself is very diffuse.}

\revision{Although the use of an \emph{extended} Skyrme EDF allows us to explore a far wider range of density dependencies of the PNM EOS, it still constitutes a choice of model. The fourth term in the symmetry energy expansion at saturation density, $Q_{\rm sym}$ can also influence the crust-core transition pressure and hence the mass of the crust \citep{Carreau:2019aa,Antic:2019tk}. We do not explore the $Q_{\rm sym}$ dependence independently.} 

 
\revision{We have made some steps towards a more consistent description of neutron skins and neutron star crusts - including the nuclear pasta phases - necessary for consistent propagation of experimental results into the astrophysical domain and vice versa. In particular, we have shown how such modeling allows examination of correlations between neutron skins, symmetry energy parameters and crust properties including the extend of the nuclear pasta phases in a statistically meaningful way.  There is still some way to go, but we have tried to highlight all the ways inconsistency and model uncertainty enters into the model, and assess how important they are. It is clear that the surface energy treatment is the dominant modeling uncertainty, and needs to be addressed in future work.}

\acknowledgements
LEB and WGN acknowledge support from NASA grant 80NSSC18K1019. This work benefited from discussions at the 2018 Frontiers in Nuclear Astrophysics Conference supported by the National Science Foundation under Grant No. PHY-1430152 (JINA Center for the Evolution of the Elements).

\bibliographystyle{aasjournal}
\bibliography{paper}

\end{document}